\documentclass[a4paper,fleqn]{cas-sc}

\usepackage[sort&compress,numbers]{natbib}   
\usepackage{color,soul} 		% Highlighting
\usepackage{caption} 			% Subfigures
\usepackage{subcaption} 		% Subfigures
\usepackage[toc,page]{appendix} % Appendix referencing
\usepackage[ansinew]{inputenc} 	% Symbols
\usepackage{mathtools,nccmath}  % Alternative for subequations

\usepackage{tabularx}           % Use to make a table fit the page width
\usepackage{placeins}           % Introduces a way to clear all floats without a page break

\usepackage{enumitem}
\newlist{steps}{enumerate}{1}
\setlist[steps, 1]{label = Step \arabic*}

\usepackage{enumitem}
\newlist{ASsteps}{enumerate}{1}
\setlist[ASsteps, 1]{label = AS-Step \arabic*}

\usepackage[nameinlink]{cleveref}   

\crefname{appsec}{appendix}{appendices}
\Crefname{appsec}{Appendix}{Appendices}

\usepackage{caption,graphicx,newfloat}
\DeclareCaptionType{Box}
\crefname{Box}{box}{boxes} % <-- new
\Crefname{Box}{Box}{Boxes} % <-- new

\begin{document}

\let\WriteBookmarks\relax
\def\floatpagepagefraction{1}
\def\textpagefraction{.001}
\shorttitle{Design loads for wave impacts}
\shortauthors{van Essen, Seyffert}

% Title
\title [mode = title]{Design loads for wave impacts -- introducing the \textit{Probabilistic Adaptive Screening} (PAS) method for predicting extreme non-linear loads on maritime structures}

% Authors
\author[1,2]{Sanne M. {van Essen}}[orcid=0000-0001-8239-0724] 
\ead{s.m.vanessen@tudelft.nl / s.v.essen@marin.nl}
\cormark[1]

\author[1]{Harleigh C. Seyffert}[orcid=0000-0003-0323-2096]
\ead{h.c.seyffert@tudelft.nl}

\cortext[cor1]{Corresponding author}

\address[1]{Department of Maritime \& Transport Technology, Delft University of Technology (TU Delft), Delft, The Netherlands}
\address[2]{Ships Department, Maritime Research Institute Netherlands (MARIN), Wageningen, The Netherlands}

%%%%%%%%%%%%%%%%%%%%%%%%%%%%%%%%%%%%%%%%%%%%%%%%%%%%%%%%%%%%%%%%%%%%%%
\begin{abstract}
Wave impact loads on maritime structures can cause casualties, damage, pollution of the sea and operational delays. Consequently, their extreme values should be accounted for in the design of these structures. However, this is challenging, as wave impact events are both rare and highly complex, requiring both high-fidelity simulations and long analysis durations to reliably quantify the associated design loads. Moreover, existing extreme value prediction methods are neither specifically developed nor adequately validated for wave impact phenomena. We therefore introduce the new \textit{Probabilistic Adaptive Screening} (PAS) method for predicting extreme non-linear loads on maritime structures. The method integrates copula-based statistical dependence modelling with multi-fidelity screening and adaptive sampling. This framework enables efficient extreme value prediction by statistically mapping low-fidelity indicator variables to high-fidelity impact loads. The method allows for efficient linear potential flow indicators to be used in the low-fidelity stage, even for strongly non-linear load cases. The statistical framework of the method is validated against four weakly and strongly non-linear test cases, including non-linear waves, ship vertical bending moments, green water impact loads, and slamming loads. It is concluded that PAS with optimal settings accurately estimates both the short-term distributions and extreme values in these test cases, with most probable maximum (MPM) values within 2-15\% of the reference brute-force Monte-Carlo Simulation (MCS) results. In addition, PAS achieves this performance very efficiently, requiring in the order of 1-3\% of the high-fidelity simulation time needed for conventional MCS. These results demonstrate that PAS can reliably reproduce the statistics of both weakly and strongly non-linear extreme load problems, while significantly reducing the associated computational cost compared to MCS. 
\end{abstract}

% While PAS is similarly accurate and slightly less efficient than its predecessor AS for the simpler cases (non-linear waves and vertical bending moments), it proves significantly more reliable for the more complex cases (green water and slamming loads).

\begin{keywords}
Extreme value prediction \sep Design loads \sep Wave impacts \sep Ships \sep Offshore structures \sep Coastal structures \sep Non-linear loads \sep Probabilistic design \sep Reliability \sep Green water \sep Slamming 
\end{keywords}

\maketitle

%%%%%%%%%%%%%%%%%%%%%%%%%%%%%%%%%%%%%%%%%%%%%%%%%%%%%%%%%%%%%%%%%%%%%%
%%%%%%%%%%%%%%%%%%%%%%%%%%%%%%%%%%%%%%%%%%%%%%%%%%%%%%%%%%%%%%%%%%%%%%
%%%%%%%%%%%%%%%%%%%%%%%%%%%%%%%%%%%%%%%%%%%%%%%%%%%%%%%%%%%%%%%%%%%%%%

\section{Introduction and background}
\label{sec:intro}

%%%%%%%%%%%%%%%
\subsection{Wave impacts}
\label{sec:introimpacts}
% Why this problem matters (in general and in complex scenarios)

\begin{figure}[t!]
	\centering
    \includegraphics[height=4.5cm]{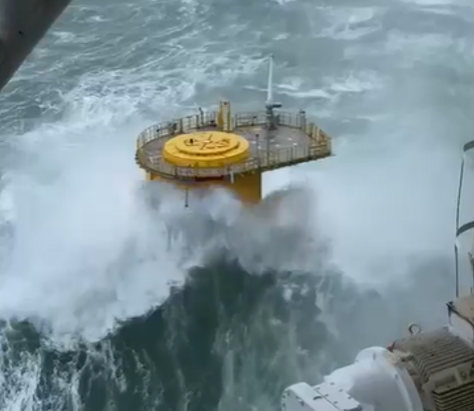} % gravity-based foundation
    \includegraphics[height=4.5cm]{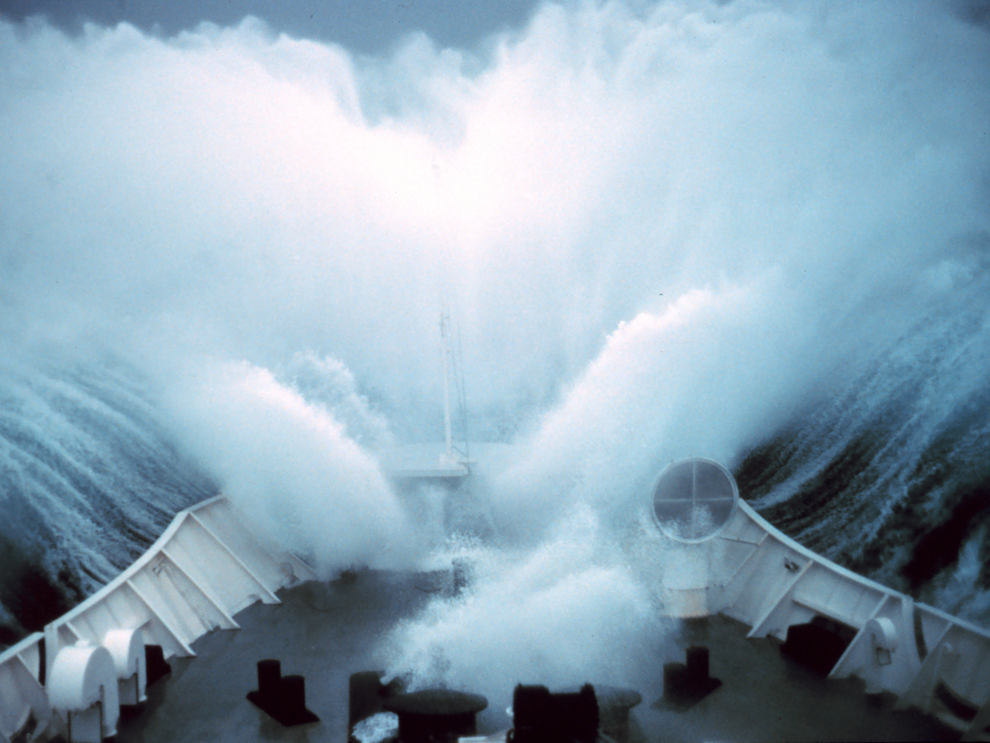} % CC BY 2.0
	\caption{Two examples of wave impacts on marine structures: a wind turbine foundation near F\'{e}camp in 2023 (left, photo: K. King) and research vessel \textit{Discoverer} on the Bering Sea in 1979 (right, photo: R. Behn / NOAA).}     
	\label{fig:pictures}
\end{figure}

A \textit{wave impact} on a ship, offshore, or coastal structure is a dynamic loading event that occurs when a water wave hits the structure, causing a rapid transfer of momentum. It may occur when such a maritime structure encounters large and steep waves, experiences large wave-induced motions, or some combination of both (see \Cref{fig:pictures}). The resulting loads can cause significant damage, endanger crew or inhabitants, cause pollution of the sea by lost cargo, decrease performance of the structure, or (in rare occasions) endanger the structure itself. Examples include green water and slamming on ships, wave-in-deck loads on offshore structures, and impacts on wind turbines, breakwaters, dams, jetties, bridges, and other coastal infrastructure. Severe wave impact accidents are documented on sailing ships \cite{DG2001,K2018,ABCViking,GuardianMaud} and offshore structures \cite{HOELAB2022,EK2000,ZXZC2025,VAOZ2016,Havtil2025}. Wave impacts also affect the structural reliability of various coastal structures, ranging from breakwaters \cite{CLST2011,ZYMCZ2023}, offshore wind turbines \cite{HWZHSL2025}, bridges \cite{CST2009}, port infrastructure \cite{LLTL2024}, lighthouses \cite{ABDRBPD2021}, to residential buildings \cite{MECRSLK2024,SES2015}. This makes accurate prediction of the distributions and extreme values of strongly non-linear loads, such as those due to wave impacts, essential. Waves and wave-induced processes are inherently stochastic, necessitating probabilistic design approaches. Even if a high-fidelity (HF) model could perfectly predict the loads associated with a specific wave event, this information alone has limited practical value. Without knowledge of the event's probability of occurrence, it is impossible to assess the overall risk or to make reliable design decisions. In other words, the statistics of wave events and wave-induced loads are at the core of the problem. %  The aim of such methods is not to avoid all accidents, but to ensure that the probability of failure of the considered structure remains below an acceptable level.

%%%%%%%%%%%%%%%
\subsection{Requirements for accurate prediction of extreme wave impact loads}
\label{sec:requirements}
As also discussed in \cite{VS2025}, obtaining extreme values for wave impact loads is particularly challenging because such events are both rare and highly non-linear. Accurate and converged statistics of rare events require long-duration analyses, while HF simulations are needed to resolve the detailed physics. HF models in this context can include Computational Fluid Dynamics (CFD) simulations or basin experiments. HF models in this context include CFD simulations and basin experiments. Although CFD can increasingly reproduce wave impact loads for individual events (see \Cref{sec:LFHF}), simulating durations in the order of the lifetime of the ship remains computationally infeasible for practical design applications. This motivates the use of dedicated extreme value prediction methods (EVPM), which estimate design loads without assessing very long time series with an HF model. Here, EVPM refers to any method that can obtain extreme values of strongly non-linear responses or loads. By combining probabilistic methods with low- and high-fidelity (LF and HF) simulations, EVPMs focus computational effort on events that matter most for design. While the statistics of wave events remain a key aspect of the problem, the choice of LF and HF models and the physical processes they represent is also crucial, as these decisions directly influence the accuracy of the predicted extreme loads. In this way, an EVPM connects the stochastic nature of waves, model selection, and design considerations. As concluded from \cite{VS2023}, an EVPM for strongly non-linear loads should then:

\begin{enumerate}[nosep]
\item Integrate multiple fidelity levels, combining suitable HF models for accurate physical modelling with LF models to efficiently capture rare-event statistics;
\item Account for the practical limitation that only a few wave events can be simulated with an HF model during the design stage; 
\item Allow for some wave non-linearity in the LF model, since linear models neglect important factors such as wave breaking and higher-order interactions that drive wave impact complexity;
\item Preserve consistent LF and HF time profiles, because peak loads are not always most relevant for structural dynamics (rise times or impulses can be more critical); 
\item Have a flexible framework that allows case-specific LF and HF models while preserving the overall methodology, enhancing general applicability;
\item Provide accurate extreme values for strongly non-linear loads, validated using real-world cases;
\item Be efficient and practical for use in design. 
\end{enumerate}

%%%%%%%%%%%%%%% Nomenclature box
\fboxsep1mm
\begin{Box}[t!]
\centering
    \fbox{\scriptsize\noindent
    \begin{centering}
    \begin{tabular}{ll|ll|ll} 
        ${x}'$ / ${x}''$  & LF / HF version of variable $x$  & RWE  & Relative Wave Elevation  & $n_w$  & \# wave encounters in MCS  \\ 
        $\widehat{x}$  & MPM of variable $x$  & VBM & Vertical Bending Moment & $N$  & \# seeds in MCS  \\ 
        AS & Adaptive Screening method & $C$ & Wave crest height & $P_{\text{exp}}$  & PoE level corr. to $T_{\text{exp}}$ and $T_{p,e}$  \\ 
        CFD  & Computational Fluid Dynamics   & $C(u,v)$ & Fitted copula model & $R_{19}$ / $R_{\text{bow}}$ & RWE at station 19 or bow \\ 
        EVPM  & Extreme Value Prediction Method   & $\left[\mathbf{d}^*,\mathbf{h}^*\right]$ & Predicted HF distrib. & $T_{\text{exp}}$  & Target / exposure duration \\ 
        HF  & High-Fidelity   & $\left[\mathbf{d}^{\text{cop}}_{H,b},\mathbf{h}_b^{\text{cop}}\right]$ & Distrib. HF copula draw $b$ & $T_1$ / $T_{1,e}$  & Mean wave (encounter) period  \\ 
        LF  & Low-Fidelity   & $\left[\mathbf{d}^{\text{mcs}}_{L},\mathbf{l}^{\text{mcs}}\right]$ & Distrib. all LF samples in MCS & $T_p$ / $T_{p,e}$  & Peak wave (encounter) period  \\ 
        MCS  & Monte-Carlo Simulation   & $\left[\mathbf{d}^{\text{sel}}_{L},\mathbf{l}^{\text{sel}}\right]$ & Distrib. present LF samples & $V_{s}$  & Forward speed of the ship \\ 
        MD & Acquisition function & $F_s$  & Bow-flare slamming force & $V_{hog}$ / $V_{sag}$  & Hogging / sagging VBM \\ 
        MVPD  & Acquisition function & $F_x$  & Green water force & $V_{r,19}$  & RVV at station 19  \\ 
        MPM & Most Probable Maximum   & $\mathbf{h}^{\text{sel}}$ & Present HF samples & $z$  & \# bootstrapping copula draws \\ 
        PAS & Probabilistic Adaptive Screening & $l_{0,\text{init}}$ / $l_{0}$ & Input / updated LF threshold  & $\epsilon_1$ / $\epsilon_2$  & Case-specific stopping criteria limits \\ 
        POT & Peak-over-threshold & $m$  & \# present HF samples  & $\mu$  & Wave heading w.r.t. structure \\ 
        RVV & Relative Rise Velocity & $n$  & \# LF indicator peaks in MCS  & $\omega$ / $\omega_e$ & Wave (encounter) frequency \\ 
    \end{tabular}
    \end{centering}}
    \caption{Most important nomenclature, where distrib.~=~distribution.
    \label{box:nomen}}
\end{Box}

% RRV = relative rise velocity
% All indicator symbols
% All HF symbols
% $T_1$ = mean wave period

%%%%%%%%%%%%%%%
\subsection{Low- and high-fidelity models}
\label{sec:LFHF}

Multi-fidelity EVPMs are only as reliable as the underlying physical modelling and the selected combination of LF and HF models. The HF model must resolve all relevant physical processes with sufficient numerical or experimental fidelity, including appropriate spatial and temporal resolution and geometric representation. For wave impact problems, the available HF modelling approaches are currently CFD simulations or basin experiments. Historically, experiments were the only reliable source of wave impact load data and they remain the primary reference in most existing standards and guidelines \cite{DNV2019,DNVGL-CG-0130,OTG14,BV2015,BV2019,ABS2020,ABS2021,ITTC2017Ev}. However, several studies have demonstrated that CFD can accurately replicate experimental breaking wave events (e.g., \cite{DSHTKB2020}) and wave impact loads when both wave kinematics and ship motions are properly represented (e.g., \cite{BH2018,BHV2020,PEtAl2022,GJL2023,BHD2015,SH2023,VMSHKSG2021}). While numerous HF modelling choices and numerical settings exist, a detailed discussion is beyond the scope of the present paper. 

As discussed in the next section, many existing EVPMs also rely on an LF \textit{indicator} variable that is used to identify critical events for subsequent HF evaluation. This LF variable may, but does not need to, represent the same physical quantity as the HF response at a lower resolution. The first requirement to define an LF indicator is that it exhibits similar \textit{order statistics} (ranked values arranged in ascending or descending order) to those of the target HF variable. While this requirement is more easily satisfied when LF and HF models represent the same variable, this study demonstrates that physically related but different indicators can be effective, particularly for strongly non-linear wave impact problems. A second essential requirement is that the LF model should be computationally inexpensive, as it must be evaluated over durations corresponding to several times the lifetime of the structure. This does not mean that the LF model has to be linear; it can be linear or weakly non-linear, and be an analytical model, potential flow model, higher-order spectral wave model, or even a coarse-mesh CFD model. Identifying a suitable LF indicator model that satisfies both requirements is challenging and typically relies on engineering judgement and prior experience, as the suitability of an indicator cannot be assessed beforehand. This reliance on expert model selection is common to most EVPMs. Guidance is therefore usually drawn from previous studies. \cite{VS2023} provides a detailed review of suitable indicator variables for different types of wave impact loading. Most of these studies employ either peak values of (relative) wave elevation or related kinematic measures obtained from linear or weakly non-linear potential flow simulations, or wave impact loads computed using coarse-mesh CFD as LF indicators. Furthermore, \cite{VMSHKSG2021} demonstrated that LF indicator results can provide effective inputs for HF CFD simulations.

%The present study builds on these concepts within a new multi-fidelity prediction framework described in the following sections.

%%%%%%%%%%%%%%%
\subsection{Existing EVPMs}
\label{sec:existing}
Extreme value prediction is similar to reliability analysis; the former focuses on extreme loads, while the latter considers failure probability $P_f = P(R-S \leq 0)$, where $R$ is resistance and $S$ loads. Since they are related, we review both, but structural resistance is not considered further. A comprehensive literature review of maritime EVPMs is given in \cite{VS2023}, and \cite{VS2025} added recent literature on both EVPMs and reliability methods. Here we summarise the key studies relevant to the present work and briefly discuss developments since the review paper. Response-conditioning methods generate critical wave events conditioned on a given response, using transfer functions, wave spectra, and phase assumptions. They produce single \cite{ABL1998} or multiple wave profiles \cite{TJH1997,TWB1998,J2009}, usually via FORM/SORM. They are efficient and event-based but mostly assume linear Gaussian waves; extensions using higher-order models \cite{TFHM2023,DLKDBD2025,DBDBB2025} are promising but have not been validated for strongly non-linear impact cases. Screening methods use LF indicator variables, as described above, to identify critical events for HF analysis by ranking waves according to their LF responses \cite{S2008,BSPFS2018,BSD2019,S2020,VMSHKSG2021}. Such methods meet most EVPM criteria and could suit strongly non-linear cases, but are associated with a non-conservative bias when the indicator is poorly chosen. This comes from assuming a monotonic relation between the order statistics of the LF and HF variables; when a non-perfect LF indicator seeks the largest magnitude event in a time record, it will invariably sometimes indicate the second, or third (or so on) largest event instead of the true extreme. This always leads to a biased underestimation of the extremes. Such a bias can be mitigated with a safety factor, but this is not ideal. Different types of rare event sampling methods also reduce the number of HF simulations compared to Monte-Carlo simulation (MCS). Examples are importance sampling \cite{TJG2022,CGMD2023,CFPO2023}, subset sampling \cite{CPS2022}, adaptive sampling, or combinations \cite{CPLZW2023,YSQD2024,CPS2025}. Adaptive sampling (or sequential sampling / Bayesian model updating / Bayesian experimental design / active learning) seems most suitable for an EVPM, as it can guide exploration towards the extremes by learning from previous iterations. Adaptive sampling uses surrogates, which can be polynomials, polynomial chaos expansion \cite{NM2024a,NM2024b}, support vector regression \cite{RMC2018,FXLL2024}, neural networks \cite{CTNCG2015,BSGT2024}, or especially Gaussian Process Regression (GPR, or Kriging) \cite{FFN2021,FSK2007,KO2000}. Applications of adaptive sampling with GPR include extreme weakly non-linear wave-induced loads on maritime structures \cite{MS2018,GEtAl2020,GS2022,GKS2023,TDEtAl2023,AJSB2024}. Adaptive sampling has to stop at convergence; criteria are studied in \cite{HANM2006,ZDF2024}. Multi-fidelity sampling for ship motions was studied by \cite{KOPS2025}. Response Surface Methods (RSM) use similar surrogates for reliability problems (see e.g., \cite{KL2015,ZFW2017,KJYN2011,WG2021,KRGV2011,SEXL2022}), with optimisation strategies discussed in \cite{MI2024a,MI2024b,LW2022,LSHH2020}. Adaptive sampling methods with GPR satisfy most EVPM criteria, but their implementations remain case-specific and untested for strongly non-linear wave-induced loads. Summarising, some existing methods meet multiple EVPM criteria, but their implementations remain case-specific and untested for strong wave impacts. Consequently, industry practice relies on standards (e.g., \cite{DNV2019,DNVGL-CG-0130,OTG14,BV2015,BV2019,ABS2020,ABS2021,ITTC2017Ev}), specifying plate thicknesses or using experiments with multiple 0.5-3~h wave seeds to estimate short-term extreme responses. 

% \hl{Does this summary answer all reviewer questions?} 
% \hl{Following reviewer 1, add some specific additional NN-surrogate applications?}

To address these limitations of the existing methods, we previously developed the Adaptive Screening (AS) method \cite{VS2025}. AS is a multi-fidelity approach for predicting extreme non-linear HF wave-induced loads, combining elements from screening, adaptive sampling and GPR. It uses an LF indicator to select critical events for HF evaluation, and builds a surrogate HF distribution from which the design loads are derived, iteratively adding HF samples until convergence. The steps of the method are briefly discussed in \Cref{app:AS}. AS was validated with good results for a weakly non-linear case (second-order waves) and a moderately non-linear case (vertical bending moments), and further tested in a pilot study of a strongly non-linear green water impact on a containership.\footnote{In \cite{VS2025}, we used the term \textit{non-linear responses} to emphasise the method's applicability to hydrodynamic motions, accelerations, green water loads etc. However, as discussed in \Cref{sec:existing}, extreme value prediction aligns with reliability analysis, where \textit{loads} is the standard term. We therefore now use \textit{loads} for all hydrodynamic responses.}. Further work \cite{VS2025prads} studied the effect of the \textit{acquisition function}, one of the elements of AS. Most data and all scripts are available in \cite{VSdata2025,VSdata2025PRADS}. The strength of AS for strongly non-linear loads lies in its flexibility, allowing a wide range of LF indicators, including non-linear ones if necessary, and in its efficiency, as it minimises the number of HF simulations. This makes it well-suited for problems with sparse datasets. However, for complex cases, the LF indicator may need to be a relatively expensive model, such as the coarse-mesh CFD used in Case 3 of \cite{VS2025}, which increases computational cost. As with other screening methods, a poorly chosen indicator can introduce a systematic non-conservative bias. Consequently, AS meets all requirements in \Cref{sec:requirements} for simple cases, but in complex problems it cannot always be both accurate and efficient. AS is used as reference method throughout this study.

Existing methods, including AS, generally do not capture the full probabilistic dependence or joint extremes of LF and HF variables, which may explain some limitations. Copula models provide a flexible framework for multivariate distributions, separating marginal behaviour from dependence. They can be used to interpolate the joint distribution learned from a limited number of samples. Further details are given in \Cref{app:copulabasics}. Copula models have recently been applied to reliability analysis of slopes in spatially variable soils with limited samples \cite{XL2026,WWZCJ2025}, systems of dependent equipment components \cite{QXSH2025,BCLJ2025}, reliability-based design optimization for correlated variables \cite{ZZYBW2026,LFFT2025}, and assessment of joint risks for breakwater overtopping \cite{MVM2024}. In extreme value analysis, copulas have been used to model joint tail distributions of multiple variables \cite{KJ2019} and the dependence between mooring line loads and sea state parameters \cite{SACM2025}. Copula models have long been used by oceanographers to capture the joint statistics of dependent environmental variables \cite{VFDKCG2024,DWLSWZ2024,FXLL2024,DV2005} and to account for spatial dependence between extreme environmental variables \cite{BGJM2020,GM2010,DPR2012}. For wave impact loading, copula models have been used to describe the joint distribution of quasi-static and impulsive components of measured deck pressures \cite{WZX2022}, wave impact maxima on a coastal bridge conditioned on rise time \cite{SC2011}, extreme stresses at multiple deck locations of a container vessel \cite{GSN2016}, interactions among parameters in experimental dam break impacts \cite{SL2024}, maxima and rise times of breakwater loads \cite{CPA2011}, and to derive the most likely wave impact load profile on a floating wind turbine based on wave conditions \cite{YXSZB2025}. Copula models have previously been integrated with multi-fidelity approaches: \cite{KNPVW2019,PCMW2016} developed multi-fidelity importance sampling, and \cite{Proppe2019} combined multi-fidelity modelling with a copula to establish a reliability model hierarchy. Copulas have also been paired with adaptive sampling: \cite{LNDH2025,MZCLGG2026} combined Bayesian model updating with MCS and copulas for reliability analysis, while \cite{BHRM2018} used a Bayesian model with a copula to capture joint annual extremes of time-varying climate variables. 

Despite their promise, copula-based approaches have not yet been combined with multi-fidelity or adaptive sampling frameworks to improve the efficiency of extreme value prediction for strongly non-linear loads such as wave impacts. The closest work \cite{PCMW2016,KNPVW2019} relies on importance sampling rather than adaptive sampling and has not been tested on real-world strongly non-linear cases. Adaptive sampling is generally considered more flexible and robust for complex tail behaviour.

%%%%%%%%%%%%%%%
\subsection{Paper objectives \& contribution}
\label{sec:objectives}

As discussed above and in \cite{VS2025}, a new EVPM is needed for wave impact loads, as serious maritime incidents still occur and existing methods are insufficiently validated for this problem. This study is also motivated by the observation that statistical aspects in maritime applications are often treated less rigorously than they could be, despite the inherently stochastic nature of waves and their responses. This work therefore focuses on the statistical framework of the EVPM rather than on LF or HF model details. We propose a new variant of AS that addresses its limitations for complex, strongly non-linear cases. Unlike classical AS, which relies heavily on LF order statistics as surrogates for HF extremes, the proposed method introduces a probabilistic coupling between LF indicators and HF loads. This reduces dependence on expensive LF models and enables the use of simpler, cheaper indicators while maintaining robust and accurate extreme-value predictions. The related objectives of this paper are to:

\begin{enumerate}[nosep]
\itemsep=0pt
\item Introduce a modification of AS, called \textit{Probabilistic Adaptive Screening} (PAS), which improves the statistical handling of extreme values of strongly non-linear loads, and complies with the requirements in \Cref{sec:requirements}.
\item Validate that the method can (more) accurately and efficiently predict short-term extreme values (than AS and MCS) for a range of realistic strongly non-linear applications. 
\item Demonstrate that the approach can be applied using practical LF indicator models derived from linear potential flow calculations, instead of relying on coarse-mesh CFD.
\end{enumerate}

The novelty of PAS lies in integrating existing method elements into a unified multi-fidelity framework. By combining screening and adaptive sampling with probabilistic LF-HF modelling, PAS enables efficient and statistically consistent prediction of extreme wave impact loads. While PAS can in principle predict both long- and short-term extremes, this work focuses on short-term extremes due to validation data availability. Overall, PAS offers a new, efficient, and accurate EVPM tailored to wave impact loads.

\Cref{sec:PAS} presents the new PAS method, and \Cref{sec:cases} introduces four case studies for its validation and comparison with AS and MCS. In addition to wave impact cases, simpler problems are included to assess robustness and illustrate performance across varying levels of non-linearity and modelling complexity. \Cref{sec:valmet} describes the performance metrics, \Cref{sec:Discussion} presents and discusses the results, and \Cref{sec:Conclusions} concludes the study.

%%%%%%%%%%%%%%%%%%%%%%%%%%%%%%%%%%%%%%%%%%%%%%%%%%%%%%%%%%%%%%%%%%%%%%
%%%%%%%%%%%%%%%%%%%%%%%%%%%%%%%%%%%%%%%%%%%%%%%%%%%%%%%%%%%%%%%%%%%%%%
%%%%%%%%%%%%%%%%%%%%%%%%%%%%%%%%%%%%%%%%%%%%%%%%%%%%%%%%%%%%%%%%%%%%%%

\begin{figure}[t!]
	\centering
	   \includegraphics[width=\linewidth]{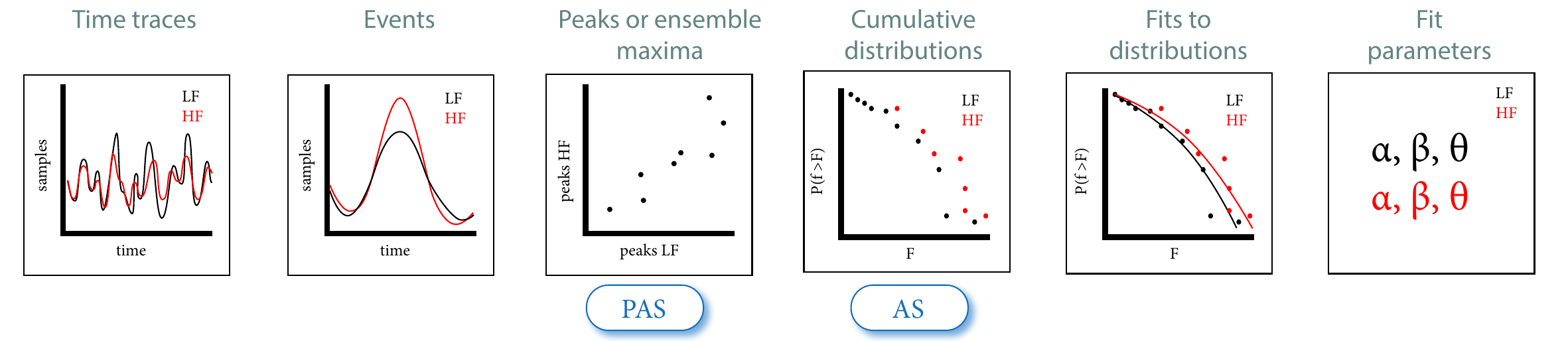}
	\caption{Some of the possible \textit{statistical levels} where multi-fidelity methods can derive or learn the relation between an LF indicator variables (black) and HF non-linear loads (red), including the location of AS and PAS. Modified from \cite{VS2025}.}
	\label{fig:plan}
\end{figure}

\section{The Probabilistic Adaptive Screening (PAS) method}
\label{sec:PAS}

Multi-fidelity extreme value methods link LF indicators and HF loads at different statistical levels (see \Cref{fig:plan}). Retaining more information (left) improves accuracy but increases complexity and simulation cost, while simplification (right) reduces cost at the risk of losing critical information. AS uses LF order statistics as surrogates for HF extremes, linking them at the level of the marginal distributions in a deterministic way. While efficient, it struggles with strongly non-linear problems, where a single LF value can be associated with a range of HF values. This makes its performance sensitive to outliers in the LF indicator, and costly LF models seem to be required for good wave impact results. The proposed \textit{Probabilistic Adaptive Screening} (PAS) method replaces the deterministic link between LF and HF marginal order statistics with a probabilistic connection between paired LF-HF peaks (as indicated on \Cref{fig:plan}). PAS retains the screening and adaptive sampling of AS, but replaces the order-statistics surrogate with a probabilistic copula model fitted to the LF-HF pairs. This captures their variability, enabling accurate HF distribution estimates from few HF samples and reducing sensitivity to the LF indicator in strongly non-linear cases. This allows for flexible dependence modelling and efficient prediction of wave impact design loads from simple linear LF indicators.

%%%%%%%%%%%
\subsection{Steps}
\label{sec:pas_steps}
\begin{figure}[t!]
    \centering
       \includegraphics[width=\linewidth]{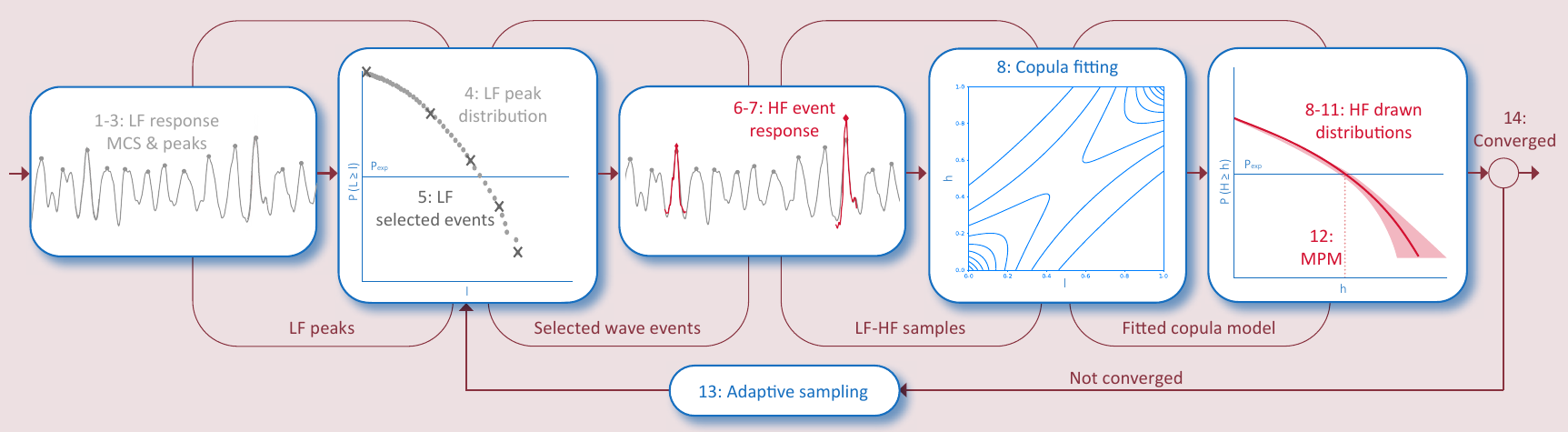}
    \caption{Schematic illustration of PAS, where the numbers roughly correspond to the method steps in \Cref{sec:pas_steps}. The left plot only shows a small part of the MCS time traces, and only a few HF samples are included in the middle and right distributions to illustrate the principle.}
    \label{fig:processPAS}
\end{figure}

The goal of PAS in the present implementation is to predict the most probable maximum (MPM) value in exposure duration $T_{\text{exp}}$ of a target HF non-linear variable. The appropriate $T_{\text{exp}}$ depends on the problem and user requirements, and is typically between 20 minutes and 3 hours (depending on the ergodicity of the operating conditions at sea). PAS uses a peak-over-threshold (POT) approach. The method consists of the steps outlined below, numbered according to the schematic overview of PAS in \Cref{fig:processPAS}. 

\begin{steps}
\itemsep=0pt
    \item\label{st:defind} \textbf{LF indicator:} Define an LF indicator variable with an expected strong order-statistics relationship with the target HF response, as in other response-conditioning and screening methods (see \Cref{sec:LFHF}). The indicator signal is not necessarily the same physical quantity as the target HF response at a lower fidelity level; it may also be a different but physically related signal. An ideal indicator exhibits a monotonic correspondence between the highest LF indicator values and the largest HF response events. A review of suitable wave impact indicators is provided in \cite{VS2023}. 
    
    \item\label{st:MCS} \textbf{LF Monte-Carlo simulation:} Perform LF MCS for a large number $N$ of wave seeds, each of duration $T_{\text{exp}}$. See \cite{VSS2023,SVS2023} for an example of the required $N$ for converged extreme values of wave crests, green water impact forces, and wave-in-deck impact forces. The total MCS duration $T_{\text{tot}}$ follows from $T_{\text{tot}} = N \cdot T_{\text{exp}}$. Also determine the total number of zero up-crossing encountered wave crests $n_w$ within $T_{\text{tot}}$. This can be estimated as $n_w \approx N T_{\text{exp}}/T_{1,e}$, or counted directly from the underlying wave record. The mean wave period is taken as is taken as $T_1 = T_p/1.98$ (for a JONSWAP spectrum with $\gamma = 3.3$). The corresponding mean encounter period for a sailing ship follows from \Cref{eq:ompe}, with wave frequency $\omega = 2\pi/T$, ship forward speed $V_s$, and wave heading $\mu$ relative to the vessel (with $\pi$ denoting head seas).

    \begin{equation}
    \label{eq:ompe}
      \begin{aligned} 
       \omega_{e}=\omega-\omega^2 V_s \cos\mu /g \qquad \rightarrow \qquad T_{1,e} = T_1 \big/ \bigl(1 - (2\pi V_s \cos\mu)/(T_1 g)\bigr)
      \end{aligned}
    \end{equation}
    
    \item\label{st:peaks} \textbf{LF peaks over threshold:} Identify the $n$ ordered LF indicator peaks $\mathbf{l}^{\text{mcs}}$ that exceed the threshold $l_{0,\text{input}}$ in the MCS duration $T_{\text{tot}}$, using a POT crossing procedure (see \Cref{eq:LFmcspeaks}). Such a threshold can be straightforward to define for some problems, for example when only relative wave elevation (RWE) values above a deck level lead to an impact. By ignoring all LF samples below this threshold, excessive sampling in the region where the HF response is likely to be zero can be avoided.   

    \begin{equation}
    \label{eq:LFmcspeaks}
      \begin{aligned}
      \mathbf{l}^{\mathrm{mcs}} = \left( l_1, l_2, \ldots, l_n \right), \qquad l_j > l_{0,\text{input}}, \;\; j=1,\ldots,n
      \end{aligned}
    \end{equation}
    
    \item\label{st:lfdnr} \textbf{LF distribution:} Calculate the LF exceedance probability for all indicator peaks $\mathbf{d}^{\text{mcs}}_{L}=\{d^{\text{mcs}}_{L,i}|i=1,2,...,n\}$, related to the number of wave encounters, by applying \Cref{eq:DNR}. $n_w$ from \ref{st:MCS} forms the common basis for all exceedance distributions in PAS. By thus normalising both LF indicators and HF responses, we correct their exceedance probabilities for differing numbers of peaks and express them on a common exposure scale, which enables consistent relations between LF and HF distributions and physical durations. The largest LF indicator peak value has an exceedance probability of $1/n_w$ and the smallest $n/n_w$. This yields LF indicator peak exceedance probability distribution $\left[\mathbf{d}^{\text{mcs}}_{L},\mathbf{l}^{\text{mcs}}\right]$, with $\mathbf{d}^{\text{mcs}}_{L}$ in ascending order. 

    \begin{equation}
    \label{eq:DNR}
      \begin{aligned}
      \mathbf{d}^{\text{mcs}}_{L} &= \frac{n}{n_w} \cdot P(\mathbf{l}^{\text{mcs}} \geq l)
      \end{aligned}
    \end{equation}
    
    \item\label{st:select} \textbf{Initial LF samples:} Select $m$ initial samples from the LF exceedance distribution. The selected set is called $\left[\mathbf{d}^{\text{sel}}_{L},\mathbf{l}^{\text{sel}}\right]$, where $\mathbf{d}^{\text{sel}}_{L}=\{d^{\text{mcs}}_{L,k}|k=1,2,...,m\}$ and $\mathbf{l}^{\text{sel}}=\{l^{\text{mcs}}_{k}|k=1,2,...,m\}$. Different sampling strategies can be used to choose the indices $k$ and the associated exceedance probability levels $[p_1,p_2,...,p_m]$, as explained in \Cref{sec:pas_init}. Using \Cref{eq:close}, we select the LF MCS exceedance distribution elements closest to these levels and assemble them in subset $\left[\mathbf{d}^{\text{sel}}_{L},\mathbf{l}^{\text{sel}}\right]$. 

    \begin{equation}
    \label{eq:close}
      \begin{aligned}
      d^{\text{sel}}_{L,k} = \left[ \arg\min_{d \in \mathbf{d}^{\text{mcs}}_{L}} | p_k - d | \right] \text{for } k=1,2,...,m
      \end{aligned}
    \end{equation}

    \item\label{st:hfresp} \textbf{Corresponding HF samples:} Obtain the HF loads for the selected indicator peaks in $\left[\mathbf{d}^{\text{sel}}_{L},\mathbf{l}^{\text{sel}}\right]$ by running an HF model for the corresponding wave events. The HF model must capture all relevant physical processes, including appropriate grid resolution, time stepping, and numerical settings. For wave impacts, this typically requires fine-mesh CFD or model experiments, with robust initialisation of HF events based on LF simulations (approaches for this are discussed in \Cref{sec:res_fut}). The selected samples may turn out to be \textit{true positives} (the LF indicator correctly predicted an impact) or \textit{false positives} (the LF indicator predicted an impact that did not occur). Only true positives (with HF load values greater than zero) are included in the matched LF-HF dataset, to avoid introducing a strong non-conservative bias for rare events. This new dataset of HF samples is called $\mathbf{h}^{\text{sel}}=\{h_{k}|k=1,2,...m\}$, here $h_k$ is the HF non-linear response value maximum for event $k$. False positives are also removed from $\left[\mathbf{d}^{\text{sel}}_{L},\mathbf{l}^{\text{sel}}\right]$ to keep the LF-HF datasets consistent, while still being counted in the total number of evaluated HF samples and iterations. 
   
    \item\label{st:outliers} \textbf{Temporary outlier removal:} Temporarily remove LF-HF pairs whose HF responses strongly deviate from the overall trend. A quadratic model $\hat{h}_k=\mathrm{poly}_2(l_k)$ is fitted to the matched data, residuals $r_k=h_k-\hat{h}_k$ are computed, and samples with $|r_k|>3,\mathrm{RMS}(r)$ are identified as outliers. These samples are excluded only for the current iteration and are reconsidered later as new LF-HF pairs are added. The remaining data are then used for copula fitting, which improves the stability of the fit.
    
    \item\label{st:copulafit} \textbf{Copula fitting:} Fit a joint copula probability model to the LF-HF paired samples $[\mathbf{l}^{\text{sel}},\mathbf{h}^{\text{sel}}]$. Before fitting the model, the samples are converted to empirical pseudo-observations $[\mathbf{u},\mathbf{v}]$ on the unit interval [0,1]. This is expressed in \Cref{eq:pseuobs}, where $\mathrm{rank}()$ denotes the statistical rank of the LF or HF elements, and $F_{\mathbf{l}^{\text{sel}}}$ and $F_{\mathbf{h}^{\text{sel}}}$ are the empirical marginal distributions of the LF and HF samples, respectively. This is necessary because copula models are defined on $[0,1]^2$. The copula model selection and fitting procedures are explained in \Cref{sec:pas_copsel}, and the fitting is illustrated in \Cref{fig:copproc}. This yields a fitted copula $C$ for the joint exceedance probability $P(H \ge h, L \ge l \mid H>h_0, L>l_0)$, where $L$ and $H$ are random variables corresponding to the LF and HF samples. The copula is only defined over the sample range, so $l_0 = \min(\mathbf{l}^{\text{sel}})$ (which may equal or exceed the initial $l_{0,\text{input}}$) and $h_0 = \min(\mathbf{h}^{\text{sel}})$.
    
    \begin{equation}
    \label{eq:pseuobs}
      \begin{aligned}
      u_i &= \frac{1}{m+1}\,\mathrm{rank}(l^{\text{sel}}_i) \approx F_{\mathbf{l}^{\text{sel}}}(l^{\text{sel}}_i). \\
      v_i &= \frac{1}{m+1}\,\mathrm{rank}(h^{\text{sel}}_i) \approx F_{\mathbf{h}^{\text{sel}}}(h^{\text{sel}}_i), 
      \qquad i=1,\dots,m
      \end{aligned}
    \end{equation} 
    
    \item\label{st:drawsamp} \textbf{Draw conditional HF samples:} Draw $z$=10 Monte Carlo realisations of HF values from the fitted copula model, conditioned on all LF values in $\textbf{l}^{\text{mcs}}$ from \ref{st:lfdnr} above $l_0$ (because the copula is not defined below this value). We call this new conditioning set $\textbf{l}^{\text{mcs}}_0$ (with elements $l^{\text{mcs}}_{0,j}|j=1,...,n_0$). Multiple realisations are drawn to quantify sampling variability. First, theoretical marginal distributions $F_{l,t}$ and $F_{h,t}$ are fitted to the LF and HF samples, as explained in \Cref{sec:pas_margfit}. Next, the LF pseudo-observations $\textbf{u}^{\text{mcs}}$ are computed using the fitted LF marginal CDF (see \Cref{eq:copdraw1}). For each realisation $b=1,2,...,z$ and each conditioning point $j$, a uniform random variable $r_{j,b}$ is drawn, and the corresponding conditional copula quantile $v^{\text{cop}}_{j,b}$ is computed from the fitted copula model $C$ (see \Cref{eq:copdraw2}, where $\varepsilon$ is a small constant to avoid numerical instabilities). $v^{\text{cop}}_{j,b}$ represents the rank of the HF value conditional on the LF value. The HF samples $h^{\text{cop}}_{j,b}$ (assembled in $\mathbf{h}_b^{\text{cop}}$) in real space are then obtained using the inverse of the fitted HF marginal, as formulated in the last part of \Cref{eq:copdraw2}.

    \begin{equation}
    \label{eq:copdraw1}
    \begin{aligned}
        \textbf{u}^{\text{mcs}} = F_{l,t}(\textbf{l}^{\text{mcs}}_0) \\
    \end{aligned}
    \end{equation}
    
    \begin{equation}
    \label{eq:copdraw2}
    \begin{aligned}
        &r_{j,b} \sim \mathrm{U}(\varepsilon,1-\varepsilon) \\
        &v^{\text{cop}}_{j,b} = C^{-1}_{V|U}\left( r_{j,b} \mid u^{\text{mcs}}_j \right) \\
        &h^{\text{cop}}_{j,b} = F_{h,t}^{-1} \left(v^{\text{cop}}_{j,b} \right), \qquad j=1,\dots,n_0,\; b=1,\dots,z, \quad \varepsilon = 10^{-6}
    \end{aligned}
    \end{equation}
    
    \item\label{st:hfdnr} \textbf{HF distribution realisations:} Compute an exceedance probability distribution for each set of drawn HF samples $\mathbf{h}_b^{\text{cop}}$. The drawn samples from the copula model are biased, as we only fitted the copula to samples above a threshold. They are de-biased by relating the exceedance probabilities to the number of wave encounters instead of the number of LF conditioning values (similar as for the LF distribution in \ref{st:lfdnr}). This is formulated in \Cref{eq:DNRhf}, which defines exceedance probabilities $\mathbf{d}^{\text{cop}}_{H,b}$ corresponding to $\mathbf{h}_b^{\text{cop}}$. The largest HF predicted sample drawn from the copula then has an exceedance probability of $1/n_w$ and the smallest $n_0/n_w$. For each realisation we now have an HF exceedance probability distribution dataset $\left[\mathbf{d}^{\text{cop}}_{H,b},\mathbf{h}_b^{\text{cop}}\right]$.

    \begin{equation}
    \label{eq:DNRhf}
      \begin{aligned}
      \mathbf{d}^{\text{cop}}_{H,b} &= \frac{n_0}{n_w} \cdot P(\mathbf{h}_b^{\text{cop}} \geq h)
      \end{aligned}
    \end{equation}
    
    \item\label{st:meanhfdnr} \textbf{Mean HF distribution:} Compute the mean distribution $[\mathbf{d}^*,\mathbf{h}^*]$ and estimate its U95\% sampling uncertainty $\mathbf{h}^{*U95}$ over these realisations (see \Cref{eq:meanp,eq:meanu95}, where $\boldsymbol{\sigma}_h$ is the standard deviation over the realisations). Since the exceedance probability range $\mathbf{d}^{\text{cop}}_{H,b}$ is fully determined by the conditioning values, it is the same for every realisation $b$.
   
    \begin{equation}
    \label{eq:meanp}
      \begin{aligned}
        \mathbf{d}^* &= \mathbf{d}^{\mathrm{cop}}_H, \quad \text{with} \quad \mathbf{d}^{\mathrm{cop}}_H = \mathbf{d}^{\mathrm{cop}}_{H,b} \ \forall b
      \end{aligned}
    \end{equation}

    \begin{equation}
    \label{eq:meanu95}
      \begin{aligned}
        \mathbf{h}^* &= \frac{1}{z}\sum_{b=1}^z \mathbf{h}^{\text{cop}}_{b} \qquad \text{and} \qquad \mathbf{h}^{*U95} \approxeq \mathbf{h}^* \pm 1.96\,\boldsymbol{\sigma}_h
      \end{aligned}
    \end{equation}

    % \begin{equation}
    % \label{eq:u95dist}
    %   \begin{aligned}
    %     \mathbf{h}^{*U95} \approxeq \mathbf{h}^* \pm 1.96\,\boldsymbol{\sigma}_h %, \textrm{where: } \qquad \boldsymbol{\sigma}_h = \sqrt{\frac{1}{z-1}\sum_{b=1}^z \left(\mathbf{h}^{\text{cop}}_{b}-\mathbf{h}^*\right)^2}
    %   \end{aligned}
    % \end{equation}

    \item\label{st:mpm} \textbf{Estimate MPM:} Estimate the $T_{\text{exp}}$-duration MPM $\widehat{H}$ from the mean predicted distribution $[\mathbf{d}^*,\mathbf{h}^*]$. The exposure duration and prevailing wave conditions define the target exceedance probability $P_{\text{exp}}$ (\Cref{eq:pexp}), where $n_{\text{exp}}$ is the average number of wave encounters during $T_{\text{exp}}$. The MPM follows from the POT exceedance distribution by evaluating it at $P_{\text{exp}}$ (\Cref{eq:MPMfromDNR}). For rare events, this provides a close approximation of the probability that the maximum within the exposure period exceeds the specified threshold. 
    
    \begin{equation}
    \label{eq:pexp}
      \begin{aligned}
      P_{\text{exp}} &= 1/n_{\text{exp}} \approxeq T_{1,e}/T_{\text{exp}}
      \end{aligned}
    \end{equation}
        
    \begin{equation}
    \label{eq:MPMfromDNR}
      \begin{aligned}
      \mathbf{d}^*(\widehat{H}) &= P_{\text{exp}},  
      \quad \text{ therefore:} \quad
      \widehat{H} &= \mathbf{d}^{*-1}(P_{\text{exp}})
      \end{aligned}
    \end{equation}
    
    \item\label{st:adaptive} \textbf{Adaptive sampling:} Start the adaptive sampling procedure, iterating over \ref{st:hfresp} to \ref{st:adaptive}. Each iteration, an acquisition function is applied to define a new sample. The new samples are selected from $\left[\mathbf{d}^{\text{mcs}}_{L},\mathbf{l}^{\text{mcs}}\right]$ defined in \ref{st:lfdnr}, without replacement. The acquisition function is discussed in \Cref{sec:pas_adap}. If the new HF sample is not a false positive, it is added to the LF selected set $\left[\mathbf{d}^{\text{sel}}_{L},\mathbf{l}^{\text{sel}}\right]$ of \ref{st:select} and the number of selected samples $m$ is updated, after which \ref{st:hfresp} to \ref{st:adaptive} are repeated. If the new sample is a false positive (with HF value zero), it is only counted towards to total numbers of evaluated iterations and HF samples. A new sample is then selected using the acquisition function, and the procedure is repeated.
    
    \item\label{st:end} \textbf{Convergence:} When convergence is reached according to a stopping criterion (\Cref{sec:pas_stop}), the result is the converged prediction for the HF distribution $\mathbf{h}^*$ over prediction range $\mathbf{d}^*$, the associated MPM value $\widehat{H}$, and the MPM uncertainty.
\end{steps}

\begin{figure}[t!]
	\centering
        \includegraphics[width=\linewidth]{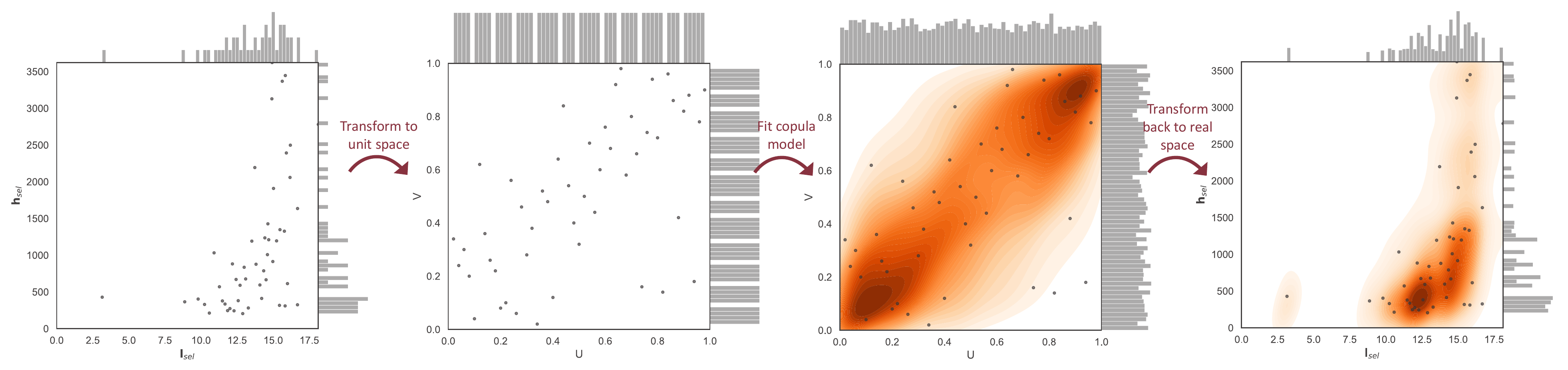}
	\caption{Example copula-fitting procedure, with from left to right: scatter diagram of LF-HF data in real space, the corresponding empirical pseudo-observations in uniform U,V-space, the fitted copula model in U,V-space, and the transformed data from the copula model in real space (transformed back using fitted marginal distributions).} 
	\label{fig:copproc}
\end{figure}

%A key advantage of both AS and PAS is the integration of adaptive sampling with screening, which enables efficient generation of new HF samples at targeted low exceedance probability values and avoids the need for extensive HF simulations, which is not possible with adaptive sampling alone. This feature makes the methods especially effective for extremes of strongly non-linear responses. The additional advantage of PAS over AS is that the copula fitting provides a stronger statistical foundation and makes the method less sensitive to the choice of indicator. This enables the use of linear potential flow indicators to predict design loads for wave impacts, which is very computationally efficient.

The steps of PAS are presented in general terms, making the method flexible and adaptable to any strongly non-linear problem with suitable LF and HF analysis tools. The following sections describe the specific choices made for the applications in this study: copula fitting and selection (\Cref{sec:pas_copsel}), marginal distributions (\Cref{sec:pas_margfit}), acquisition functions (\Cref{sec:pas_adap}), initial sampling (\Cref{sec:pas_init}), and stopping criteria (\Cref{sec:pas_stop}). The selection of LF and HF models as well as the applied LF thresholds are case-specific and are discussed in the corresponding case descriptions. Assumptions are discussed in \Cref{sec:pas_assump}, and implementation details are given in \Cref{sec:pas_implem}.

PAS is not a physics-free method, as both the LF and HF solvers are based on hydrodynamic principles. PAS establishes a statistical mapping between multi-fidelity models that already embed the relevant physics. Physical consistency is inherited from the governing equations, while the statistical component efficiently transfers information across fidelity levels. Similar multi-fidelity principles are used in many alternative EVPMs, as discussed in \Cref{sec:intro}. PAS should therefore be applied with appropriate LF and HF models, selected on a case-by-case basis.

The brief description of AS in \Cref{app:AS} indicates which steps are inherited from AS and which steps are new or modified. The similarity requirements between the LF indicator order statistics and the HF load order statistics are less stringent in PAS than in AS. This simplifies indicator selection and, by permitting lower-fidelity indicators, accelerates \ref{st:MCS} compared to AS.

%%%%%%%%%%%%%%%%%%%%%%%%%%%%%%
\subsection{Copula fitting and selection}
\label{sec:pas_copsel}
As mentioned in the introduction, a copula model can be used to model the dependence between two variables, here the LF and HF variables. The basics of copula theory are provided in \Cref{app:copulabasics}. In \ref{st:copulafit}, the best-fitting copula for the paired LF-HF data is not known a priori and may vary between iterations. We therefore fit several candidate copulas at each iteration and select the best model. We consider Gaussian, Frank, Clayton, Gumbel, and Student-T copulas (\Cref{app:copulabasics_form}), which together cover symmetric and asymmetric dependence, with no, one-sided, or symmetric tail dependence. These relatively simple models are robust for small sample sizes. This makes them suitable for repeated fitting within the iterative framework. When an LF threshold is applied, the data are treated as conditional on LF tail events, and all candidate copulas are fitted accordingly. Extreme-value copulas \cite{GS2010} were also tested, but showed unstable fits for small samples and no performance gain for larger samples. This is consistent with their reliance on asymptotic tail behaviour \cite{Lavaud2018}. First, each candidate copula was fitted to the paired empirical pseudo-observations $[\mathbf{u},\mathbf{v}]$ from \Cref{eq:pseuobs}. We used the default fitting methods from the used copula packages; see \Cref{sec:pas_implem} for implementation and fitting details.

Formal goodness-of-fit tests such as Cram\'{e}r--von Mises or Kolmogorov--Smirnov do not work well for small datasets, as we have in PAS. The goal is to select the most plausible copula at each iteration, not to identify the true model. We therefore use the Akaike Information Criterion (AIC, \cite{Akaike1974}) in \Cref{eq:aic}) to choose among the five candidates. The log-likelihood $\ell$ in \Cref{eq:ll} is computed from the copula density $c$ at pseudo-observations $[\mathbf{u},\mathbf{v}]$, using fitted parameters $\hat{\boldsymbol{\theta}}$ (via maximum likelihood or rank-based inversion of Kendall's tau). AIC balances fit and complexity; most copulas have one parameter $k$, except Student-T with two. To limit volatility in model selection across iterations, a new copula is adopted only if its AIC is at least 4.0 lower than the previous model; otherwise, the previous model is retained (following \cite{BA2002}). % Since some copulas are not fitted by maximum likelihood, the resulting AIC values are technically ``pseudo-AIC''; however, they still provide a valid criterion for relative comparison of candidate models. 

\begin{equation}
\label{eq:aic}
\text{AIC} = 2k - 2\ell(\hat{\boldsymbol{\theta}}) 
\end{equation}

\begin{equation}
\label{eq:ll}
\ell(\hat{\boldsymbol{\theta}}) = \sum_{i=1}^{m} \log c(u_i, v_i; \hat{\boldsymbol{\theta}})
\end{equation}

%%%%%%%%%%%%%%%%%%%%%%%%%%%%%%
\subsection{Marginal distribution fitting}
\label{sec:pas_margfit}
In \ref{st:copulafit}, the LF-HF paired samples are transformed to empirical pseudo-observations, without requiring parametric marginals. Full marginal distributions are needed only in \ref{st:drawsamp} to interpolate conditional copula samples back to physical space. For small samples, interpolation using parametric fits is preferred, as linear interpolation can create kinks and kernel smoothing may distort tails. The copula is valid only over the fitted LF-HF ranges, so these marginals are used solely for interpolation.

For this purpose, we fit three-parameter Weibull cumulative distributions (\Cref{eq:Weibull3}) to the available LF and HF samples separately. Weibull-type tails correspond to a negative shape parameter ($\xi$ < 0) in the generalised extreme value (GEV) or generalised Pareto (GP) framework. While the GP distribution represents the asymptotic limit for threshold exceedances, its maximum likelihood estimates are highly variable and sensitive to the threshold in small samples (e.g., \cite{DS1990,JRWT2020}). The three-parameter Weibull provides a more stable and accurate interpolation of the empirical tail over the observed range and is commonly used in the field (see e.g., \cite{ITTC2017_Glob,DNVGL-CG-0130,Haver2017}). We apply least-squares fitting, as recommended for tail modeling (e.g., \cite{DNV-RP-C205}). 

\begin{equation}
\label{eq:Weibull3}
  P(X \leq x) = 1-\exp\left(-\left(\frac{x-\theta}{\alpha}\right)^\beta\right)
\end{equation}

Because the Weibull fit is formally valid only for independent POT values, an optional upper-tail fitting mode is provided when no threshold is set. For very small samples (<10), all data are used; for large samples (>30), only the upper 80\% is fitted, with a linear transition from 100 to 80\% for intermediate sizes (10-30). When a threshold is specified, the full dataset is always used.

%%%%%%%%%%%%%%%%%%%%%%%%%%%%%%
\subsection{Adaptive sampling and acquisition function}
\label{sec:pas_adap}
\ref{st:adaptive} of PAS uses an acquisition function to define the next best sample to evaluate. We select one new HF sample per iteration. In \cite{VS2025prads}, AS was evaluated with seven acquisition functions. Those distributing samples uniformly along the logarithmic exceedance probability axis or concentrating near $P_{\text{exp}}$ were most effective. However, the HF distribution in PAS is constructed differently, and copula fitting requires samples across a wider range. Accordingly, PAS uses the maximum-distance (MD) acquisition function, which selects new samples based on their largest separation from existing HF points on the logarithmic exceedance probability axis.

For each iteration $i$, we have the set of LF selected samples $\mathbf{d}^{\text{sel}}_{L}$ and the set of HF selected samples $[\mathbf{d}^{\text{sel}}_{H},\mathbf{h}^{\text{sel}}]$ in \ref{st:select}, with $\mathbf{d}^{\text{sel}}_{H} = \{ d^s_1, d^s_2, \ldots, d^s_m \}$ ordered such that \( d^s_1 < d^s_2 < \cdots < d^s_m \). Here we drop the iteration index $i$ for convenience. We now define $p_{\text{new}}$ as the optimal exceedance probability for adding a new LF sample to the set. The new point is located halfway between \( \text{ln}(d^s_k) \) and \( \text{ln}(d^s_{k+1}) \), where the gap \( [\text{ln}(d^s_{k+1}) - \text{ln}(d^s_k)]\) is the largest among all pairs of consecutive selected samples. The function operates on the logarithm of exceedance probability to focus on the tail of the distribution. 

\begin{equation}
\label{eq:af_MD}
\begin{aligned}
    p^{\text{md}}_{\text{new}} = \underset{k \in \{1, \ldots, m-1\}}{\arg\max} \, \left( \frac{\text{ln}(d^s_{k+1}) - \text{ln}(d^s_k)}{2} + \text{ln}(d^s_k) \right)
\end{aligned}
\end{equation}

This would be the best new sample value to add to $\mathbf{d}^{\text{sel}}_{L,i}$ on a continuous scale (including iteration subscript $i$ again). However, in PAS, we only have the discrete LF samples in the available LF MCS sample pool of \ref{st:lfdnr}. We therefore select the sample closest to \(p^{\text{md}}_{\text{new},i}\) 
from the available pool \(\mathbf{d}^{\text{mcs}}_L \setminus \mathbf{d}^{\text{sel}}_{L,i}\), 
without replacement (i.e., excluding already selected samples; see \Cref{eq:close2}). The new sample $d_{L,i}$ is added to the existing LF pool $\left[\mathbf{d}^{\text{sel}}_{L,i},\mathbf{l}^{\text{sel}}_i\right]$ in \ref{st:select} for the next iteration.

\begin{equation}
    \label{eq:close2}
    \begin{aligned}
    d_{L,i} &= \underset{d \in \mathbf{d}^{\text{mcs}}_L \setminus \mathbf{d}^{\text{sel}}_{L,i}}{\arg\min} \, \big|\text{ln}(p^{\text{md}}_{\text{new},i}) - \text{ln}(d)\big| 
    \end{aligned}
\end{equation}

The risk associated with this MD function when the physical threshold in the input is not properly set is that it may oversample zero HF values, which do not contribute to characterising the HF extremes. To mitigate this, we implemented a stabilising mechanism: if 5 consecutive zero HF values (or false positives) occur and the exceedance probability of a new sample exceeds that of any existing sample in $\left[\mathbf{d}^{\text{sel}}_{L},\mathbf{l}^{\text{sel}}\right]$, the LF threshold $l_0$ is increased by 2\% of the maximum LF value in $\textbf{l}^{\text{mcs}}_0$ for the next iteration.

Choosing MD as the acquisition function results in a semi-adaptive scheme: the next sample is determined by the previously selected conditioning points rather than the outcomes of the previous iteration, so adaptivity only enters through the stopping criterion and the adaptive threshold explained above. Fully adaptive sampling functions can still be employed in PAS, but in practice they tend to perform worse than MD.

%%%%%%%%%%%%%%%%%%%%%%%%%%%%%%
\subsection{Initial sampling}
\label{sec:pas_init}
Bayesian and adaptive sampling methods can be sensitive to initial samples, as early data shape the first predictions and subsequent updates. Poor initial coverage may slow convergence or bias results. Although PAS combined with MD is not fully Bayesian, initial sampling can still influence outcomes. To reduce this sensitivity, we use an initial sampling strategy that is consistent with the MD acquisition function: four samples selected at uniformly spaced indices over $\ln(\mathbf{d}^{\text{mcs}}_{L})$ from \ref{st:lfdnr}.

%%%%%%
\subsection{Stopping criterion}
\label{sec:pas_stop}
We stop the iterative adaptive sampling procedure when new iterations no longer produce significant changes in the predicted exceedance probability distributions. This is done using a stopping criterion consisting of two parts. The first part sets a limit $\epsilon_1$ for the mean absolute difference between each set of subsequently predicted distributions, averaged over a number of the last iterations. The second part sets a limit $\epsilon_2$ for the coefficient of variation (COV) of the MPM value, averaged over a number of the last iterations. The complete formulations for this stopping criterion are in \Cref{app:pas_stop}. This appendix also explains how this criterion is different from the criterion used for AS in \cite{VS2025}.  
% We considered including a variation criterion for the copula parameter in the stopping criterion, but decided against it, as the current criteria yield reliable results and the copula parameter changes when a different model is selected, as described in \Cref{sec:pas_copsel}. 

%%%%%%
\subsection{Assumptions}
\label{sec:pas_assump}
The most important assumption in PAS is the similarity between the order statistics of the LF indicator and the HF response in \ref{st:hfdnr}. Most alternative methods, such as response-conditioning methods and other screening techniques discussed in \cite{VS2023}, also rely on this assumption. Its validity depends on the indicator choice; a poorly chosen indicator can significantly reduce the result accuracy. However, introducing the copula fitting technique in PAS reduces the weight of this assumption, as it makes the LF-HF coupling probabilistic instead of deterministic. This reduces the impact of outliers in the order statistics comparison. Another critical assumption in \ref{st:hfresp} is that the HF tool accurately calculates the true HF event response. Previous studies have shown that CFD can effectively predict wave impact loads if wave kinematics and ship motions are modelled well (as discussed in \Cref{sec:LFHF}). Additionally, \cite{VMSHKSG2021} demonstrated that screening results could serve as effective inputs for such calculations. Finally, PAS assumes that the LF-HF mapping of each problem fits one of the candidate copula models discussed in \Cref{sec:pas_copsel}. The dynamic best-fitting selection procedure in \ref{st:copulafit} already reduces the method's reliance on a single copula model. The present study shows that these assumptions work for a wide range of considered test problems.

%%%%%%
\subsection{Implementation}
\label{sec:pas_implem}
The PAS framework, including its iterative procedure, acquisition functions, and stopping criteria, was implemented in \texttt{Python} (v3.8.20). It supports the integration of copula models from various sources. In this study, we use Gumbel, Clayton and Frank copula models from the \href{https://sdv.dev/Copulas/}{\texttt{Python copulas}} package (v0.12.0), and Gaussian and Student-T copula models from the \texttt{R} (v4.5.2) library \texttt{copula} (v1.1.6) \cite{RcopPackage,RcopBook} via an embedded \texttt{Python} interface. \texttt{R} provides access to a wider range of models than \texttt{Python}, including extreme-value copulas. For the copulas from the \texttt{Python} \texttt{copulas} package, model parameters were estimated using the default rank-based procedure based on inversion of Kendall's tau (which has a fixed relation with the copula parameters, see \Cref{app:copulabasics_form}). For the copulas from the \texttt{R} \texttt{copula} package, model parameters were estimated using maximum likelihood via the \texttt{fitCopula} function (method~=~`ml'). All scripts are available in the 4TU repository: \cite{VSdata2025pas}. Although the code is still research-grade and not yet optimised for speed, the computational time required to run the PAS scripts for a single iteration is negligible compared to the LF and, in particular, the HF physical models. Running PAS for one iteration typically takes only a few seconds to one minute on a single laptop core.

% Overview cases (not yet with results - add that later).
\begin{table}[t]
\caption{Summary of considered cases, where AC~=~accommodation, BF~=~bow flare, GW~=~green water, PF~=~potential flow, RWE~=~relative wave elevation, RRV~=~relative rise velocity, VBM~=~vertical bending moment and $T_{\text{MCS}}$~=~MCS duration. The variable $P_{\text{exp}}$ for an exposure duration ($T_{\text{exp}}$) of 1~hour for all cases is defined in \Cref{eq:pexp}, and $n_{w}$ in \ref{st:MCS} of PAS. Cases 1a,b and 2a are identical to those evaluated with AS in \cite{VS2025}.} 
\label{tab:cases}
\scriptsize % Enable to make font smaller
\begin{tabularx}{\textwidth}{l|llllllccc} % X = auto-stretching column type
    Case  & Structure & HF variable & HF source & LF indicator & LF source & Conditions & $T_{\text{MCS}}$ & $n_{w}$ & $P_{\text{exp}}$  \\  
    &  &  & & &  & \Cref{tab:conditions} & [h] &  [-] & [-] \\ 
    \endthead
    1a & - & 2$^{\text{nd}}$-order waves & Analytical & Lin. waves & Analytical & A & 50:00  & 16364 & $2.55\times10^{-3}$ \\  
    1b & - & 2$^{\text{nd}}$-order waves & Analytical & Lin. noisy waves & Analytical & A & 50:00  & 16364 & $2.55\times10^{-3}$ \\
    2a & Ferry & Hogging VBM & Non-linear PF & Hogging VBM & Linear PF & B  & 30:00 &  14359 & $1.74\times10^{-3}$ \\ 
    2b & Ferry & Sagging VBM & Non-linear PF & Sagging VBM & Linear PF & B  & 30:00  & 14359 & $1.74\times10^{-3}$ \\  
    3a & Ferry & GW force AC & Experiments & RWE bow & Linear PF & Heavy  & 34:49  & 15198 & $1.91\times10^{-3}$ \\  
    3b & Ferry & GW force AC & Experiments & RWE bow & Linear PF & Extreme  & 23:43  & 10900 & $1.82\times10^{-3}$ \\ 
    4a & Ferry & Slam force BF & Experiments & RRV st19 & Linear PF & Heavy  & 34:49 & 15198 &  $1.91\times10^{-3}$ \\ 
    4b & Ferry & Slam force BF & Experiments & RRV st19 & Linear PF & Extreme  & 23:43  & 10900 & $1.82\times10^{-3}$ 
    % 3a & Ferry & RWE st20 & Experiments & Zbow & Linear PF & C  & 34:49 & 15198 &  $1.91\times10^{-3}$ \\ 
    % 3b & Ferry & RWE st20 & Experiments & Zbow & Linear PF & D  & 23:43  & 10900 & $1.82\times10^{-3}$ \\ 
\end{tabularx}
\end{table}

% Overview wave / ship conditions
\begin{table}[t]
\caption{Summary of irregular wave conditions (all with a JONSWAP spectrum) and operational conditions used in the cases, where $H_s$ is significant wave height, $T_p$ is peak wave period, $s=2\pi H_s/(g T_p^2)$ is non-dimensional wave steepness, $\gamma$ is the JONSWAP peak enhancement factor, $h$ is water depth, $\mu$ is the wave direction w.r.t. the ship (180 deg is head waves), $V_x$ is the forward speed of the ship or structure, $V_y$ is its transverse drift speed, and $T_{p,e}$ is the corresponding peak wave encounter period (\Cref{eq:ompe}). The names of the last two conditions were defined in the corresponding experimental campaign.}
\label{tab:conditions}
\begin{tabular}{l|ccccc|cccc|l}
    Name &$H_s$ [m] & $T_p$ [s] & $s$ [-] & $\gamma$ [-] & $h$ [m] & $\mu$ [deg] & $V_x$ [m/s] & $V_y$ [m/s] & $T_{p,e}$ [s] &  Comment \\ 
    \endthead
    A & 10.0 & 11.0 & 0.053 & 3.3 & 30.0 & - & - & - & - &  No ship \\ 
    B & 13.2 & 10.0 & 0.085 & 3.0 & 1000.0 & 180 &  5.1 & 0.0 & 7.5  &  \\ 
    Heavy & 8.1 & 9.4 & 0.059 & 3.3 & 179.9 & 150 & 2.4 & 0.4 & 8.2  & \\ 
    Extreme & 8.3 & 10.0 & 0.053 & 3.3 & 179.9 & 150 & 5.0 & 0.6 & 7.8  & \\ 
\end{tabular}
\end{table}

%%%%%%%%%%%%%%%%%%%%%%%%%%%%%%%%%%%%%%%%%%%%%%%%%%%%%%%%%%%%%%%%%%%%%%
%%%%%%%%%%%%%%%%%%%%%%%%%%%%%%%%%%%%%%%%%%%%%%%%%%%%%%%%%%%%%%%%%%%%%%
%%%%%%%%%%%%%%%%%%%%%%%%%%%%%%%%%%%%%%%%%%%%%%%%%%%%%%%%%%%%%%%%%%%%%%
\section{Test cases}
\label{sec:cases}
We applied PAS (and AS for reference) to four cases, each with a variation. \Cref{tab:cases} summarizes the cases, with the corresponding wave and ship operating conditions listed in \Cref{tab:conditions}, and the ground-truth MPM values from the validation dataset provided in \Cref{tab:results}. The cases increase in complexity. Case 1 (waves) is a simple analytical wave-only, weakly non-linear problem, where the LF and HF models resolve the same variable at different fidelity levels. Case 2 (vertical bending moments, VBM) is a fully numerical, weakly non-linear ship response problem, again with LF and HF models resolving the same variable. Case 3 (green water) introduces strongly non-linear HF wave impact loads and combines numerical LF indicators with experimental HF measurements, demonstrating that PAS can link LF and HF models even when they resolve different physical variables. Additional complexity is due to threshold behaviour and the presence of false positives. Finally, case 4 (slamming) represents an even more non-linear case, again combining numerical LF indicators with experimental HF measurements, adding the complexity of large outliers in the LF-HF relation. The magnitude of slamming loads strongly depends on relative wave-ship timing and shape, explaining the occurrence of such outliers. The green water and slamming cases include two sea states, while the wave and VBM cases each use other wave conditions. PAS accounts for these wave conditions in the underlying LF and HF models. By considering eight sub-cases involving in total four wave conditions, we evaluate the robustness and accuracy of PAS across a range of wave environments, including steep and strongly non-linear waves. Together, these cases cover a large spectrum of non-linearity, complexity, and LF-HF model combinations. The focus is on strongly non-linear wave impact problems, where alternative extreme value prediction methods are limited or inapplicable. Accordingly, wave impact cases 3 and 4 are most relevant for PAS. The simpler cases, which may also be solvable with more efficient methods or even direct HF MCS, are included to demonstrate the robustness and performance of PAS across increasingly complex scenarios. 

\subsection{Cases 1 (second-order waves) and 2 (vertical bending moments)}
\label{sec:case12}
Cases 1 and 2 are identical to those used in \cite{VS2025} to validate AS, so their description is only briefly repeated in \Cref{app:rescases12}.

\noindent Summarising case 1:
\begin{itemize}[nosep]
\itemsep=0pt
\item HF target extreme value: one-hour MPM of second-order wave crest height $\widehat{C''}$.
\item LF indicator: linear wave crest height $C_{\text{good}}'$ (case 1a), linear noisy wave crest height $C_{\text{worse}}'$ (case 1b).
\end{itemize}
\noindent Summarising case 2:
\begin{itemize}[nosep]
\itemsep=0pt
\item HF target extreme value: one-hour MPM of weakly non-linear hogging $\widehat{V_{hog}''}$ (case 2a) and sagging $\widehat{V_{sag}''}$ (case 2b) VBM.
\item LF indicator: linear hogging $V_{hog}'$ (case 2a) and sagging $V_{sag}'$ (case 2b) VBM.
\end{itemize}

%%%%%%%%%%%%%%%%%%%%%%%%%%%%%%%%%%%%%%%%%%%%%%%%%%%%%%%%%%%%%%%%%%%%%%
\subsection{Case 3 (strongly non-linear): green water impact forces}
\label{sec:case3}
\begin{figure}[t!]
	\centering
        \includegraphics[height=4.25cm]{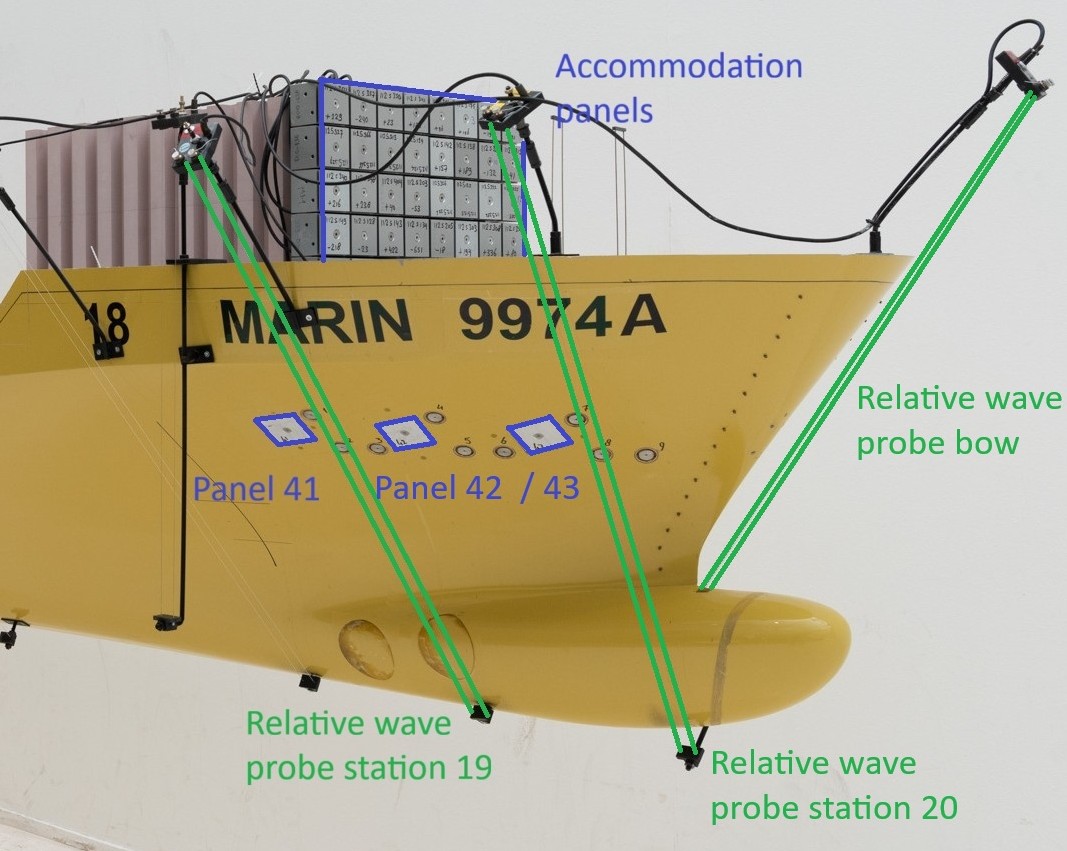}
        \includegraphics[height=4.25cm]{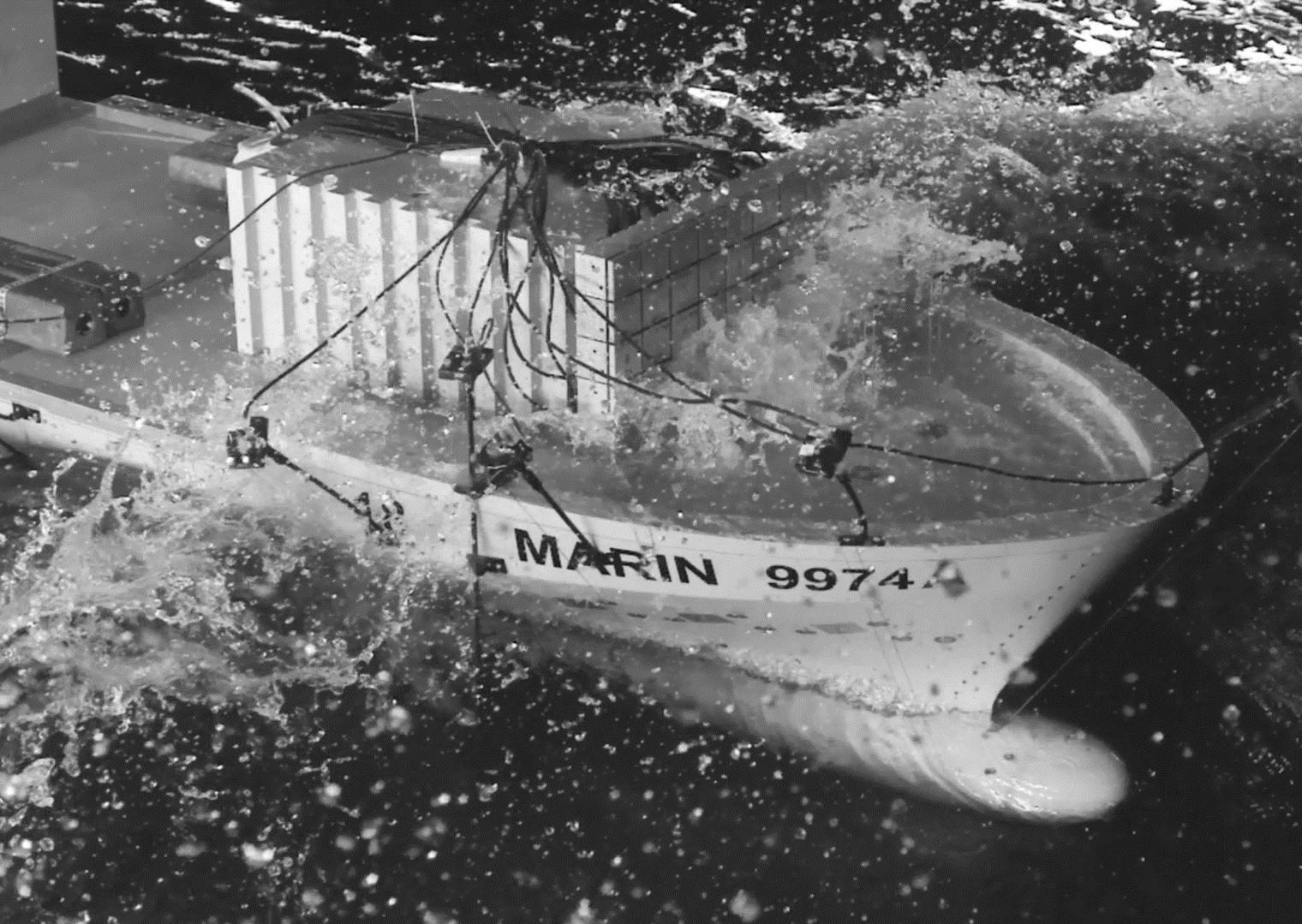}  
        \includegraphics[height=4.25cm]{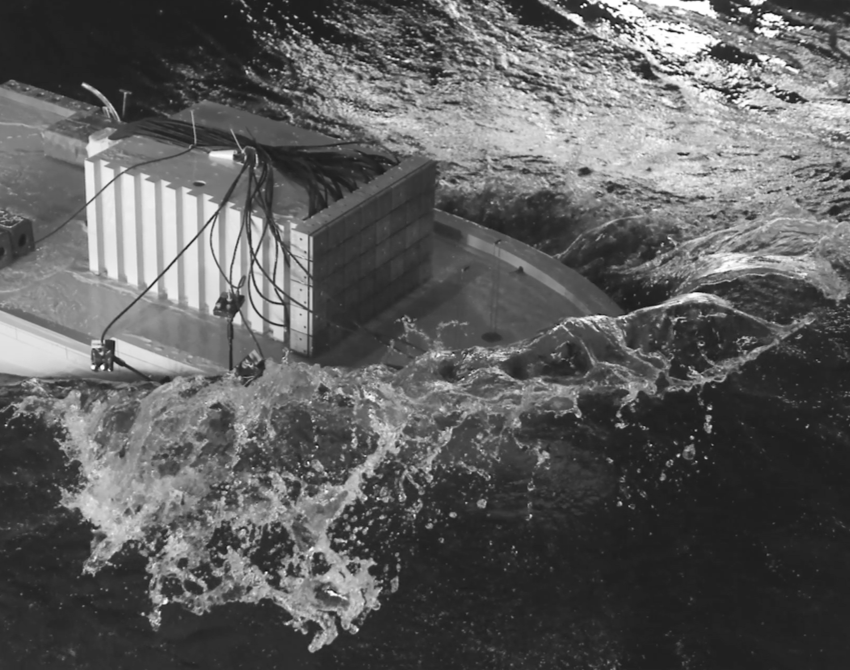}  
	\caption{MARIN ferry 2 and the relevant instrumentation around the bow before the \textit{CRS SCREAM} experiments (left), example green water impact (middle) and example bow-flare slamming impact (right).}
	\label{fig:screamexp}
\end{figure}

Case 3 considers a strongly non-linear problem: predicting extreme green water impact forces on the accommodation of a 190~m ferry (the same vessel as in Case 2) in extreme irregular waves, under two severe, steep bow-quartering wave conditions (Heavy and Extreme in \Cref{tab:conditions}). The HF green water impact forces in longitudinal ship direction on the accommodation, called $F_x''$, were obtained from long-duration model experiments. Details of the experiments are described in \cite{VSS2023,VBS2024}, with selected aspects (e.g., a discussion of potential scale effects) summarised in \Cref{app:case_exp}. Details of the test conditions are provided in \Cref{tab:cases}, showing test durations of 34:49~hours for the Heavy condition and 23:43~h for the Extreme condition. As discussed in \cite{VSS2023}, this is sufficiently long to have a converged HF validation reference for the one-hour MPM. Some pictures of the experiments are shown in \Cref{fig:screamexp}. The impact forces were measured with 40 force panels (\cite{VDSJ2021}) on the front of the accommodation (total area 9.0$\times$14.4~m full-scale), with $F_x''$ representing the instantaneous sum of all panel measurements. Previous studies have shown that the relative wave elevation (RWE) around the ship's bow can be an effective indicator for green water events (see e.g., \cite{VMSHKSG2021,HM1970,HLG1993,OMTKMH2002,VV2014}). Accordingly, the selected LF indicator for this case is $R_{\text{bow}}'$, generated using a linear potential flow simulations. The location of the RWE probe in the simulations mirrors that in the experiments (see \Cref{fig:screamexp}). \cite{VS2025} demonstrated that accurate green water predictions were achieved for a pilot case using AS in combination with an LF indicator generated from coarse-mesh CFD. If linear potential flow can be used instead to generate an indicator for green water impact loads (in combination with PAS), this can have a large advantage in computational time. The LF-HF peak scatter plots are shown in \Cref{app:scatters}. The plot shows that, as expected, the LF and HF peaks for this case have less similar order statistics than those of case 2. Because the experimental green water force peaks (ground truth HF validation data) have a large tail variability, we fitted a three-parameter Weibull distribution (\Cref{eq:Weibull3}) to the 30\% highest force peaks, to derive the reference `true' distribution. \Cref{app:case_imp} provides more details on the potential flow simulations used to generate the LF data. We used an LF threshold of 12.0~m in \ref{st:peaks} of PAS. This value was chosen slightly above the local freeboard at the bow (10.5 m). RWE values below this level are expected to generate many false positives. Due to the ferry's large bow flare angle, events with RWEs only marginally exceeding the freeboard are often deflected to the sides, preventing water from flowing onto the deck and thus avoiding green water loads on the accommodation. Summarising:
\begin{itemize}[nosep]
\itemsep=0pt
\item HF target extreme value: one-hour MPM of green water impact force $\widehat{F_x''}$ in the Heavy (case 3a) and Extreme (case 3b) condition.
\item LF indicator: linear potential flow RWE at the bow $R_{\text{bow}}'$.
\end{itemize}

%%%%%%%%%%%%%%%%%%%%%%%%%%%%%%%%%%%%%%%%%%%%%%%%%%%%%%%%%%%%%%%%%%%%%%
\subsection{Case 4 (strongly non-linear): bow-flare slamming forces}
\label{sec:case4}
Case 4 studies another strongly non-linear problem, based on the same experiments as case 3. Here we predict extreme values of the bow-flare slamming forces on the ferry, sailing in the same two irregular Heavy and Extreme bow-quartering wave conditions. The HF slamming impact forces, denoted $F_{s}''$, were obtained by summing the measured peak forces for each wave event across three force panels located in the bow flare (see \Cref{fig:screamexp}). The size of each of these panels 41, 42 and 43 is 1.8$\times$1.8~cm full-scale. Previous studies have shown that the relative velocity between the wave crest and the ship's bow flare can be an effective indicator for slamming events (see e.g., \cite{BSPFS2018,BSD2019,K2018,O1964a}). Motivated by this, we adopt the relative rise velocity (RRV) at station 19, denoted $V_{r,19}'$, as the LF indicator. For each wave event, the RRV is defined as the peak RWE divided by its rise time, with the rise time measured from the RWE's zero up-crossing to the subsequent peak. The RWE used to compute the RRV is obtained from linear potential flow simulations on the weatherside at station 19 (located at 19/20 of the ship length from the stern; see \Cref{fig:screamexp}). The LF-HF peak scatter plots are shown in \Cref{app:scatters}, which shows that the LF and HF peaks for this case have a distinctly different relation than those of the other cases. The true HF MPM values from the experiments in \Cref{tab:results} were again derived from a 3-parameter Weibull fit to the 30\% highest force peaks. \Cref{app:case_imp} provides more details on the potential flow simulations used to generate the LF data. We did not use an LF threshold value in \ref{st:peaks} of PAS for the RRV, because no clear threshold can be identified for the occurrence of slamming events at the panels. This would differ if, for example, peaks in RWE would be used as the indicator, as this would naturally introduce a threshold at some distance above the panel underside measured from the calm waterline (2.875~m). Summarising:
\begin{itemize}[nosep]
\itemsep=0pt
\item HF target extreme value: one-hour MPM of bow-flare slamming force $\widehat{F_{s}''}$ in the Heavy (case 4a) and Extreme (case 4b) condition.
\item LF indicator: linear potential flow relative rise velocity at station 19 $V_{r,19}'$.
\end{itemize}

%%%%%%%%%%%%%%%%%%%%%%%%%%%%%%%%%%%%%%%%%%%%%%%%%%%%%%%%%%%%%%%%%%%%%%
%%%%%%%%%%%%%%%%%%%%%%%%%%%%%%%%%%%%%%%%%%%%%%%%%%%%%%%%%%%%%%%%%%%%%%
%%%%%%%%%%%%%%%%%%%%%%%%%%%%%%%%%%%%%%%%%%%%%%%%%%%%%%%%%%%%%%%%%%%%%%

\section{Method used to validate PAS}
\label{sec:valmet}

In a practical design procedure for a maritime structure, HF CFD simulations or experiments would need to be performed for new wave events at every iteration in \ref{st:hfresp} to obtain the required HF loads. In the present validation study, however, we circumvent this requirement: rather than running CFD simulations, we extract the corresponding HF load responses directly from the validation datasets for each selected wave event. This allows us to validate the statistical framework of PAS without mixing it with CFD load prediction accuracy (following the same approach used in \cite{VS2025} for AS). Earlier studies have shown before that accurate HF wave event impact loads can be obtained with CFD (see \Cref{sec:LFHF}). 

For cases 1 and 2, obtaining identical deterministic wave inputs for the LF and HF variables was straightforward. For cases 3 and 4, this was achieved by combining experimentally measured waves recorded next to the ferry (propagated to the vessel's centre of gravity), with the linear frequency-domain potential flow database of RWE response functions. This ensured that the wave phasing of the LF time traces closely matched that of the experimental waves. These simulations were previously performed in \cite{VBS2024}, where the full procedure is described in detail.

The performance of PAS compared to the validation material is assessed using three metrics: $M_1$ for efficiency, and $M_2$ and $M_3$ for accuracy. $M_1$ (\Cref{eq:M1M2}) is the ratio of HF samples required to reach convergence, $n_c$, to the total number of HF samples without PAS, $n_w$ (the total number of wave events in the validation dataset). $M_2$ measures the mean deviation of the MPM prediction over the last 10 true positive iterations at convergence, $\widehat{H}_{i_c,10}$, from the ground truth $\widehat{H_t}$ (\Cref{eq:M1M2}), where $i_c$ is the converged iteration number. Finally, $M_3$ quantifies the maximum deviation of the predicted distribution from the ground truth over an exceedance probability range $[0.5P_{\text{exp}},2P_{\text{exp}}]$ (\Cref{eq:M3}), again averaged over the last 10 true positive iterations at convergence, with $\mathbf{h^*}_{\text{ran},i_c,10}(k)$ and $\mathbf{h_{t,\text{ran}}}(k)$ denoting the mean predicted and true distributions over this range with elements $k$, respectively. Convergence of PAS can be somewhat erratic; the averaging over a number of iterations reduces this noise. Smaller absolute values of all three metrics indicate better performance. Note that $M_2$ and $M_3$ may also be negative, indicating whether predictions are conservative. In theory, the convergence criteria should take care of the accuracy of the prediction. However, $M_2$ and $M_3$ are used to check whether the results converge towards the ground truth and not towards a biased value. These criteria are similar, but not identical, to those used for AS in \cite{VS2025prads}.

\begin{equation}
    \label{eq:M1M2}
    M_1 = \frac{n_c}{n_w} \qquad  M_2 = \frac{ \widehat{H}_{i_c,10} - \widehat{H_{t}} } {\widehat{H_{t}}}
\end{equation}

\begin{equation}
\label{eq:M3}
\begin{aligned}
    \mathbf{\Delta}(k) = \mathbf{h^*}_{\text{ran},i_c,10}(k)-\mathbf{h_{t,\text{ran}}}(k) \quad \rightarrow \quad
    k^{\textrm{absmax}} = \arg\max_{k} |\mathbf{\Delta}(k)| \quad \rightarrow \quad 
    M_3 = \frac{\mathbf{\Delta}(k^{\textrm{absmax}})}{\widehat{H_{t}}},
\end{aligned}    
\end{equation}

The performance of PAS is compared to that of AS and the brute-force MCS approach. An overview of the steps in AS that differ from PAS and some relevant formulations and settings are included in \Cref{app:AS}. It is important to note that throughout this study, PAS and AS are evaluated using the same analytical and linear potential-flow LF indicators. This represents a deviation from the approach used for pilot-case 3 in \cite{VS2025}, where the AS method was used in combination with an LF indicator derived from coarse-mesh CFD simulations. That AS-coarse mesh CFD combination yielded accurate distributions of green water impact loads on a containership in the pilot study, but using this in practice would be associated with a high computational cost.

%%%%%%%%%%%%%%%%%%%%%%%%%%%%%%%%%%%%%%%%%%%%%%%%%%%%%%%%%%%%%%%%%%%%%%
%%%%%%%%%%%%%%%%%%%%%%%%%%%%%%%%%%%%%%%%%%%%%%%%%%%%%%%%%%%%%%%%%%%%%%
%%%%%%%%%%%%%%%%%%%%%%%%%%%%%%%%%%%%%%%%%%%%%%%%%%%%%%%%%%%%%%%%%%%%%%

\begin{figure}[t]
	\centering
        % ---------- Left column ----------
        \begin{subfigure}[t]{0.4\textwidth}
            \centering
            \begin{subfigure}[t]{\textwidth}
                \centering
                \includegraphics[width=\textwidth]{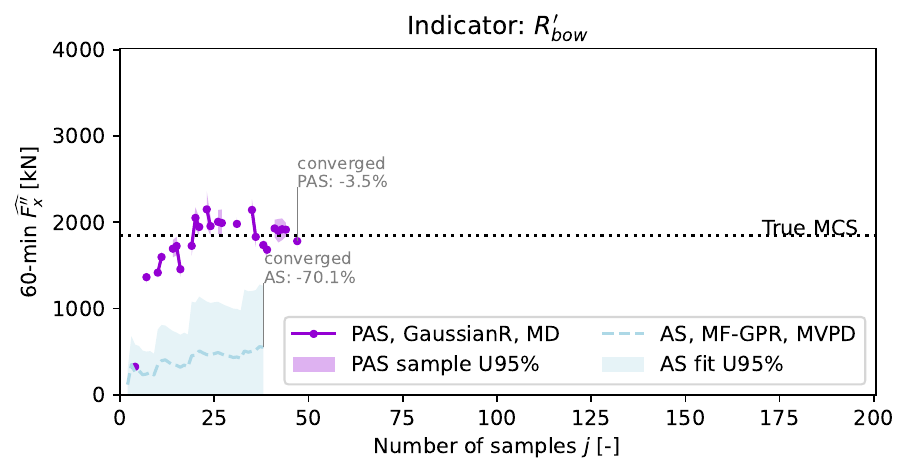}
            \end{subfigure}
            \begin{subfigure}[t]{\textwidth}
                \centering
                \includegraphics[width=\textwidth]{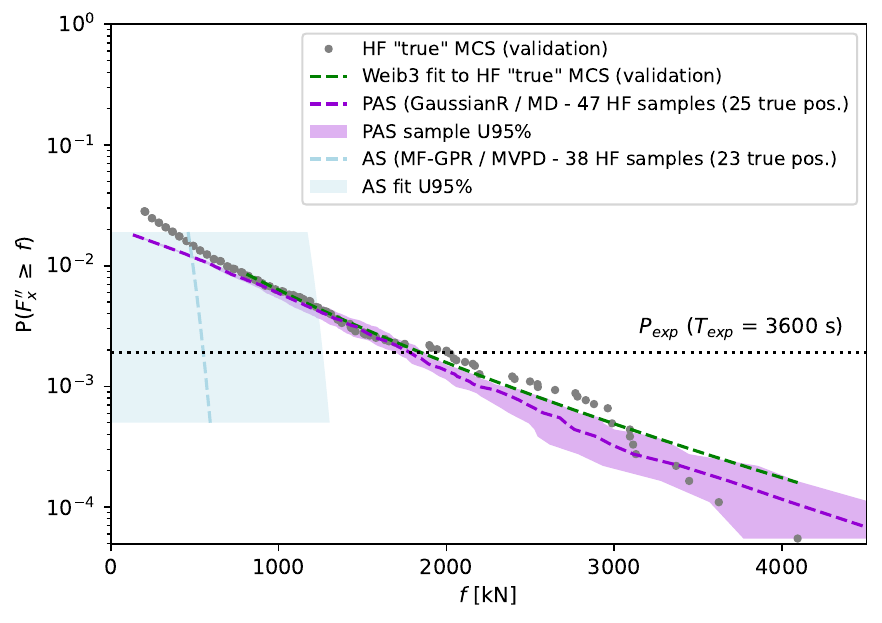}
            \end{subfigure}
            \subcaption{Case 3a, Heavy condition}
            \label{fig:res3a}
        \end{subfigure}
        % ---------- Right column ----------
        \begin{subfigure}[t]{0.4\textwidth}
            \centering
            \begin{subfigure}[t]{\textwidth}
                \centering
                \includegraphics[width=\textwidth]{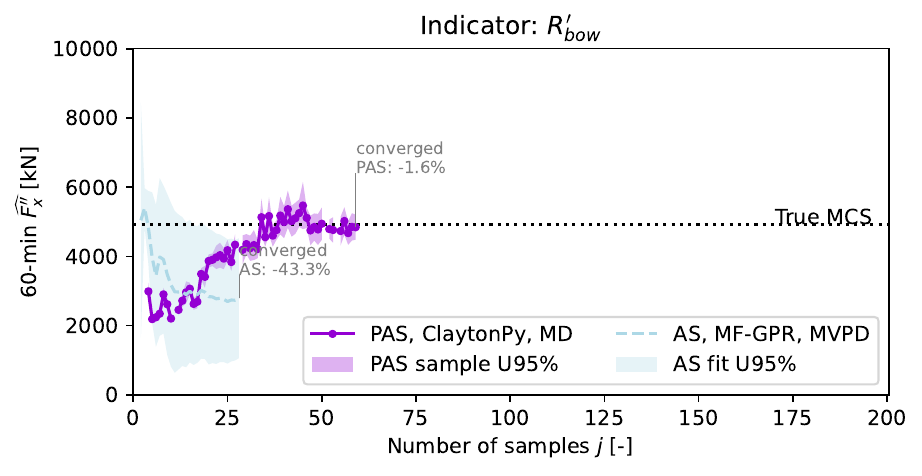}
            \end{subfigure}
            \begin{subfigure}[t]{\textwidth}
                \centering
                \includegraphics[width=\textwidth]{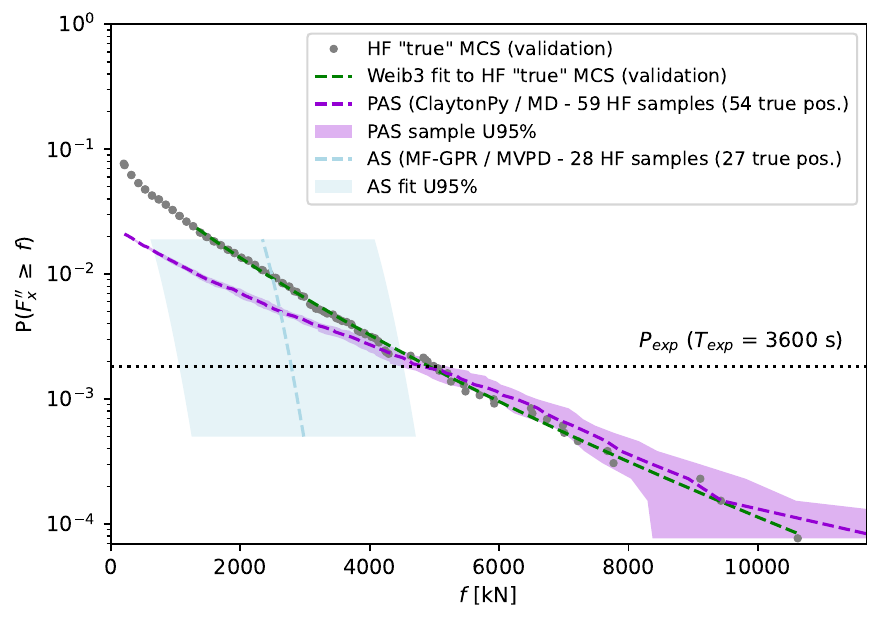}
            \end{subfigure}
            \subcaption{Case 3b, Extreme condition}
            \label{fig:res3b}
        \end{subfigure}
	\caption{Case 3 - green water: convergence of one-hour MPM as a function of number of samples (top) and final converged distributions (bottom) from AS and PAS, both with linear potential flow indicator $R_{\text{bow}}'$. The copula in the name of the PAS results is the utilised model in the last iteration.}
	\label{fig:res3}
\end{figure}

\begin{figure}[t]
	\centering
        % ---------- Left column ----------
        \begin{subfigure}[t]{0.4\textwidth}
            \centering
            \begin{subfigure}[t]{\textwidth}
                \centering
                \includegraphics[width=\textwidth]{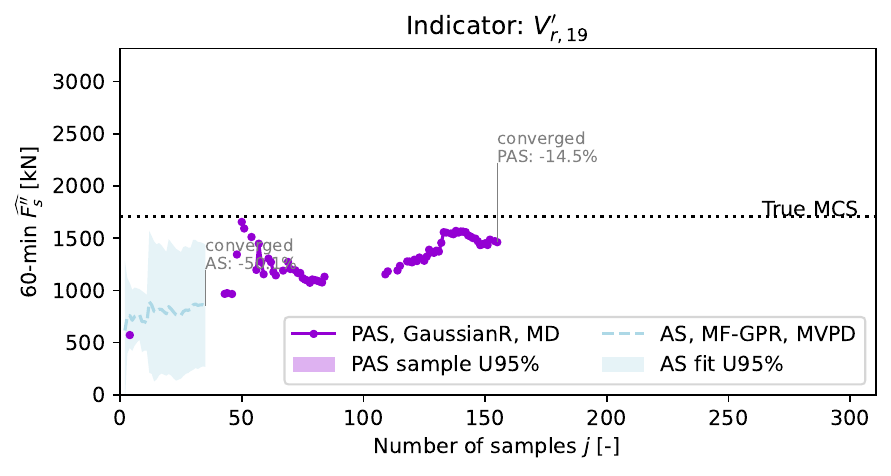}
            \end{subfigure}
            \begin{subfigure}[t]{\textwidth}
                \centering
                \includegraphics[width=\textwidth]{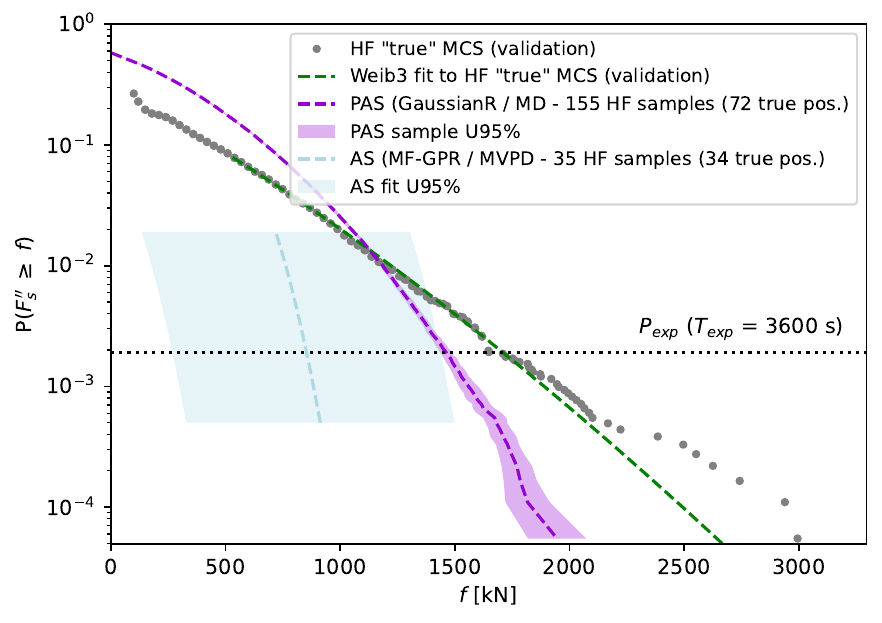}
            \end{subfigure}
            \subcaption{Case 4a, Heavy condition}
            \label{fig:res4a}
        \end{subfigure}
        % ---------- Right column ----------
        \begin{subfigure}[t]{0.4\textwidth}
            \centering
            \begin{subfigure}[t]{\textwidth}
                \centering
                \includegraphics[width=\textwidth]{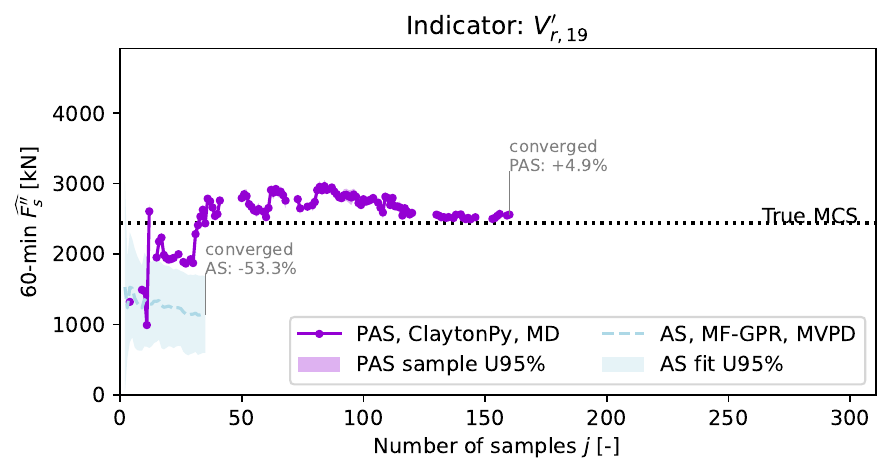}
            \end{subfigure}
            \begin{subfigure}[t]{\textwidth}
                \centering
                \includegraphics[width=\textwidth]{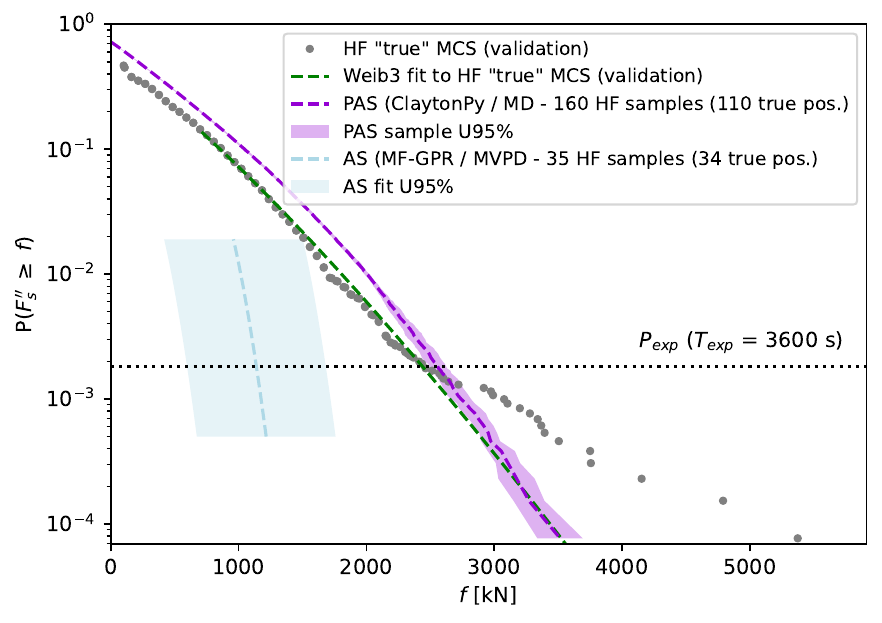}
            \end{subfigure}
            \subcaption{Case 4b, Extreme condition}
            \label{fig:res4b}
        \end{subfigure}
	\caption{Case 4 - slamming: convergence of one-hour MPM as a function of number of samples (top) and final converged distributions (bottom) from AS and PAS, both with linear potential flow indicator $V_{r,19}'$. The copula in the name of the PAS results is the used model in the last iteration.}
	\label{fig:res4}
\end{figure}

%%%%%%%%%%%%%%%%%%
\section{Results and discussion}
\label{sec:Discussion}
Before comparing PAS and AS results to each other and to the MCS ground truth, we define stopping criteria for $S(j)$ in \Cref{eqPAS:convcrit}: $\epsilon_1$ for the maximum absolute distribution difference between iterations, and $\epsilon_2$ for the coefficient of variation (COV) of the one-hour MPM. We used $\epsilon_2~=~0.1$ for all cases (standard deviation 10\% of mean). The $\epsilon_1$ value is case-specific, as it shares the units of the predicted quantity. For consistency across cases, we selected round numbers approximately 1-2\% of the maximum HF value in each validation dataset: $\epsilon_1 = 0.2$~m wave crest height for Case~1, $2\times10^7$~Nm VBM for Case~2, 200~kN green-water force for Case~3, and 20~kN slamming force for Case~4. These limits are more lenient than those used for case 1 and 2 with AS in \cite{VS2025}. To ensure a fair comparison, we applied these to both AS and PAS. \Cref{sec:res_sens} includes a brief discussion of the sensitivity of the results for these stopping criteria.

\begin{figure}[t]
	\centering
        \includegraphics[width=\linewidth]{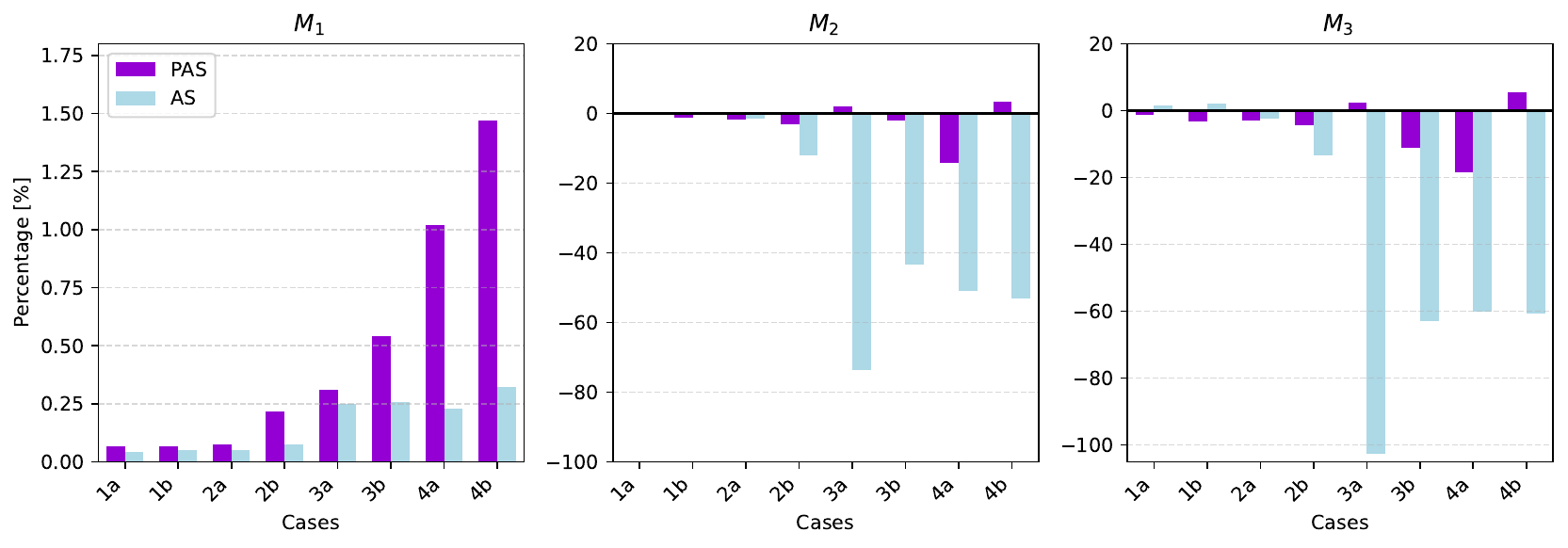}
	\caption{Performance metrics $M_1$, $M_2$ and $M_3$ defined in \Cref{sec:valmet} for all considered cases.}
	\label{fig:m123}
\end{figure}

The resulting converged distributions from AS and PAS for the wave impact cases are presented in \Cref{fig:res3} (cases 3a,b) and \Cref{fig:res4} (cases 4a,b), together with the iterative convergence of the MPM for each case. The corresponding figures for the simpler cases 1 and 2 can be found in \Cref{app:rescases12}. The performance of the methods across all cases is summarised in \Cref{fig:m123}, using the metrics defined in \Cref{sec:valmet}. As a reminder, $M_1$ measures efficiency relative to the HF ground truth, $M_2$ the deviation of the predicted HF MPM, and $M_3$ the maximum deviation between predicted and ground-truth distributions. The ground truth for each case is always derived from the HF validation MCS datasets. The predicted MPM at convergence is also included in \Cref{tab:results}. The MPM data underlying \Cref{fig:m123} differs slightly from the table, as the table shows the results for the converged iteration, whereas the MPM used for $M_2$ is averaged over the last 10 iterations at convergence (see \Cref{sec:valmet}).

%The following sections discuss the accuracy, uncertainty and efficiency of these results, their sensitivity to different input parameters, and future work. 

\begin{table}[t]
{\scriptsize % Enable to make font smaller
\begin{tabular}{|l|c|c|c|c|c|c|c|c|c|c|}
        \hline
        & Case 1a & Case 1b & Case 2a & Case 2b & Case 3a & Case 3b & Case 4a & Case 4b \\
        \hline
        \multicolumn{9}{|c|}{\textbf{Predicted one-hour MPM at convergence}} \\
        \hline
        unit & m & m & Nm & Nm & kN & kN & kN & kN \\
        HF MCS* & \multicolumn{2}{c|}{10.53 (U95\% 0.10)} & $1.10 \times 10^9$ & $9.00 \times 10^8$ & $1.85\times10^3$ & $4.92\times10^3$ & $1.71\times10^3$ & $2.44\times10^3$ \\
        AS & 10.55 & 10.76 & $1.06 \times 10^9$ & $8.06 \times 10^8$ & $0.55\times10^3$ & $2.79\times10^3$ & $0.85\times10^3$ & $1.14\times10^3$ \\
        PAS & 10.59 & 10.53 & $1.11 \times 10^9$ & $9.44 \times 10^8$ & $1.78\times10^3$ & $4.84\times10^3$ & $1.46\times10^3$ & $2.56\times10^3$ \\
        \hline
        \multicolumn{9}{|c|}{\textbf{Required number of HF samples for convergence}} \\
        \hline
        HF MCS** & 16364 & 16364  & 14359  & 14359 & 15198  & 10900  & 15198  & 10900 \\
        AS & 7 & 8 & 7 & 11 & 38 & 28 & 35 & 35 \\
        PAS & 15 & 15 & 15 & 31 & 47 & 59 & 155 & 160 \\
        \hline
        \multicolumn{9}{|c|}{\textbf{LF duration to simulate (analytical / linear potential flow)}} \\
        \hline
        HF MCS & - & - & - & - & - & - & - & -  \\
        AS & 50 h & 50 h & 30 h & 30 h & 34:49 h & 23:43 h & 34:49 h & 23:43 h  \\
        PAS & 50 h & 50 h & 30 h & 30 h & 34:49 h & 23:43 h & 34:49 h & 23:43 h  \\
        \hline
        \multicolumn{9}{|c|}{\textbf{Expected HF duration to simulate (CFD / experiments)}} \\
        \hline
        HF MCS & 50 h & 50 h & 30 h & 30 h & 34:49 h & 23:43 h & 34:49 h & 23:43 h  \\
        AS (est.***) & $\sim$2 min & $\sim$3 min & $\sim$2 min & $\sim$4 min & $\sim$13 min & $\sim$9 min & $\sim$12 min & $\sim$12 min \\
        PAS (est.***) & $\sim$5 min & $\sim$5 min & $\sim$5 min & $\sim$10 min & $\sim$16 min & $\sim$20 min & $\sim$52 min & $\sim$53 min \\
        \hline
        \multicolumn{9}{l}{\textit{* Ground truth.}} \\
        \multicolumn{9}{l}{\textit{** All waves.}} \\
        \multicolumn{9}{l}{\textit{*** Estimate, assuming HF events with a duration of 20 s each as discussed in \Cref{sec:Discussion}.}} \\
    \end{tabular}}
    \caption{Summary of the results for all cases.} 
    \label{tab:results}
\end{table}

%%%%
\subsection{Accuracy}
\label{sec:res_acc}
The figures in \Cref{app:rescases12} and \Cref{tab:results} show that AS and PAS both provide accurate predictions for cases 1 and 2. PAS outperforms AS for both cases, but still does not predict the true tail of the sagging distribution (case 2a in \Cref{fig:res2b}) well. $M_2$ in \Cref{fig:m123} shows that PAS predicts MPM values within 2\% error for these two cases. Similar results were found in \cite{VS2025} for AS, where the USMV acquisition function produced slightly more accurate results than the current AS with MVPD for Case~2. However, the results for cases 3 and 4 clearly highlight the advantage of PAS over AS when combined with a linear potential flow indicator. Although PAS also relies on a screening indicator, the inclusion of probabilistic copula fitting substantially reduces the bias associated with indicator quality. This is particularly important when the LF and HF order statistics are not very similar, a situation in which AS performs poorly while PAS remains robust. The $M_2$ values show that PAS is able to predict the target extreme MPM values with an accuracy (averaged over the last 10 iterations before convergence) within 2\% for both variations of case 3, and within 15\% for both variations of case 4, whereas AS shows consistently non-conservative deviations of up to 75\%. With the current settings, PAS predicts the one-hour MPM over all four cases within 15\% accuracy at convergence. Similar results are observed for $M_3$; the maximum deviations of the PAS distributions compared to the ground truth are within 19\% of the MPM for all test cases. It should be kept in mind that the present results rely on combining both PAS and AS with linear potential flow indicators, offering a computationally cheap way to examine these extreme responses.

% \hl{Mention number of removed outliers?}

%\cite{VS2025} demonstrated that AS, when combined with a coarse-mesh CFD indicator, can yield accurate green water load distributions. However, obtaining indicators from 3D coarse-mesh CFD is far more computationally expensive than using linear potential flow, which highlights the advantage of PAS over AS.

%%%%
\subsection{Uncertainty}
\label{sec:res_unc}
When comparing the extreme values from the validation datasets with those obtained using (P)AS, their respective uncertainties must be considered. The distributions in \Cref{fig:res1,fig:res2,fig:res3,fig:res4} also show the U95\% sample uncertainty from the fitted copula model for PAS, and the U95\% MF-GPR uncertainty for AS. These uncertainty measures are not directly comparable, but they show that the sample uncertainty from PAS is small around $P_{\text{exp}}$ for all eight case variations (whereas the MF-GPR uncertainty band from AS is much larger). 

For cases 1 and 3, the sample variability across multiple realisations of the same wave condition is available for the validation datasets. This is not directly applicable to the present validation, as the LF simulations reproduce only one wave realisation (see \Cref{sec:valmet}); instead, the relevant uncertainty is the repeatability of the experimental HF measurements. Although not directly available, previous studies (e.g., \cite{Vprads2019} on wave impact loads measured in a seakeeping basin) indicate that this uncertainty is comparable to or smaller than the sample variability. We therefore compare the PAS uncertainty with the HF sample variability across multiple realisations of the same wave condition to assess whether our validation is meaningful. 

For case~1, the HF validation sample uncertainty is 0.1~m ($\sim$1\% of the true MPM value, see \Cref{tab:results} and \cite{VS2025}). The U95\% sample uncertainty of the PAS results at the converged iteration is 0.0~m (case 1a) and 0.07~m (case 1b), which corresponds to $\sim$0-0.7\% of the true MPM values. For green water case 3, the root-mean-square errors (RMSE) of extreme values in the HF validation data are detailed in \cite{VSS2023,VBS2024}, with the latter showing convergence to well below 0.1\% of the one-hour MPM in both test conditions. Assuming U95\% $\approx1.96$ RMSE, this corresponds to an uncertainty below 0.2\% of the MPM. By comparison, the U95\% uncertainty of the one-hour MPM predicted by PAS is 176~kN (Case~~3a) and 722~kN (Case~3b), roughly 10\% and 15\% of the true MPM values, respectively. The uncertainty of extreme values in the HF validation data was not assessed for Cases~2 and 4, but it is expected to lie between that of Cases~1 and 3 for Case~2, and to be comparable to Case~3 for Case~4.

Because the uncertainty in the extreme values of the validation data is small and of similar or lower order than that of the PAS results, the validation comparison is meaningful.

%%%%
\subsection{Efficiency}
\label{sec:res_eff}
When accuracy would be the only metric for the performance of an EVPM, MCS would always be preferred. The objective of PAS is to make accurate predictions \textit{in a more efficient way}. $M_1$ in \Cref{fig:m123} shows that both AS and PAS reduce the required HF samples to under 1.5\% of those needed for MCS across all cases. PAS is the more reliable method for accurate and efficient predictions, especially for complex cases. It requires fewer than 0.2\% of HF events compared to MCS for Cases~1a,b and 2a,b, and less than 1.5\% for Cases~3a,b and 4a,b. The exact number depends on the stopping criteria, which are case-specific; optimal limits for new cases depend on design requirements, acceptable risks, target quantiles, and possibly class society regulations. For simpler problems (Cases~1 and 2), AS is nearly as accurate and more efficient, making it a viable alternative.

In all cases, this greatly reduces the number of HF events compared to MCS or experiments. Assessing PAS feasibility for a new vessel requires translating these numbers into simulation durations. As \Cref{sec:valmet} notes, CFD was not performed for the selected HF events in the present validation. We can therefore only estimate the required durations. However, extensive literature shows that HF CFD can accurately reproduce wave impact loads when it closely matches experimental wave and ship motion data. The duration per event varies in these studies, ranging from $4T_p$ \cite{GJL2023}, 50~s \cite{BHD2015}, 35~s \cite{BH2018}, $\sim$20~s \cite{BHV2020}, to even as short as 3~s \cite{PEtAl2022}. There are also studies that initialise the CFD simulations from an LF tool instead of experiments (as done in a screening method); this can be done with 52~s \cite{GKS2023} or 12~s \cite{VMSHKSG2021} event durations. When fully non-linear wave kinematics and ship motions are provided by the LF tool, only short simulations per event are needed. In contrast, longer durations are required if waves and motions must be initialised from linear wave elevations or further from the structure. In our study, it is reasonable to assume that the waves in CFD can be initialised close to the structure, and that the ship motions can be imposed from the linear potential flow simulations. This should make it possible to reduce the duration per HF event to $\sim$20~s (following the procedure in the green water impact study \cite{BHV2020}). Using this assumption, we can translate the required number of HF events to HF simulation durations (see \Cref{tab:results}). This results in total durations to simulate with HF CFD or experiments in the order of 2 to 53~minutes, which seems feasible in the design stage of a ship. In comparison, the HF validation datasets used between 23 and 50 hours of HF experiments. Based on this, it is estimated that the PAS converged results can be obtained by analysing between 0.04\% and 2.2\% of the HF MCS durations with an HF tool across all cases. Note that these percentages are roughly twice the $M_1$ values, because with typical wave periods of around 10~s, a 20~s event corresponds to two wave encounters. 

The computational time required for these simulations (in CPU hours, CPUh) also depends on the HF tool used to simulate these 2-53~minutes, as well as the chosen grid size, domain, and time-step settings (in case of CFD) for each case. When using PAS, the full MCS duration must be simulated with an LF tool, but with frequency-domain potential flow or analytical methods (as in all cases here) the cost is negligible compared to HF simulations. For example in cases 2 to 4, the frequency-domain \texttt{SEACAL} calculations (see \Cref{sec:pas_implem}, 1-2 speeds and 1-2 headings) required approximately 0.25 CPUh per case on a standard laptop, with analytical linear wave simulations requiring even less. Conservatively, we estimate a maximum of 1 CPUh for LF simulations, including the transformation from frequency to linear time domain. Running PAS itself requires at most 1 min per iteration (0.017 CPUh), resulting in a total of approximately 2-3 CPUh for all iterations. Overall, the expected computational cost is roughly 1~CPUh for LF simulations, 2-3~CPUh for PAS, plus the time for HF CFD or experimental simulations, which typically cover up to one hour of simulated time divided over 7-160 events. The HF simulations dominate the total cost. A future study implementing the full PAS workflow with HF simulations will enable a more precise estimate.

%%%%
\subsection{Possible error sources}
\label{sec:res_error}
PAS is not guaranteed to converge to the exact HF exceedance distribution, even with increasing LF-HF samples. This is inherent to exceedance probability estimation, which is based on binary events (exceeds or not), and is therefore very sensitive to individual observations, especially in the tails. For instance, a single low-indicator but high-HF-load value can substantially shift the HF exceedance distribution, as also observed in AS and other screening studies \cite{BSPFS2018,BSD2019,VS2025}. Consequently, adding more samples does not always improve tail estimates monotonically. New observations change the empirical pseudo-observations, the fitted copula and marginal parameters, and redistribute exceedance probabilities. Finite-sample effects mean that these updates can introduce noise, slow convergence, or shifting tail behaviour rather than refining it. In this context, convergence means stabilisation of the fitted model under the chosen sampling and modelling assumptions, rather than asymptotic convergence to the true distribution. This risk is lower with a copula model than in sampling approaches without it, as the copula introduces a probabilistic joint dependence structure rather than a deterministic one. The outlier removal feature also dampens this effect.

A risk when combining copula modelling with adaptive sampling is bias introduced by using targeted samples. In general, adaptive sampling can produce non-\textit{i.i.d.} data (independent and identically distributed) and deform dependence estimates by over-representing certain regions. Copula models treat the transformed marginals as if they were draws from a uniform distribution (see e.g., \cite{N2006,GF2007}). Small or concentrated samples can violate this assumption and increase bias and instability. In the present case, this risk is mitigated by using an acquisition function that distributes samples approximately uniformly over the logarithm of the LF exceedance probability space, rather than focusing on tails or uncertainty. This leads to a balanced coverage over the probability levels. 

Large differences in order statistics between the LF indicator and the target HF variable (i.e., a poor indicator choice) can also distort the fitted dependence structure, especially in the tails, and lead to biased HF exceedance estimates. For example, in the green water case, a large bow flare may deflect water to the sides. A high RWE (LF indicator) would then not correspond well to the green water load on the deck (HF load). Other potential errors come from imperfect copula choices and uncertainty in picking the best model from a limited set, errors in marginal distribution fitting and interpolation, Monte-Carlo sample variability in conditional simulation, and the fact that the stopping criterion reflects stability between iterations rather than consistency with the true exceedance distribution (necessary because the true distribution is unknown in a real design case). 

\begin{figure}[t!]
	\centering
     \begin{subfigure}[b]{0.195\textwidth}
        \centering
        \includegraphics[width=\textwidth]{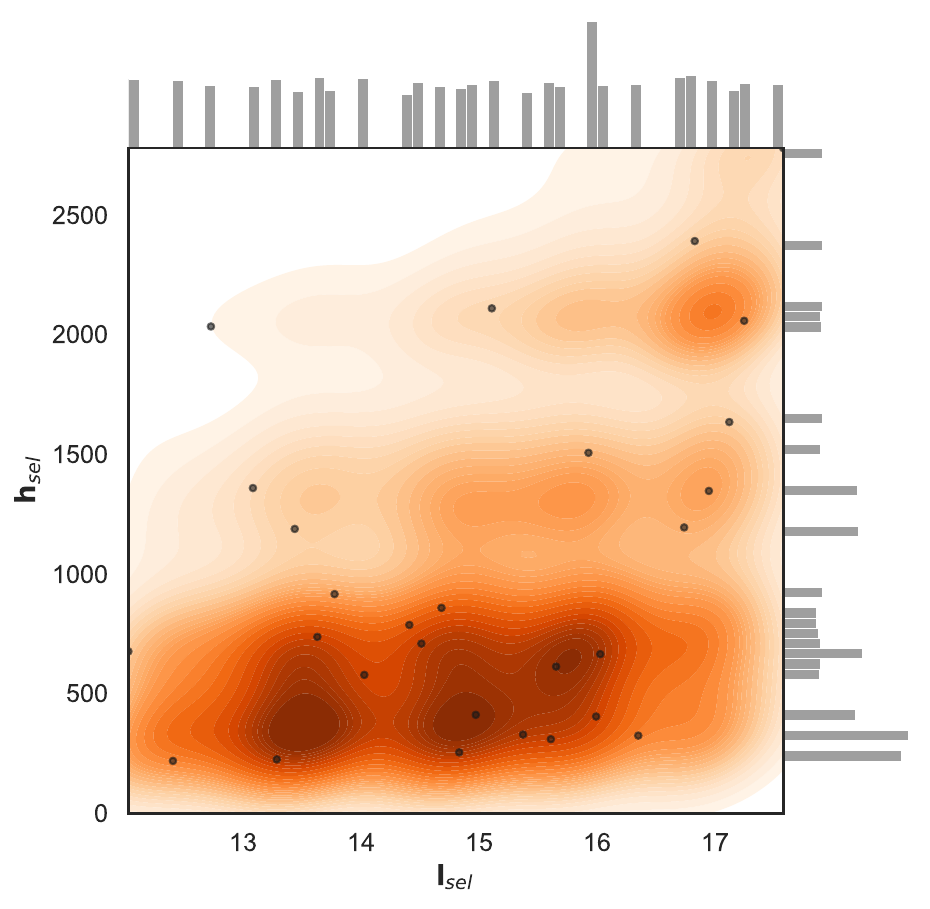}
        \caption{Gumbel (AIC -5.1)}
        \label{fig:copvar2}
    \end{subfigure}
    \begin{subfigure}[b]{0.195\textwidth}
        \centering
        \includegraphics[width=\textwidth]{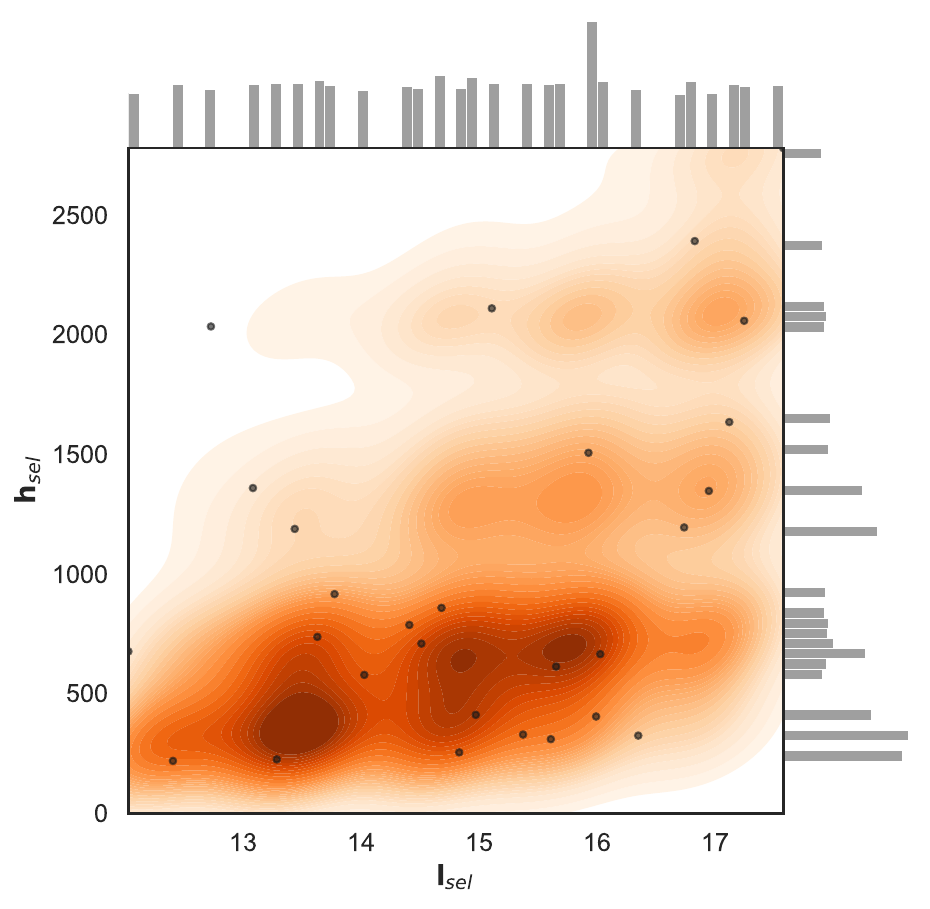}
        \caption{Gaussian (AIC -3.9)}
        \label{fig:copvar1}
    \end{subfigure}
    \begin{subfigure}[b]{0.195\textwidth}
        \centering
        \includegraphics[width=\textwidth]{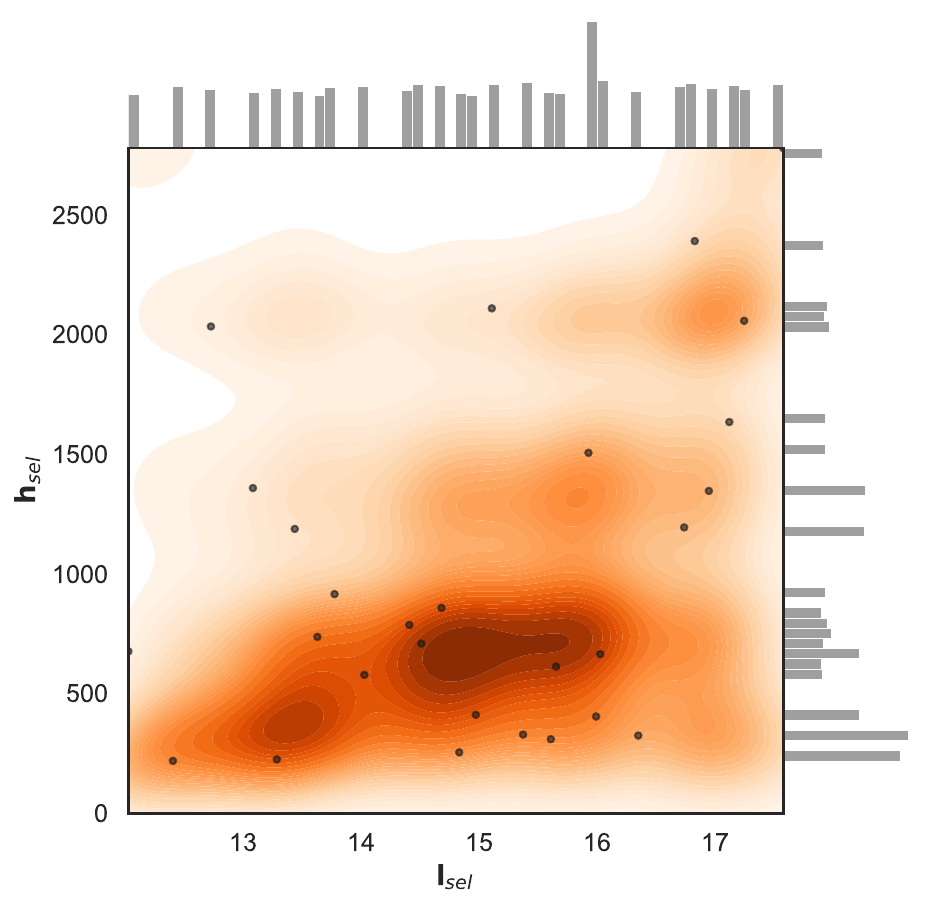}
        \caption{Student-T (AIC -2.8)}
        \label{fig:copvar5}
    \end{subfigure}
    \begin{subfigure}[b]{0.195\textwidth}
        \centering
        \includegraphics[width=\textwidth]{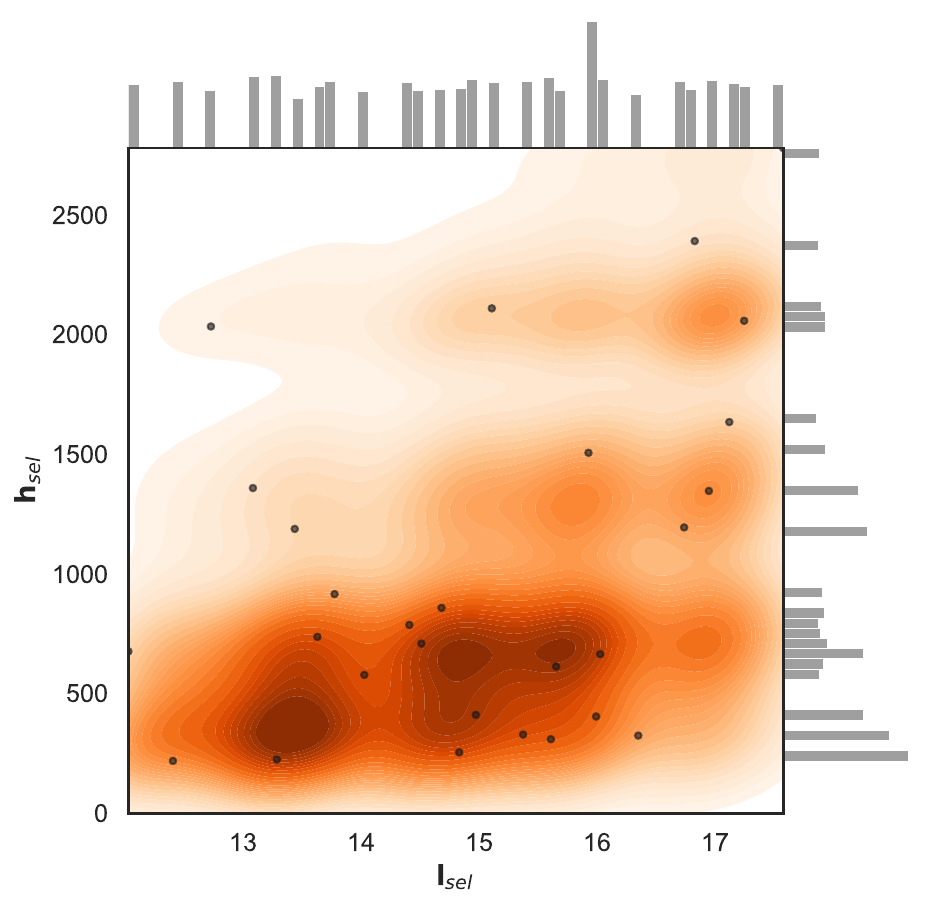}
        \caption{Frank (AIC -1.7)}
        \label{fig:copvar4}
    \end{subfigure}
    \begin{subfigure}[b]{0.195\textwidth}
        \centering
        \includegraphics[width=\textwidth]{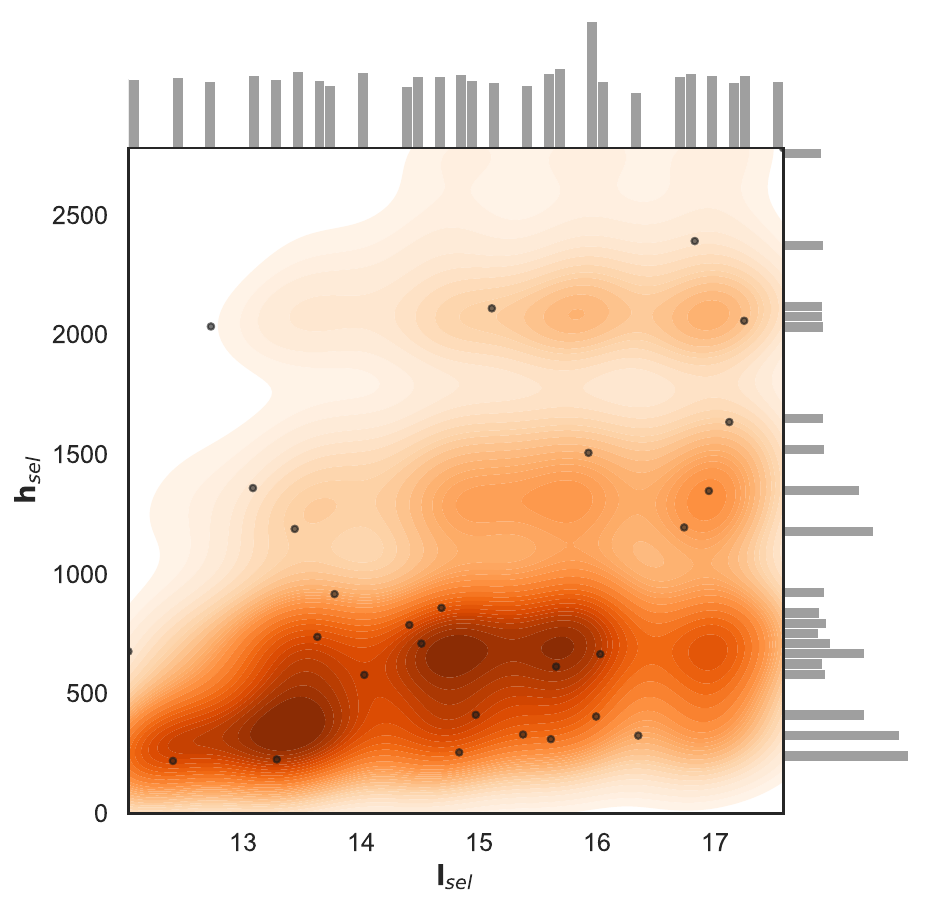}
        \caption{Clayton (AIC -0.7)}
        \label{fig:copvar3}
    \end{subfigure}    
    \caption{Different copula models fitted to the LF-HF sample data at converged PAS iteration 47 (51 HF samples of which 28 true positives), for green water case 3a. The plots are ordered by lowest AIC (best-fitting) on the left to highest AIC on the right.}
	\label{fig:copvar}
\end{figure}

\begin{figure}[t!]
	\centering
        \begin{subfigure}[b]{0.45\textwidth}
            \centering
            \includegraphics[height=4cm]{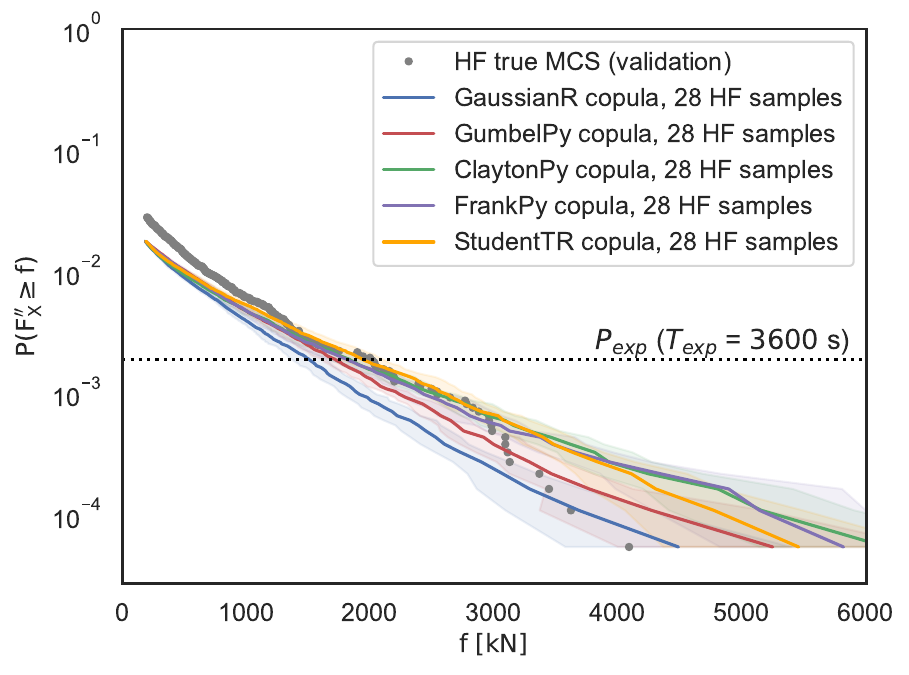}
            \caption{It. 47 (51 HF samples of which 28 true positives)}
            \label{fig:copvardist1}
        \end{subfigure}
        % \hfill
        \begin{subfigure}[b]{0.45\textwidth}
            \centering
            \includegraphics[height=4cm]{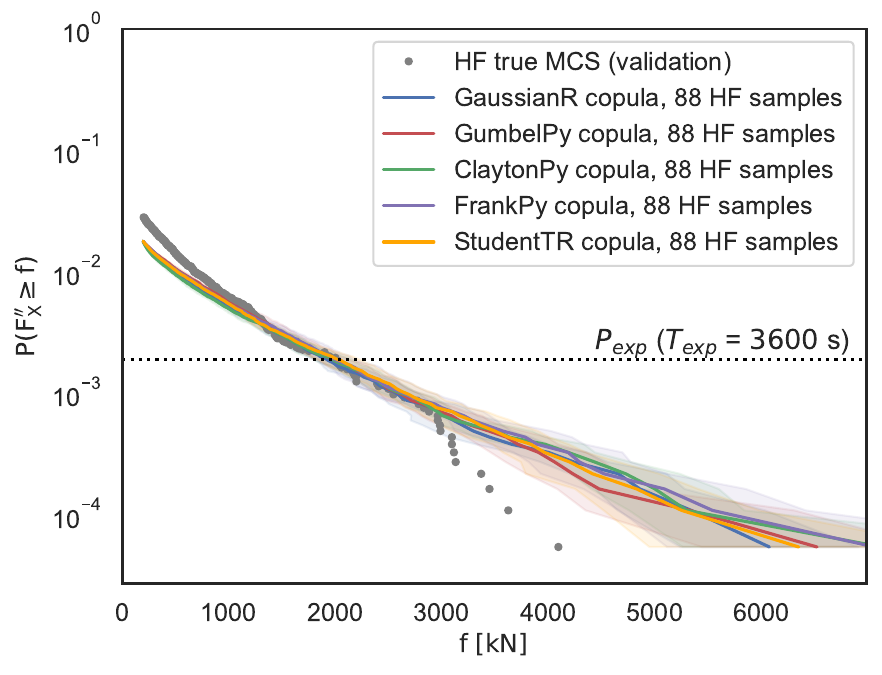}
            \caption{It. 150 (154 HF samples of which 88 true positives)}
            \label{fig:copvardist2}
        \end{subfigure}
	\caption{Exceedance probability distributions (mean and U95\%) resulting from 10 conditional draws from the copula models fitted to the LF-HF samples at iteration 47 and 150, for green water case 3a.}
	\label{fig:copvardist}
\end{figure}

%%%%
\subsection{Sensitivity to copula model}
\label{sec:res_senscop}
Since PAS results are largely driven by the fitted copula, and this step differs from AS, we evaluated the sensitivity of the results to the selected copula model in more detail. 

Firstly, convergence of the copula selection process is discussed in \Cref{app:copulafittingdetails}. As explained in \Cref{sec:pas_copsel}, the AIC-based copula selection in PAS adapts to the currently available HF samples rather than seeking a single `true' copula model. The preferred copula may therefore change as new samples are added. The appendix shows that, over all cases, the Gaussian copula is most often selected. This is likely due to its robustness with sparse data combined with the stabilising effect of the AIC threshold. When an alternative model is chosen, it is typically the Clayton copula, consistent with the observed dominance of lower-tail dependence (especially for cases 3 and 4, see \Cref{fig:scatters}). 

Secondly, we assess the sensitivity of predicted distributions to the copula model, by analysing green water Case~3a at iteration 47. \Cref{fig:copvar} shows the 28 true positive samples at this iteration, together with the five candidate copula models fitted to these samples. This shows that the fitted copulas are quite similar overall, but there are also differences. The plots are ordered by AIC value from low (left, best-fitting) to high (right); the Gumbel and Gaussian copulas fit the dataset best at this iteration. \Cref{fig:copvardist1} shows the drawn conditional distributions from each of these models at the same iteration. This shows that the distribution differences are limited at $P_{\text{exp}}$, but the tails can be significantly different. Selecting the most-likely copula model per iteration instead of \textit{a priori} selecting one model can therefore pay off in the prediction of the extremes. The same is plotted for PAS iteration 150 in \Cref{fig:copvardist2}. This shows that the difference in the tail reduces with an increasing number of samples. In other words, for a large number of samples, all copula models seem to converge to a similar distribution. 

\begin{figure}[t]
	\centering
        \begin{subfigure}[t]{\textwidth}
            \centering
            \includegraphics[width=\textwidth]{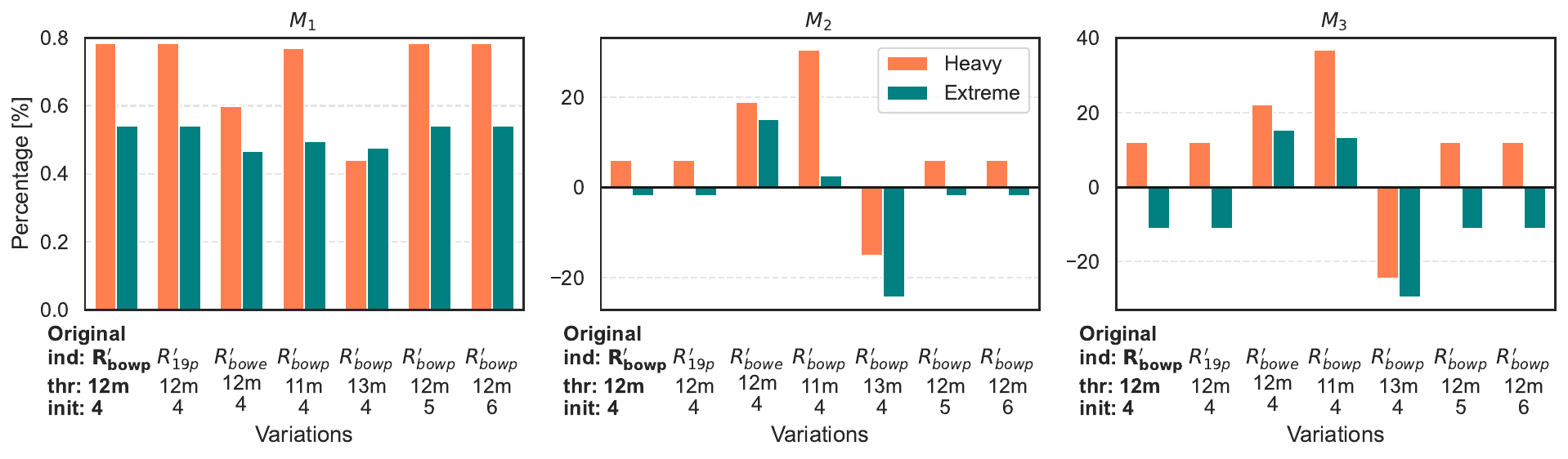}
        \subcaption{Performance metrics $M_1$, $M_2$ and $M_3$.}
        \label{fig:sensGWa}
        \end{subfigure}
        % ---------- Left bottom column ----------
        \begin{subfigure}[t]{0.4\textwidth}
            \centering
            \includegraphics[width=\textwidth]{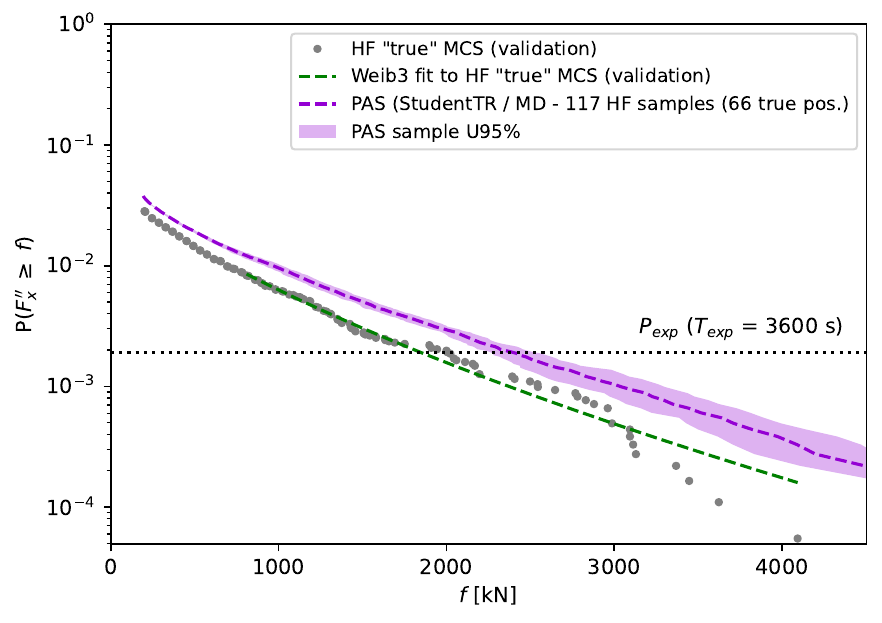}
        \subcaption{PAS Heavy distribution, threshold 11 m }
        \label{fig:sensGWb}
        \end{subfigure}
        % ---------- Right bottom column ----------
        \begin{subfigure}[t]{0.4\textwidth}
            \centering
            \includegraphics[width=\textwidth]{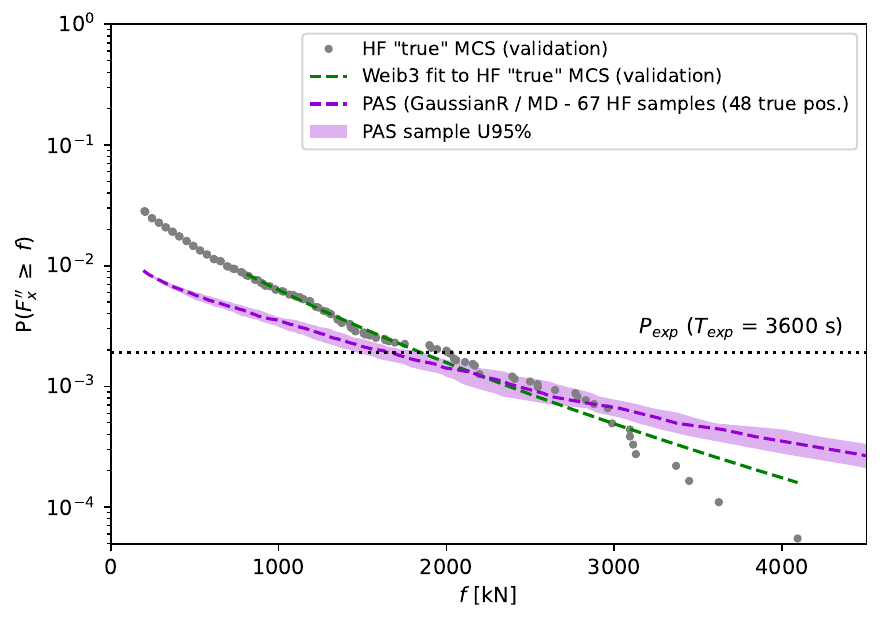}
        \subcaption{PAS Heavy distribution, threshold 13 m}
        \label{fig:sensGWc}
        \end{subfigure}
	\caption{Sensitivity of the green water results for indicator choice, LF threshold choice and number of initial samples (with modified COV criterion 5\%, so results of the original case differ slightly from \Cref{fig:res3}), expressed as performance metrics, with some example distribution plots. Ind = indicator, thr = LF threshold, init = number of initial samples, and $R_{Xp}$ = RWE at location X from potential flow, $R_{Xe}$ = RWE at location X from experiments.} 
	\label{fig:sensGW}
\end{figure}

\begin{figure}[t]
	\centering
        \begin{subfigure}[t]{\textwidth}
            \centering
            \includegraphics[width=.75\textwidth]{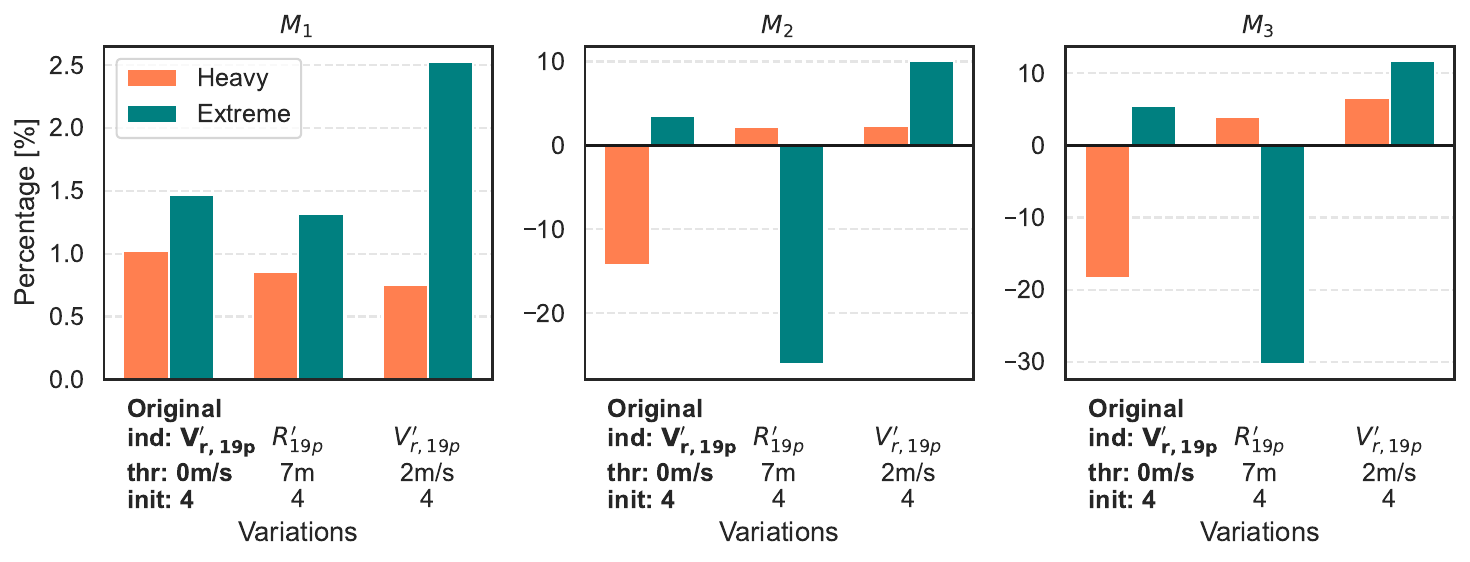}
        \subcaption{Performance metrics $M_1$, $M_2$ and $M_3$.}
        \label{fig:sensSLa}
        \end{subfigure}
        % ---------- Left bottom column ----------
        \begin{subfigure}[t]{0.4\textwidth}
            \centering
            \includegraphics[width=\textwidth]{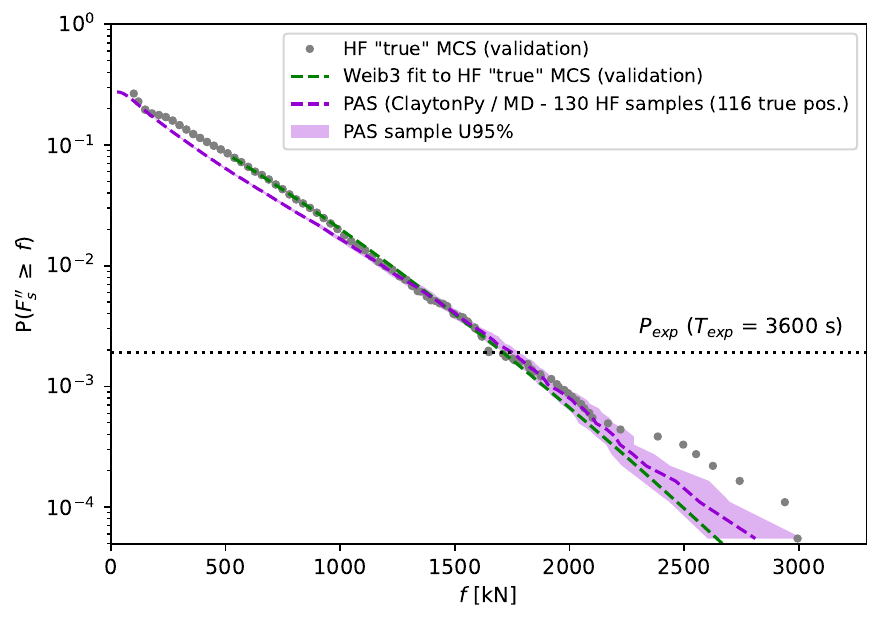}
        \subcaption{PAS Heavy distribution, indicator $R_{19p}'$}
        \label{fig:sensSLb}
        \end{subfigure}
        % ---------- Right bottom column ----------
        \begin{subfigure}[t]{0.4\textwidth}
            \centering
            \includegraphics[width=\textwidth]{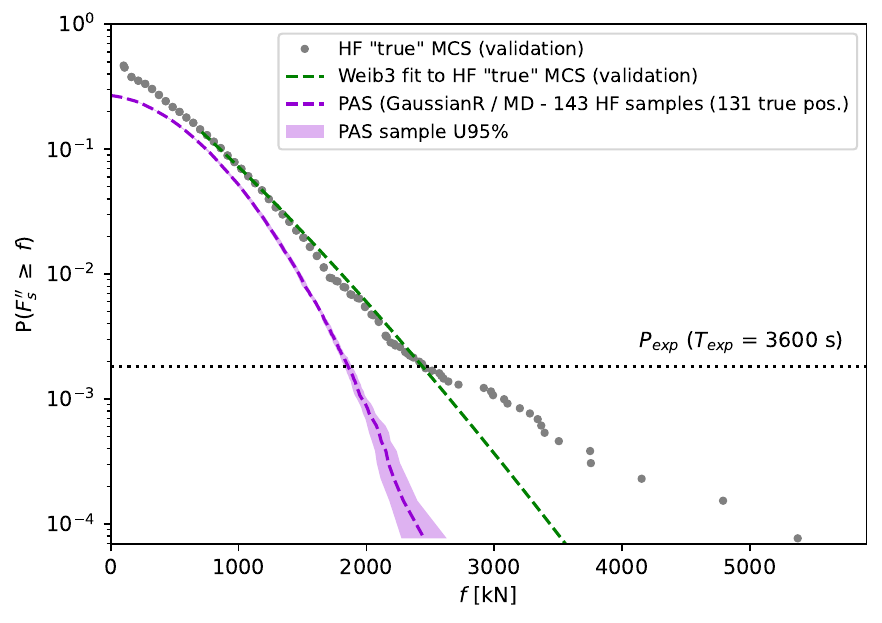}
        \subcaption{PAS Extreme distribution, indicator $R_{19p}'$}
        \label{fig:sensSLc}
        \end{subfigure}
	\caption{Sensitivity of the slamming results for indicator choice and LF threshold choice, expressed as performance metrics, with some example distribution plots. Ind = indicator, thr = LF threshold, init = number of initial samples, and $V_{r,Xp}'$ = RRV at location X from potential flow, $R_{Xp}'$ = RWE at location X from potential flow.} 
	\label{fig:sensSL}
\end{figure}

%%%%
\subsection{Sensitivity to other parameters}
\label{sec:res_sens}
In addition to the copula model, PAS includes many internal parameters and user-defined inputs that can be adjusted. For most settings, we selected values that perform robustly across the considered problems, as demonstrated by the four cases in this study. Here, we present a sensitivity analysis of the most influential choices: the LF indicator, LF threshold and initial sampling, and we further discuss the impact of the acquisition function and stopping criteria.

First, we consider the sensitivity of the green water results (Case 3) for indicator choice, LF threshold, and initial sampling. We ran PAS with three LF indicators: bow RWE from potential flow ($R_{\text{bow}p}'$, the original case with added subscript $p$ to emphasise its source), station-19 RWE from potential flow ($R_{19p}'$), and bow RWE from experiments ($R_{\text{bow}e}$). We also tested two alternative LF thresholds (11 m and 13 m, versus the original 12 m) and two additional initial sample sizes (5 and 6, versus 4), all selected as in \Cref{sec:pas_init}. Results are summarised in \Cref{fig:sensGW}, using a modified COV stopping criterion of 5\% instead of 10\%. While 10\% was identified as the optimal choice across all four cases, a stricter criterion was required for case~3 alone to ensure convergence across all its variations; consequently, the results for the original case differ slightly from those shown in \Cref{fig:res3}. A similar plot is shown for two variations of the slamming case in \Cref{fig:sensSL} (with the same COV criterion as used for \Cref{fig:res4}). Here, we ran PAS with two LF indicators: the RRV at station 19 from potential flow ($V_{r,19p}'$, the original case) and the RWE at station 19 from potential flow ($R_{19p}'$). We also varied the LF threshold from no threshold (original case) to 2 m/s. For both cases, all other settings were kept unchanged. The results shown are converged according to the combined criterion, implying some dependence on the stopping limits. Nevertheless, the comparison provides a clear indication of the sensitivities.
% Fewer than four initial samples proved unreliable in the presence of false positives, often yielding only one true positive starting point.

Based on the order-statistics assumption that is the foundation of screening-based methods, the choice of indicator is expected to strongly influence results. This was also observed in \cite{VS2025}, where AS with a coarse-mesh CFD RWE indicator produced accurate green water distributions, whereas the potential-flow RWE indicator used here did not. The copula model in PAS was introduced to reduce sensitivity to indicator choice by modelling LF-HF relation probabilistically rather than deterministically, thereby limiting the influence of outliers. The three green water indicators in \Cref{fig:sensGW} support this idea: the difference in results between the two potential-flow indicators is small, indicating that the specific choice does not strongly affect the results. This is likely because these variables are internally statistically consistent. Interestingly, using the experimental bow RWE as indicator yields poorer results than the linear potential-flow RWE, probably due to measurement noise in the experimental signal. For the two slamming indicators in \Cref{fig:sensSL}, we see something slightly different; both perform well in one case but less so in the other. This is also visible when comparing the predicted slamming distributions with indicator $R_{19p}'$ in \Cref{fig:sensSLb,fig:sensSLc} to those of the original indicator $V_{r,19(p)}'$ in \Cref{fig:res4}. This suggests that neither indicator is fully adequate and that indicator choice is more critical for slamming than for green water. As discussed before, we did not know `the best' indicators for our cases, but selected a good candidate indicator based on experience and earlier work. There could be even better indicators, with fewer false positives for these cases, especially for slamming (such as e.g., the Wagner model \cite{HM2007} or momentum theory \cite{K2018}).

\Cref{fig:sensGW} also shows the sensitivity of the green water results to the selected LF threshold. The original threshold of 12~m provides very good results under both test conditions, whereas lowering the threshold to 11~m or increasing it to 13~m both leads to deteriorated performance. Setting the threshold is challenging. A value that is too high creates unbalanced sampling that focuses on extreme events, reducing the quality of the copula fit to the full dataset. A value that is too low on the other hand complicates the definition of the HF distribution zero-crossing, and reduces the efficiency by increases the number of false negatives (though this effect is not yet pronounced for 11~m, see $M_1$ in \Cref{fig:sensGW}). Choosing a threshold slightly above the physical threshold (in this case the freeboard of 10.5~m) appears to be a good compromise, with 12~m performing well. For the slamming case, we also introduced a rise velocity threshold of 2~m/s (see \Cref{fig:sensSL}) to assess whether this sensitivity to threshold selection is also valid for other cases. This threshold similarly leads to an overestimation of the MPM and its distribution, as sampling becomes more focused on the tail of the distribution. 

\Cref{fig:sensGW} also shows that the influence of the number of initial samples is negligible, as expected since they are defined analogously to the MD function. 
% Adding an extra initial sample primarily affects the speed of convergence rather than the final results. An even-odd effect may occur, whereby even initial sample counts (2, 4, 6, ...) lead to the same selected samples (up to a small iteration delay) because the MD acquisition function places new points at midpoints on a logarithmic scale. Odd initial numbers (3, 5, 7, ...) therefore yield slightly different samples, but the figure shows that this has a negligible impact on the predicted results.

\begin{figure}[t!]
    \centering
    \begin{subfigure}[b]{0.49\textwidth}
        \centering
        \includegraphics[width=\textwidth]{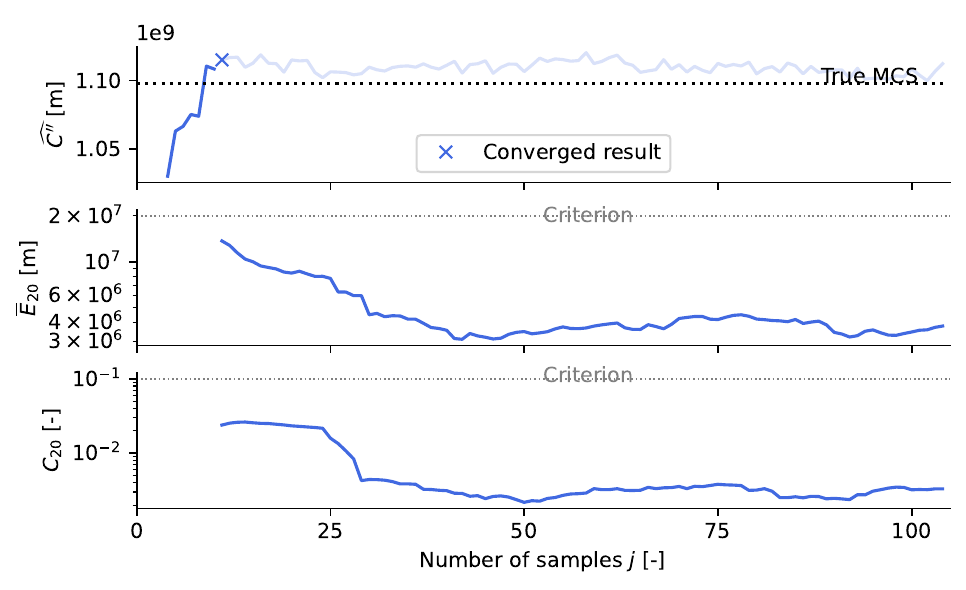}
        \caption{Hogging VBM (from \Cref{fig:res2a}).}
        \label{fig:sensStopa}
    \end{subfigure}
    \begin{subfigure}[b]{0.49\textwidth}
        \centering
        \includegraphics[width=\textwidth]{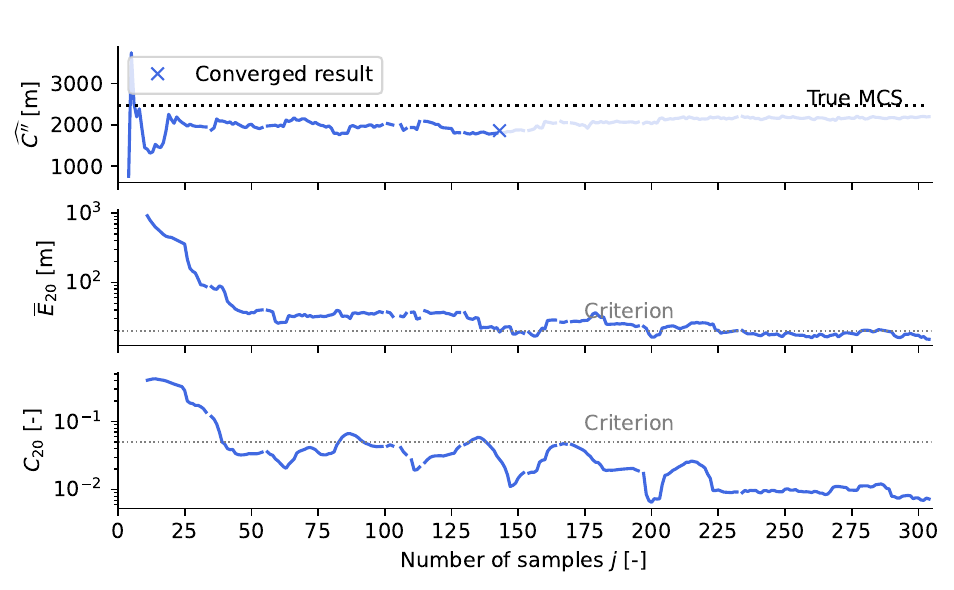}
        \caption{Slamming, Extreme condition, ind. $R_{19p}'$ (from \Cref{fig:sensSL}).}
        \label{fig:sensStopb}
    \end{subfigure}
    \caption{Stopping criteria and convergence limits for simple and complex example cases, both combined with an $\epsilon_1$ criterion in the range of 1-2\% of the maximum HF value in the validation dataset (middle plot), and $\epsilon_2=0.1$ (bottom plot). Top plot shows MPM as a function of the number of HF samples.}
	\label{fig:sensStop}
\end{figure}

In \cite{VS2025prads}, we evaluated seven acquisition functions for AS. For PAS, coverage of the full exceedance probability domain is especially important (as explained in \Cref{sec:pas_adap}). While we did not repeat the full sensitivity study, we tested most previous functions: only the MD function consistently performed well with the copula, as most other functions focus on subregions. A variant of MD aiming for uniform coverage of the full domain on linear instead of logarithmic scale also worked, but was less efficient and occasionally produced badly-behaving tail predictions. 

PAS results are sensitive to the stopping criterion settings: averaging windows $K_1$ and $K_2$ and convergence thresholds $\epsilon_1$ and $\epsilon_2$ (see \Cref{app:pas_stop}). Ideal settings depend on the case and are generally unknown: noisier LF-HF statistics may converge best with relatively lenient criteria and large averaging windows, while smoother cases quickly converge with strict limits and short windows. Long averaging windows generally improve stability of the convergence, but also slow it down (which is expensive if every extra sample requires a CFD simulation). For robustness across our four test cases, we chose $K_1 = K_2 = 20$, $\epsilon_1 \approx$1-2\% of the expected maximum HF value, and $\epsilon_2 = 0.1$ (10\% COV of the MPM). These settings typically indicate convergence once the MPM and distribution stabilise over 20 iterations, but \Cref{fig:sensStopb} shows that `stable' predictions do not always reflect the true converged value; results may still shift with more iterations. The figure also shows that the current averaging window is short for sensitive cases like slamming (\Cref{fig:sensStopb}) but more than sufficient for simpler cases (\Cref{fig:sensStopa}). The chosen settings represent a practical balance.

%%%%
\subsection{Future work}
\label{sec:res_fut}

In the present validation approach (see \Cref{sec:valmet}), the HF response is not obtained by explicitly running CFD simulations but is instead extracted directly from the validation dataset. As a result, the potential impact of CFD simulation errors on PAS performance is not accounted for. Two distinct aspects are involved: (i) the numerical accuracy of the CFD simulations themselves, and (ii) the statistical equivalence between LF wave events and their replication in HF CFD simulations. Both aspects are related to substantial research questions, beyond the scope of the present study. With respect to CFD accuracy (aspect i), some insight is available from earlier work, as discussed in \Cref{sec:LFHF}. In particular, \cite{BHV2020} showed that CFD could successfully reproduce experimental green water loads for case~3a, supporting its suitability for capturing the relevant physics. As noted in \Cref{sec:pas_steps}, initialising HF event simulations in \ref{st:hfresp} of PAS using the LF screening results is not straightforward (aspect ii). No universally accepted method exists to correlate linear (or weakly non-linear) wave events with equivalent fully non-linear wave events, or to define the corresponding acceptable tolerances in wave parameters. Potential solutions include event-matching procedures \cite{JL2018,GJL2023} or using coarse-mesh CFD as an LF tool \cite{VMSHKSG2021}. The first method is less suitable for sailing ships, and the second requires costly CFD. An alternative is to perform MCS with a fully non-linear wave-only tool (e.g., a higher-order spectral method, HOS) and then apply linear potential-flow screening to the resulting wave traces, allowing HF simulations to be initialised directly from the corresponding HOS events. A promising direction for future research is to perform the full validation of PAS including HF simulations, using a suitable CFD tool paired with this HOS-initialisation approach.

PAS could potentially be further improved and made more robust for broader applications, for example by using different marginal and copula models for tail versus body (to improve tail predictions in the simpler cases), selecting multiple new samples per iteration for parallel HF simulations, or incorporating copula fit quality over the LF range into the acquisition function. PAS could also use event frequencies instead of exceedance probabilities, or adopt a block-maximum / return-period approach instead of POT. This would improve interpretability, align with industry practice, and allow easier application to non-wave problems by referencing fixed-duration blocks rather than variable wave events. However, it would reduce efficiency, since each block spans many waves (e.g., 47 samples would require 47 block simulations instead of 47 wave events). Validating a block-maximum approach in a similar context as used here would require further research to define representative paired peaks, either as the block's maximum HF response or the HF response of the wave with the maximum LF value.

The MPM predicted in \ref{st:mpm} of PAS corresponds to the $q=e^{-1}\approx0.368$ quantile of the short-term distribution for linear Gaussian processes (see e.g., \cite{Ochi1990}). Accordingly, the probability that the maximum response over the target exposure duration exceeds the MPM is 62.3\%. In offshore design, higher quantiles are often preferred. These can in theory be estimated by replacing $P_{\text{exp}}$ in \Cref{eq:MPMfromDNR} with $P_q=1-q^{m/n}$ \cite{OTG14}, but this requires substantially longer LF MCS to ensure convergence. Further work is therefore needed to assess the method's applicability at higher quantiles.

As noted in \Cref{sec:objectives}, this study validates PAS only for short-term extremes. In practical applications, sea states and long-term statistics must also be handled, which could be addressed with PAS as well, with minor modifications (e.g., to \ref{st:MCS}). Alternatively, a version of the environmental contour method, as described in \cite{WEtAl1993,HVN2013,MMJ2025}, could be used. This approach is widely accepted in the offshore industry (e.g., \cite{DNV2019}), though it can sometimes fail to accurately characterise the joint distribution of environmental variables \cite{SRTJ2024,EnvContBenchmark,DMV2022}. Another option is to use another adaptive sampling method for the long-term screening, as proposed by \cite{GEtAl2020,WGSMV2024}. 

Interpreting the significance of PAS wave impact extreme value errors (2-15\% in the MPMs of the green water and slamming case) in an engineering context is challenging. As discussed in \Cref{sec:existing}, there is no universally accepted extreme value prediction framework for wave impact loads, and in practice the industry usually follows prescriptive standards or relies on direct analysis using experiments. Such approaches are typically combined with substantial safety margins and prediction of higher quantiles. Exact values of these margins are difficult to obtain, but they are generally understood to exceed the 2-15\% level associated with the MPM error reported here. The relationship between such margins and the statistical uncertainty of experimental extreme value estimates is, however, not clearly defined. Embedding the use of PAS and its associated uncertainties within a formal design context, including safety margins and workflow, is considered an important topic for future research.

As a framework for combining LF and HF model data, PAS has no inherent application limitations beyond those of the underlying LF and HF models. It could also be generalisable to non-linear extreme value problems in other fields (possibly with some small modifications to account for non-wave-based problems). Nonetheless, its performance should be validated again when applied to very different types of problems (e.g., in other fields).

%%%%%%%%%%%%%%%%%%%%%%%%%%%%%%%%%%%%%%%%%%%%%%%%%%%%%%%%%%%%%%%%%%%%%%
%%%%%%%%%%%%%%%%%%%%%%%%%%%%%%%%%%%%%%%%%%%%%%%%%%%%%%%%%%%%%%%%%%%%%%
%%%%%%%%%%%%%%%%%%%%%%%%%%%%%%%%%%%%%%%%%%%%%%%%%%%%%%%%%%%%%%%%%%%%%%

\section{Conclusions}
\label{sec:Conclusions}
Based on the present work, it can be concluded that the newly proposed extreme value prediction method PAS provides accurate and efficient estimates of exceedance distributions and extreme values for a range of non-linear wave-induced response problems, compared to the available full brute-force MCS datasets. The method uses a probabilistic approach with copulas to model the joint distribution of low-fidelity (LF) and high-fidelity (HF) variables. This enables the use of efficient linear potential-flow indicators in the LF stage, even for strongly non-linear loads. PAS then predicts extreme HF responses by statistically linking these LF indicators to a limited set of HF samples, effectively learning the dependence between the two fidelity levels. Results are iteratively updated using adaptive sampling, until convergence is reached. PAS with optimal settings achieves a most probable maximum (MPM) accuracy of 2-15\% for all considered cases in the present study, including non-linear waves, vertical bending moments, green water impact loads, and slamming loads. In addition, PAS achieves this performance very efficiently, requiring in the order of 1-3\% of the high-fidelity simulation time needed for conventional MCS over all cases. While PAS is similarly accurate and slightly less efficient than its predecessor AS for the simpler cases (non-linear waves and vertical bending moments), it proves significantly more reliable for the more complex cases (green water and slamming loads). These results demonstrate that PAS can reliably reproduce the statistics of both weakly and strongly non-linear extreme load problems, while significantly reducing the associated computational cost.

PAS is largely insensitive to the choice of copula, though tail predictions can vary for small sample sizes; selecting the best-fitting copula from five candidates at each iteration helps mitigate this sensitivity. PAS green water predictions are less sensitive to indicator choice than those from AS, performing well with different potential-flow indicators. Slamming predictions are more sensitive, suggesting potential for improved indicators. For cases with a physical LF threshold below which many HF loads are zero, the selected input LF threshold strongly affects results: too high reduces the quality of the copula fit across the full exceedance range, too low decreases efficiency. A value slightly above the threshold is a good compromise, while zero is best when no physical threshold exists. PAS is also sensitive to the acquisition function (only the MD function reliably covers the full exceedance probability domain), and to the stopping criteria, which must balance accuracy for complex cases with efficiency for simpler ones. The number of initial samples has negligible effect under the current selection method.

The present study validates the statistical PAS framework. PAS does not impose explicit theoretical limits on structure type or wave conditions. Its applicability is limited by the availability of a suitable LF indicator with order statistics comparable to the HF response. The cases studied here demonstrate feasibility for the investigated scenarios but do not define hard boundaries of the method. Further work should focus on validating the full procedure including CFD load simulations, assessing long-term extremes and extremes at higher quantiles, embedding the use of PAS within a formal design context, and possibly testing it against strongly non-linear problems beyond wave-related scenarios.

% \clearpage
\FloatBarrier   % all pending figures will be placed before this point

%%%%%%%%%%%%%%%%%%%%%%%%%%%%%%%%%%%%%%%%%%%%%%%%%%%%%%%%%%%%%%%%%%%%%%
%%%%%%%%%%%%%%%%%%%%%%%%%%%%%%%%%%%%%%%%%%%%%%%%%%%%%%%%%%%%%%%%%%%%%%
%%%%%%%%%%%%%%%%%%%%%%%%%%%%%%%%%%%%%%%%%%%%%%%%%%%%%%%%%%%%%%%%%%%%%%

\section*{Acknowledgements}
\label{sec:acknowledgements}
This publication is part of the project ``Multi-fidelity Probabilistic Design Framework for Complex Marine Structures'' (project TWM.BL.019.007) of the research programme ``Topsector Water \& Maritime: the Blue route'' which is (partly) financed by the Dutch Research Council (NWO). We thank the \href{https://crships.org/}{Cooperative Research Ships (CRS)} for permitting us to use the \texttt{PySeaWave} package, the \texttt{SEACAL} and \texttt{PRETTI\_R} programs, and the experimental data of cases 3 and 4. We also thank Cees de Valk of KNMI for suggesting copula modelling as a way to reduce bias in sampling methods. Finally, we thank the anonymous Ocean Engineering reviewers as well as Tormod Landet and Guillaume de Hauteclocque of the CRS SPEC working group for their useful comments on the first version of this paper.

%%%%%%%%%%%%%%%%%%%%%%%%%%%%%%%%%%%%%%%%%%%%%%%%%%%%%%%%%%%%%%%%%%%%%%
\section*{Data availability}
\label{sec:data}
All scripts underlying this publication, the full datasets of case 1 and 2, and the relevant parts of the experimental dataset for cases 3 and 4 are available in the 4TU repository: \cite{VSdata2025pas}. % \hl{Clean scripts and update!}.  % \hl{Consider restructuring and cleaning the code and putting it on Github / 4TU. > for the final version.}

%%%%%%%%%%%%%%%%%%%%%%%%%%%%%%%%%%%%%%%%%%%%%%%%%%%%%%%%%%%%%%%%%%%%%%
%%%%%%%%%%%%%%%%%%%%%%%%%%%%%%%%%%%%%%%%%%%%%%%%%%%%%%%%%%%%%%%%%%%%%%
%%%%%%%%%%%%%%%%%%%%%%%%%%%%%%%%%%%%%%%%%%%%%%%%%%%%%%%%%%%%%%%%%%%%%%

\begin{appendices} 
\crefalias{section}{appsec}
\crefalias{subsection}{appsec}

%%%%%%%%%%%%%%%%%%%%%%%%%%%%%%%%%%%%%%%%%%%%%%%%%%%%%%%%%%%%%%%%%%%%%%
%%%%%%%%%%%%%%%%%%%%%%%%%%%%%%%%%%%%%%%%%%%%%%%%%%%%%%%%%%%%%%%%%%%%%%
%%%%%%%%%%%%%%%%%%%%%%%%%%%%%%%%%%%%%%%%%%%%%%%%%%%%%%%%%%%%%%%%%%%%%%

\section{Details of the Adaptive Screening (AS) method}
\label{app:AS}

Here, the steps of AS are briefly summarised. This is a shortened reproduction of the full description and formulations in \cite{VS2025}. Steps 1 to 6 of AS are identical to \ref{st:defind} to \ref{st:hfresp} described in \Cref{sec:pas_steps}. The next three steps are different. They are described here in general terms (for details, see \cite{VS2025}). \textbf{Step 7} estimates the sample HF distribution, by assuming that the order statistics of $\mathbf{l}^{\text{sel}}$ and $\mathbf{h}^{\text{sel}}$ are identical. This is a critical screening assumption that only works if a suitable indicator signal is chosen. It indicates that the HF distribution $[\mathbf{d}^{\text{sel}}_{H},\mathbf{h}^{\text{sel}}]$ can be estimated using $\mathbf{d}^{\text{sel}}_{H} \approx \mathbf{d}^{\text{sel}}_{L}$. \textbf{Step 8} defines an exceedance probability range $\mathbf{d}^* \in [0,1]$ around $P_{\text{exp}}$, over which to estimate the HF distribution. \textbf{Step 9} constructs the surrogate HF distribution $\mathbf{h}^*$ over $\text{ln } (\mathbf{d}^*$) using single- or multi-fidelity Gaussian Process Regression (GPR). With single-fidelity GPR, we use the HF sample dataset $[\text{ln }(\mathbf{d}^{\text{sel}}_{H}),\mathbf{h}^{\text{sel}}]$ as input. With multi-fidelity MF-GPR, we use the same HF sample dataset $[\text{ln }(\mathbf{d}^{\text{sel}}_{H}),\mathbf{h}^{\text{sel}}]$ as input, \textit{and} LF dataset $\left[\text{ln }(\mathbf{d}^{\text{mcs}}_{L}),\mathbf{l}^{\text{mcs}}\right]$ from \ref{st:peaks}. After that, steps 10 to 12 of AS are identical to \ref{st:mpm} to \ref{st:end} of PAS again. As in \Cref{sec:pas_adap}, AS adds one new HF sample per iteration using an acquisition function. In \cite{VS2025}, the USMV function balanced uncertainty and upper-tail sampling. Later, \cite{VS2025prads} showed that MVPD, which balances sampling around the target probability and the upper tail, is more efficient. When AS is used in this study it is combined with two initial samples, MF-GPR and MVPD (\Cref{eqAS:af_MVPD}, where $\overline{\mathbf{h}^*_n}$ is the normalised mean HF prediction from the previous iteration and $\mathbf{d}^*$ are the corresponding exceedance probabilities). The closest MCS sample is then selected and added to the LF pool for the next iteration, as in PAS.

\begin{equation}
\label{eqAS:af_MVPD}
\begin{aligned}
    p^{\text{mvpd}}_{\text{new}} = \arg\max\left[\overline{\mathbf{h}^*_n} \cdot \left(\frac{\mathbf{f_2}}{\textrm{max}(\mathbf{f_2})}\right)\right]  \quad \text{where:} \quad 
     \begin{cases}
        \mathbf{f_1} = 1-(\text{ln}(\mathbf{d}^*)-\text{ln}(P_{\text{exp}}))^2 \\
        \mathbf{f_2} = \mathbf{f_1}-\textrm{min}(\mathbf{f_1})
    \end{cases}       
\end{aligned}
\end{equation}

%%%%%%%%%%%%%%%%%%%%%%%%%%%%%%%%%%%%%%%%%%%%%%%%%%%%%%%%%%%%%%%%%%%%%%
%%%%%%%%%%%%%%%%%%%%%%%%%%%%%%%%%%%%%%%%%%%%%%%%%%%%%%%%%%%%%%%%%%%%%%
%%%%%%%%%%%%%%%%%%%%%%%%%%%%%%%%%%%%%%%%%%%%%%%%%%%%%%%%%%%%%%%%%%%%%%

\section{Copula basics}
\label{app:copulabasics}

\subsection{General theory}
\label{app:copulabasics_gen}
Copulas are multivariate distribution functions with uniform marginals on the interval $[0,1]$. If $F$ is a joint cumulative distribution function (CDF) with marginals $F_1, F_2, \dots, F_d$, then there exists a copula $C$ defined in \Cref{eq:gencopula1}. If the marginals are continuous, then the copula $C$ is unique and can be written as \Cref{eq:gencopula2} (Sklar's theorem, \cite{S1959}). Copulas model dependence between random variables by separating marginal distributions from the dependence structure, allowing flexible joint modelling, particularly in the tails. Compared to empirical distributions, they require less data and support reliable interpolation and extrapolation, making them well suited for surrogate modelling, uncertainty quantification, and multivariate extreme value analysis.  

\begin{equation}
\label{eq:gencopula1}
F(x_1, x_2, \dots, x_d) = C(F_1(x_1), F_2(x_2), \dots, F_d(x_d))
\end{equation}

\begin{equation}
\label{eq:gencopula2}
C(u_1, u_2, \dots, u_d) = F(F_1^{-1}(u_1), F_2^{-1}(u_2), \dots, F_d^{-1}(u_d))
\end{equation}

\subsection{Copula families and formulations}
\label{app:copulabasics_form}
Copula families are commonly grouped into \textit{elliptical} and \textit{Archimedean} copulas. Elliptical copulas, derived from distributions such as the Gaussian or Student-T, model symmetric dependence, while Archimedean copulas use a generator function and allow asymmetric and tail-dependent behaviour. This work uses Gaussian and Student-T copulas from the former class, and Clayton, Gumbel, and Frank copulas from the latter, restricted to the bivariate case. Overall dependence between variables is measured by Kendall's tau $\tau$, which can be estimated empirically and related to copula parameters via the formulations below. Similarly, lower and upper tail dependence are quantified by the coefficients $\lambda_L$ and $\lambda_U$. For example, $\lambda_U=0.81$ indicates an 81\% probability that one variable is extreme given that the other is.

The \textbf{Gaussian} copula (\Cref{eq:cop_gaus}) is defined using the standard normal CDF $\Phi$ and the bivariate normal CDF $\Phi_\rho$, with correlation parameter $\rho\in[-1,1]$. It models symmetric dependence with no tail dependence; $\lambda_L=\lambda_U=0$, and $\tau=\tfrac{2}{\pi}\arcsin(\rho)$. The \textbf{Student-T} copula (\Cref{eq:cop_st}) uses the univariate and bivariate Student-T CDFs, $t_\nu$ and $t_{\rho,\nu}$, parameterised by correlation $\rho\in[-1,1]$ and degrees of freedom $\nu>0$. It exhibits symmetric upper and lower tail dependence for finite $\nu$, with $\tau = \frac{2}{\pi} \arcsin(\rho)$. The \textbf{Clayton} copula (\Cref{eq:cop_clay}) is parameterised by $\theta>0$ and captures lower-tail dependence only, with $\lambda_U = 0$, $\lambda_L=2^{-1/\theta}$ and $\tau=\theta/(\theta+2)$. The \textbf{Gumbel} copula (\Cref{eq:cop_gumb}) has parameter $\theta\ge1$ and captures upper-tail dependence only, with $\lambda_L = 0$, $\lambda_U=2-2^{1/\theta}$ and $\tau=1-1/\theta$. The \textbf{Frank} copula (\Cref{eq:cop_fran}) is defined by a dependence parameter $\theta\in\mathbb{R}\setminus{0}$, where the sign of $\theta$ determines positive or negative dependence. It has no tail dependence ($\lambda_L = \lambda_U = 0$) and $\tau = 1 - \frac{4}{\theta} \left(1 - D_1(\theta)\right)$, where $D_1(\theta) = \frac{1}{\theta} \int_0^\theta \frac{t}{e^t - 1},dt$ is the Debye function.

\begin{equation}
\label{eq:cop_gaus}
    C^{\text{gaussian}}_\rho(u, v) = \Phi_\rho(\Phi^{-1}(u), \Phi^{-1}(v))
\end{equation}

\begin{equation}
\label{eq:cop_st}
    C^{\text{Student-T}}_{\rho,\nu}(u,v) = t_{\rho,\nu}(t^{-1}_{\nu}(u), t^{-1}_{\nu}(v))
\end{equation}

\begin{equation}
\label{eq:cop_clay}
    C^{\text{clayton}}_\theta(u,v) = \left( u^{-\theta} + v^{-\theta} - 1 \right)^{-1/\theta}, \quad \theta > 0.
\end{equation}

\begin{equation}
\label{eq:cop_gumb}
    C^{\text{gumbel}}_\theta(u,v) = \exp \left( - \left[(-\ln u)^\theta + (-\ln v)^\theta \right]^{1/\theta} \right), \quad \theta \geq 1.
\end{equation}

\begin{equation}
\label{eq:cop_fran}
    C^{\text{frank}}_\theta(u,v) = -\frac{1}{\theta} \ln \left( 1 + \frac{(e^{-\theta u} - 1)(e^{-\theta v} - 1)}{e^{-\theta} - 1} \right), \quad \theta \in \mathbb{R} \setminus \{0\}.
\end{equation}

%%%%%%%%%%%%%%%%%%%%%%%%%%%%%%%%%%%%%%%%%%%%%%%%%%%%%%%%%%%%%%%%%%%%%%
%%%%%%%%%%%%%%%%%%%%%%%%%%%%%%%%%%%%%%%%%%%%%%%%%%%%%%%%%%%%%%%%%%%%%%
%%%%%%%%%%%%%%%%%%%%%%%%%%%%%%%%%%%%%%%%%%%%%%%%%%%%%%%%%%%%%%%%%%%%%%

\section{PAS stopping criterion}
\label{app:pas_stop}

The stopping criterion in PAS has two components: one monitors the convergence of the predicted distribution shape, and the other monitors the convergence of the resulting MPM value. The criterion ignores iterations where a false positive sample was added. The first component is based on the mean absolute difference $E(j)$ between successive predicted distributions over an exceedance probability range, $\mathbf{h^*_\text{ran}}(j-1)$ and $\mathbf{h^*_\text{ran}}(j)$ for iteration $j$. This range is taken between $P_{\text{exp}}/2$ and $P_{\text{exp}}\cdot2$, to focus on the distribution shape around $P_{\text{exp}}$. Part one of the stopping criterion $\overline{E}_{K_1}(j)$ is then defined in \Cref{eqPAS:avMAE}, by taking the average $E(j)$ over the last $K_1$ iterations. This is done to ensure a smooth convergence. When $j<K_1$, all available iterations are used, but we set a convergence threshold at $K_1/3$ (convergence can only be declared after this number of iterations). 

\begin{equation}
\label{eqPAS:avMAE}
\begin{aligned}
    &\overline{E}_{K_1}(j) = \frac{1}{\psi} \sum_{i=\chi}^{j} E(i) \quad \text{where:} \quad 
    \begin{cases}
        E(j) = \text{mean} \left|\mathbf{h^*_\text{ran}}(j)-\mathbf{h^*_\text{ran}}(j-1) \right| \\
        \chi = 1 \text{  and  } \psi = j & \text{for } j = 1,2,...,K_1-1 \\
        \chi = j-K_1+1 \text{  and  } \psi = K_1 & \text{for } j \geq K_1
    \end{cases}
\end{aligned}
\end{equation}

The second part is based on a convergence criterion on the coefficient of variation (COV) of the MPM value over the last $K_2$ iterations: $C_{K_2}(j)$. This is expressed in \Cref{eqPAS:cov}, where $\widehat{H}(j)$ is the MPM value predicted in iteration $j$. Again, we took the COV over the available iterations when $j<K_2$.

\begin{equation}
\label{eqPAS:cov}
\begin{aligned}
    C_{K_2}(j) = \frac{\sigma_{K_2}(j)}{\mu_{K_2}(j)} 
    \quad \text{where:} \quad 
    \begin{cases}
        \mu_{K_2}(j) = \frac{1}{\psi} \sum_{i=\chi}^{j} \widehat{H}(i) \\
        \sigma_{K_2}(j) = \sqrt{\frac{1}{\psi} \sum_{i=\chi}^{j} \left(\widehat{H}(i)-\boldsymbol{\mu}_{K_2}(i)\right)^2} \\
        \chi = 1 \text{  and  } \psi = j & \text{for } j = 1,2,...,K_2-1 \\
        \chi = j-K_2+1 \text{  and  } \psi = K_2 & \text{for } j \geq K_2
    \end{cases}
\end{aligned}
\end{equation}

The total stopping criterion $S(j)$ is provided in \Cref{eqPAS:convcrit}, where $K_1=K_2=20$ limit the influence of outliers and the minimum number of iterations for which convergence can be detected. Stopping criteria $\epsilon_1,\epsilon_2$ are chosen case-dependently, using acceptable tolerances for the problem at hand.

\begin{equation}
\label{eqPAS:convcrit}
\begin{aligned}
    S(j) &= 
    \begin{cases}
        \text{stop} & \text{if } \left( \overline{E}_{50}(j) < \epsilon_1 \right) \cap \left( C_{50}(j) < \epsilon_2  \right)\\
        \text{continue} & \text{otherwise}
    \end{cases}
\end{aligned}
\end{equation}

Compared to the stopping criterion used for AS in \cite{VS2025}, we removed the rejection criterion for distribution shapes that violate distribution assumptions (as drawing from the copula model in \ref{st:drawsamp} of PAS always produces a proper distribution), we added a range over which \Cref{eqPAS:avMAE,eqPAS:cov} are calculated (this was natural in AS because most steps considered such a range, but not in PAS), we used the mean absolute difference instead of the maximum absolute difference in \Cref{eqPAS:avMAE}, we used $K_1=K_2=50$ instead of 20 (to ensure smoother convergence), and we added the convergence threshold.

%%%%%%%%%%%%%%%%%%%%%%%%%%%%%%%%%%%%%%%%%%%%%%%%%%%%%%%%%%%%%%%%%%%%%%
%%%%%%%%%%%%%%%%%%%%%%%%%%%%%%%%%%%%%%%%%%%%%%%%%%%%%%%%%%%%%%%%%%%%%%
%%%%%%%%%%%%%%%%%%%%%%%%%%%%%%%%%%%%%%%%%%%%%%%%%%%%%%%%%%%%%%%%%%%%%%

\begin{figure}[t]
	\centering
        % ---------- Left column ----------
        \begin{subfigure}[t]{0.35\textwidth}
            \centering
            \begin{subfigure}[t]{\textwidth}
                \centering
                \includegraphics[width=\textwidth]{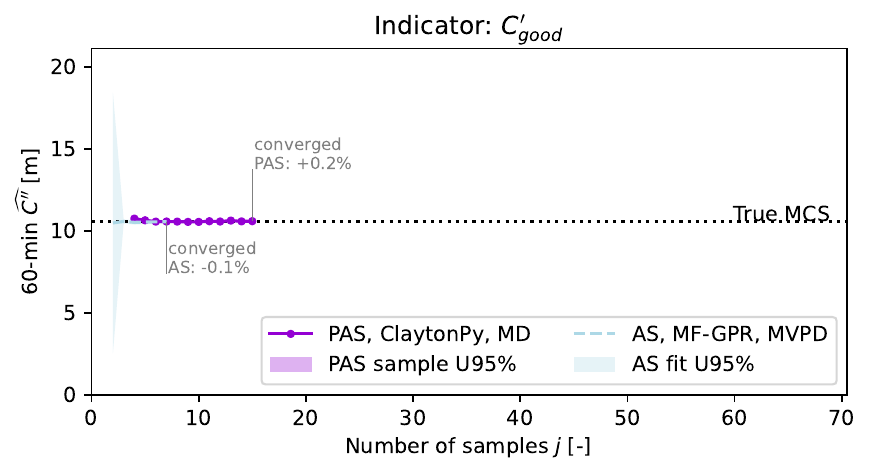}
            \end{subfigure}
            \begin{subfigure}[t]{\textwidth}
                \centering
                \includegraphics[width=\textwidth]{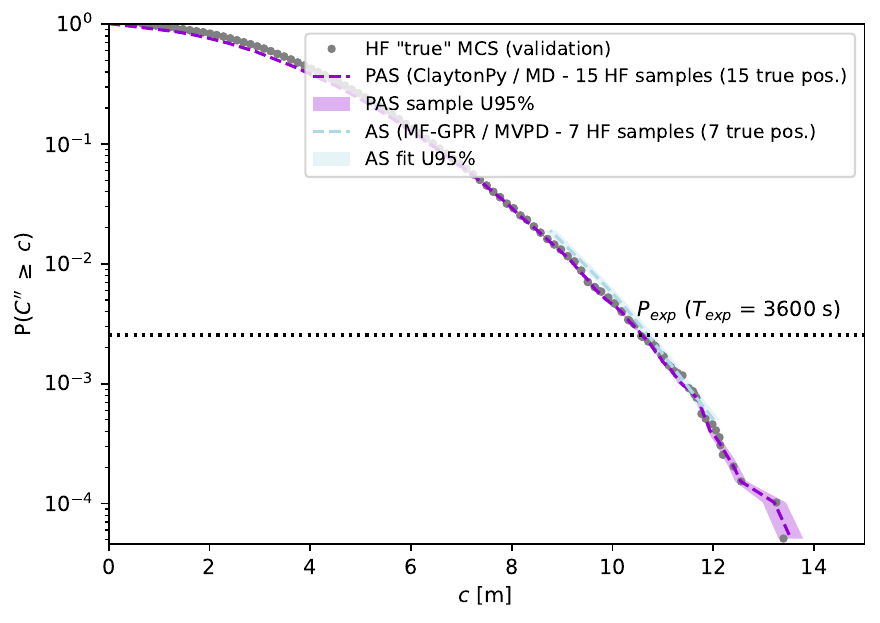}
            \end{subfigure}
            \subcaption{Case 1a, indicator GoodInd}
            \label{fig:res1a}
        \end{subfigure}
        % ---------- Right column ----------
        \begin{subfigure}[t]{0.35\textwidth}
            \centering
            \begin{subfigure}[t]{\textwidth}
                \centering
                \includegraphics[width=\textwidth]{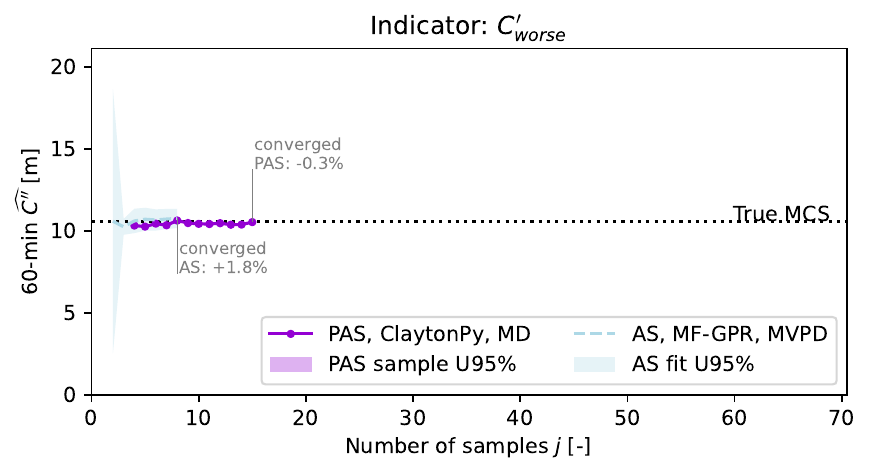}
            \end{subfigure}
            \begin{subfigure}[t]{\textwidth}
                \centering
                \includegraphics[width=\textwidth]{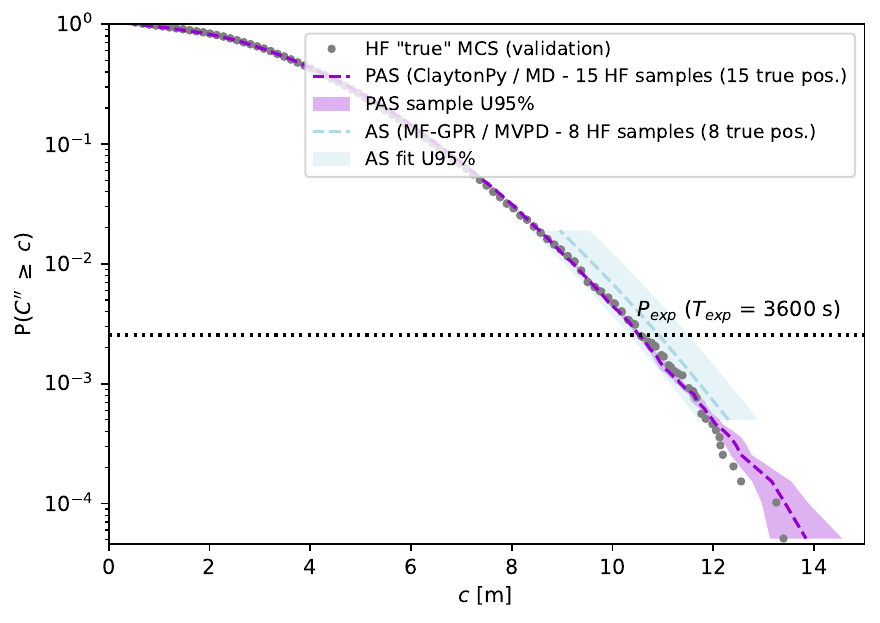}
            \end{subfigure}
            \subcaption{Case 1b, indicator WorseInd}
            \label{fig:res1b}
        \end{subfigure}
	\caption{Case 1 - waves: convergence of one-hour MPM as a function of number of samples (top) and final converged distributions (bottom) from AS and PAS. The copula in the name of the PAS results is the used model in the last iteration.} 
	\label{fig:res1}
\end{figure}

\begin{figure}[t]
	\centering
        % ---------- Left column ----------
        \begin{subfigure}[t]{0.35\textwidth}
            \centering
            \begin{subfigure}[t]{\textwidth}
                \centering
                \includegraphics[width=\textwidth]{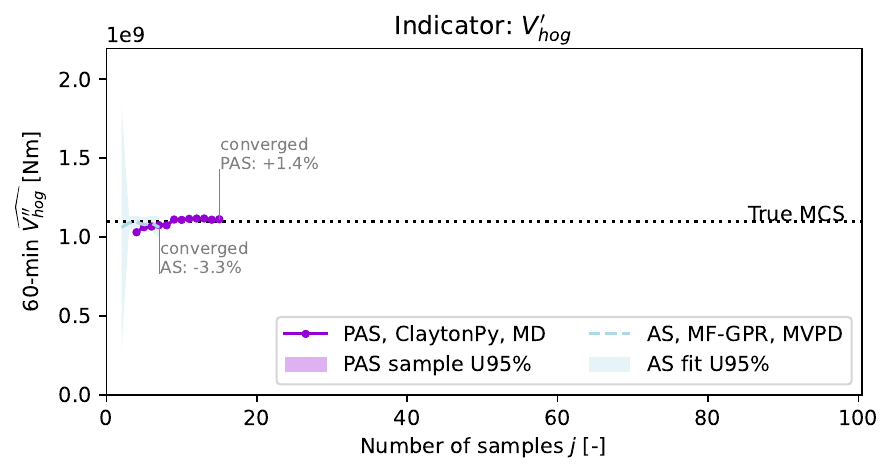}
            \end{subfigure}
            \begin{subfigure}[t]{\textwidth}
                \centering
                \includegraphics[width=\textwidth]{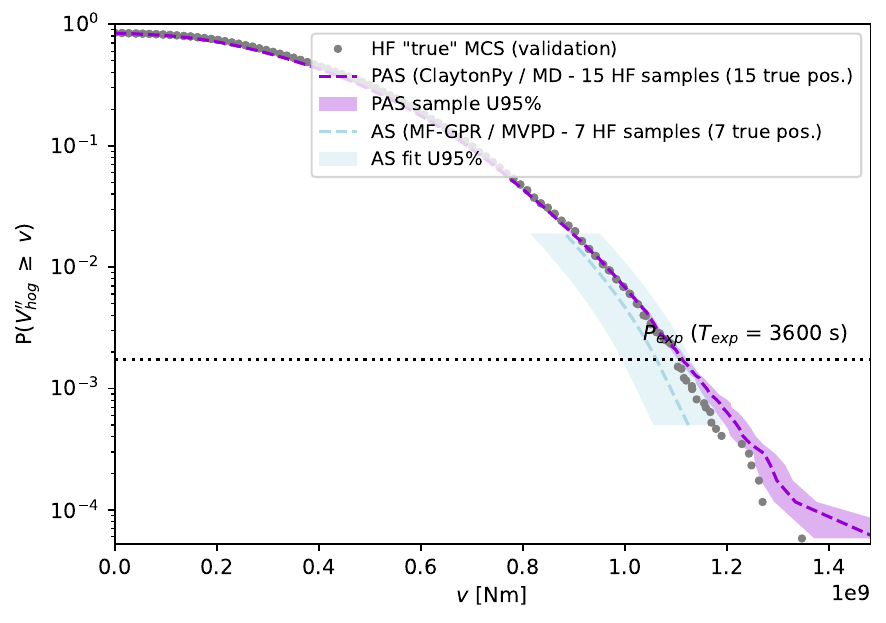}
            \end{subfigure}
            \subcaption{Case 2a, hogging mode}
            \label{fig:res2a}
        \end{subfigure}
        % ---------- Right column ----------
        \begin{subfigure}[t]{0.35\textwidth}
            \centering
            \begin{subfigure}[t]{\textwidth}
                \centering
                \includegraphics[width=\textwidth]{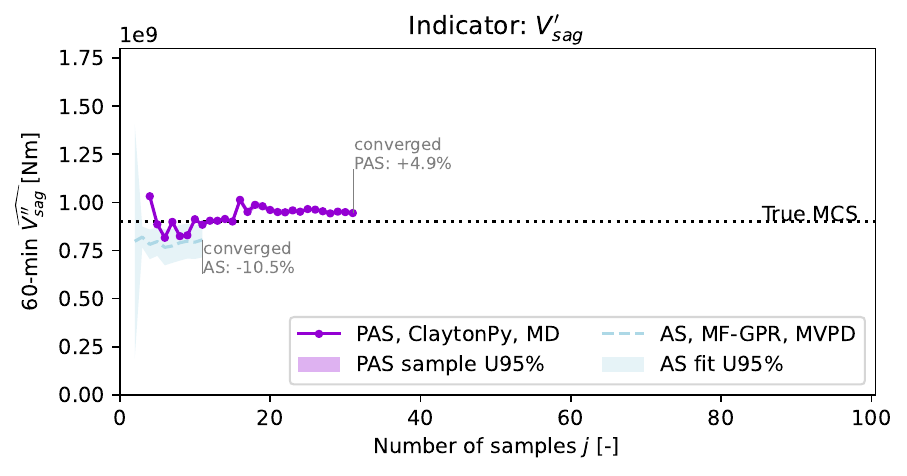}
            \end{subfigure}
            \begin{subfigure}[t]{\textwidth}
                \centering
                \includegraphics[width=\textwidth]{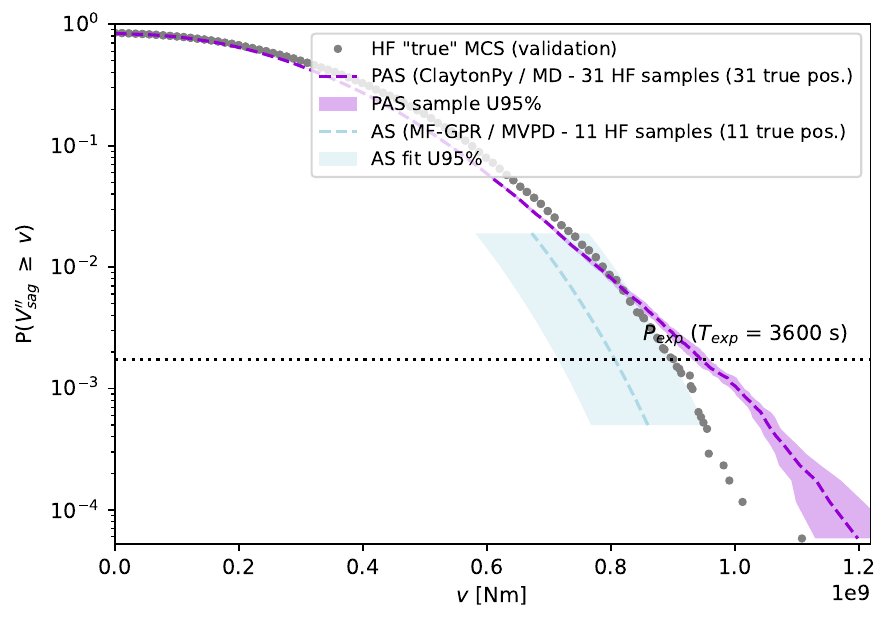}
            \end{subfigure}
            \subcaption{Case 2b, sagging mode}
            \label{fig:res2b}
        \end{subfigure}
	\caption{Case 2 - VBM: convergence of one-hour MPM as a function of number of samples (top) and final converged distributions (bottom) from AS and PAS. The copula in the names of the PAS results is the used model in the last iteration.}
	\label{fig:res2}
\end{figure}

\section{Details cases 1 and 2}
\label{app:rescases12}

\subsection{Case 1 (weakly non-linear): second-order wave crest heights}
\label{app:case1}

Case 1 from \cite{VS2025,VS2025prads}, with related datasets in \cite{VSdata2025,VSdata2025PRADS}, addresses a weakly non-linear problem. We predict extreme values of `HF' second-order wave crest heights $C''$, in long-crested steep waves (condition A in \Cref{tab:conditions}). The HF validation dataset consisted of analytical second-order wave time traces according to \cite{SD1979}. Two LF indicators were defined: linear Gaussian wave crests $C_{\text{good}}'$ (case 1a) and crests in the same linear Gaussian waves including an additional noise wave system $C_{\text{worse}}'$ (case 1b). The LF MCS of \ref{st:MCS} was done for 50 hours duration. Since the HF variable is analytically tractable, it was easy to generate long HF ground truth time traces for this example case. \Cref{app:scatters} shows a scatter plot of LF indicator versus HF validation response peaks. This plot shows that the LF and HF order statistics align closely, but not perfectly. To generate a new HF sample in each iteration, we used the matched LF and HF peaks. When an event was selected based on its LF indicator value (in \ref{st:select} or \ref{st:adaptive}), the corresponding HF value was drawn from the matched LF-HF peaks. This is only possible because we have an analytically traceable HF variable; in reality, each iteration would require a new HF event calculation. Details of the simulation methods are provided in \Cref{app:case_imp}. We aim to predict the HF one-hour MPM value $\widehat{C''}$. We did not use an LF threshold value of PAS; no false positives were expected for this case.

\subsection{Case 2 (weakly non-linear): vertical bending moments}
\label{app:case2}
Case 2 from \cite{VS2025,VS2025prads}, with related datasets in \cite{VSdata2025,VSdata2025PRADS}, studies a moderately non-linear problem. We predict extreme values of hogging vertical bending moments (VBM) on a 190~m ferry in extreme irregular head waves (condition B in \Cref{tab:conditions}). The HF VBM in hogging  $V_{\text{hog}}''$ and sagging mode $V_{\text{sag}}''$ were generated using a non-linear (Froude-Krylov) potential flow simulation, while the LF VBM in hogging $V_{\text{hog}}'$ and sagging mode $V_{\text{sag}}'$ were generated using a linear potential flow simulation. The hogging peaks were defined by the zero up-crossing troughs in these VBM simulations; the sagging peaks by the zero up-crossing peaks. Unlike in case 1, this HF response is not analytically tractable, but 30-hour HF MCS non-linear simulations to produce validation material were feasible. Details of the simulation methods are provided in \Cref{app:case_imp}. The LF-HF peak scatter plots are shown in \Cref{app:scatters}. This shows that, as expected, the LF and HF VBM peaks have less similar order statistics than those of the first- and second-order wave crests in case 1. For further details on the case, see \cite{VS2025}. We aim to predict the HF one-hour MPM values $\widehat{V_{hog}''}$ and $\widehat{V_{sag}''}$. We did not use an LF threshold value of PAS; no false positives were expected for this case.

\subsection{Cases 1 and 2: convergence and predicted distributions}
\label{app:case12res}
Similar plots as shown for the complex cases in \Cref{sec:Discussion}, are shown here for the simpler verification cases 1 and 2. \Cref{fig:res1} shows these results for cases 1a,b, \Cref{fig:res2} and for cases 2a,b. Discussion of these results is included in \Cref{sec:Discussion}.

%%%%%%%%%%%%%%%%%%%%%%%%%%%%%%%%%%%%%%%%%%%%%%%%%%%%%%%%%%%%%%%%%%%%%%
%%%%%%%%%%%%%%%%%%%%%%%%%%%%%%%%%%%%%%%%%%%%%%%%%%%%%%%%%%%%%%%%%%%%%%
%%%%%%%%%%%%%%%%%%%%%%%%%%%%%%%%%%%%%%%%%%%%%%%%%%%%%%%%%%%%%%%%%%%%%%

\begin{figure}[t!]
	\centering
        \includegraphics[width=.9\linewidth]{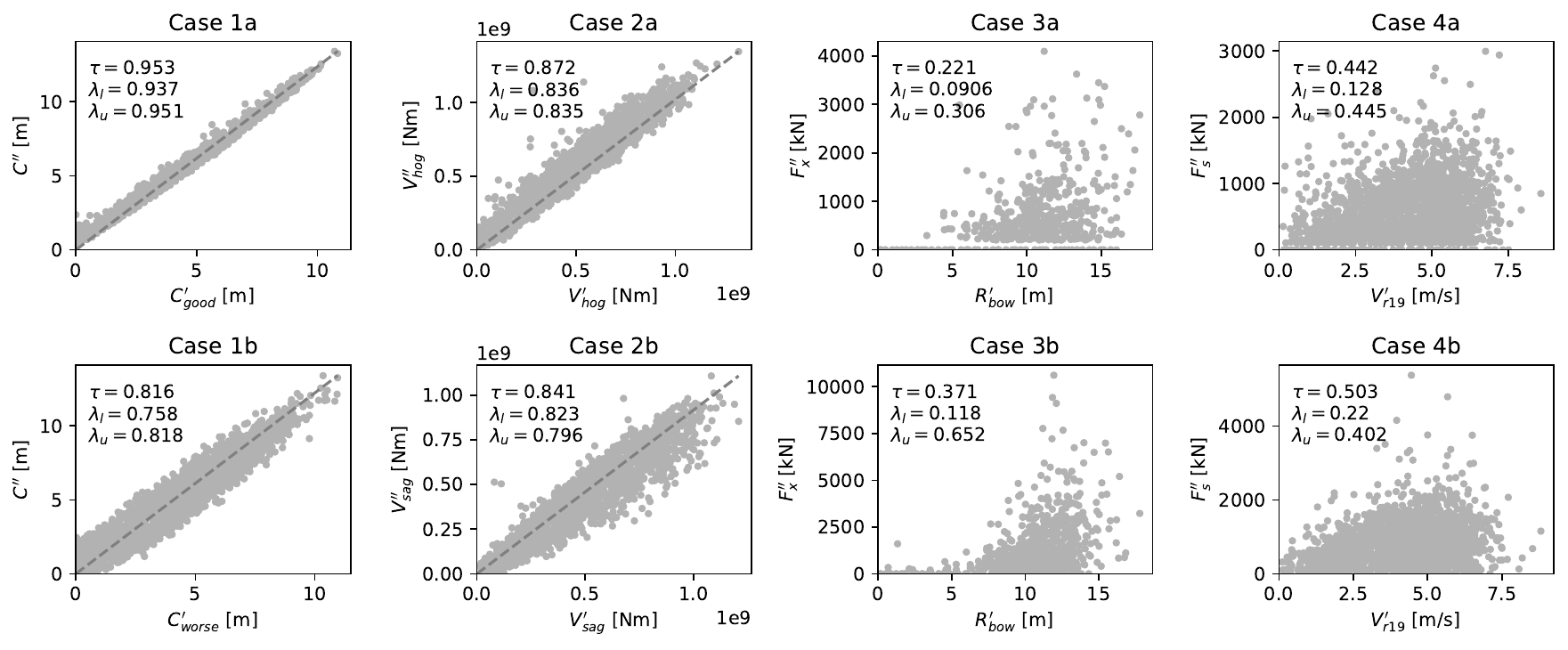}   
	\caption{Scatter plots of LF indicator peaks and matched HF validation peaks for all cases, including the associated Kendall's $\tau$ and upper / lower tail dependence coefficients $\lambda_U$ / $\lambda_L$ (defined in \Cref{app:copulabasics_gen}).}
	\label{fig:scatters}
\end{figure}

\section{Extra case information: implementation, specifications and experimental details}
\label{app:caseinfo}

\subsection{Scatter plots}
\label{app:scatters}
Scatter plots of LF indicator peaks versus matched HF response peaks from the validation data for all cases are shown in \Cref{fig:scatters}, illustrating the screening assumption that LF and HF order statistics are comparable. Each plot includes Kendall's $\tau$ and tail dependence coefficients $\lambda_U$ and $\lambda_L$ (see \Cref{app:copulabasics_form}), quantifying the rank correlation between LF and HF peaks and indicating the suitability of a copula for modelling their joint behaviour. Perfectly matched order statistics would yield a monotonically increasing line with $\tau = 1$, but the plots show varying correspondence across cases. Cases 3 and 4 additionally reflect the 100~kN per-panel threshold applied in the original experimental data to remove measurement noise and low-frequency inertial effects.

%%%%%%%%
\subsection{Experimental details and scale effects}
\label{app:case_exp}
For cases 3, 4 and 5, the HF validation datasets were generated using experiments at a scale of 1:35.986 in the Seakeeping and Manoeuvring Basin of MARIN in Wageningen, The Netherlands. More details about these experiments can be found in \cite{VSS2023,VBS2024}. We used standard Froude scaling to obtain HF full-scale RWEs, green water and slamming forces. Seakeeping responses are predominantly governed by inertia, so scale effects on most global (relative) motions and first-order loads are expected to be small. Scale effects may occur in the impact loads when air inclusions occur during wave impact events (see e.g. \cite{SH2023,ED2025}). However, no significant air inclusions were observed in the green water impacts of the present dataset. In most cases a horizontal `jet' was formed over the foredeck, which subsequently hit the accommodation; this does not easily produce air pockets (see e.g., \cite{B2002}). This is supported by the full-scale CFD calculations with single-phase code ComFLOW in \cite{BHV2020}, which were able to replicate a few of the Froude-scaled largest green water impact force events very accurately for this case. Small air inclusions may locally violate the assumptions underlying Froude scaling for the slamming impacts, but no reliable alternative scaling approach is available. Visual observations during the experiments indicated that the occurrence of such inclusions seems limited. 

%%%%%%%%
\subsection{Simulation details}
\label{app:case_imp}
For case 1, the HF validation dataset of second-order wave elevation time traces was analytically generated with the \texttt{Python} \texttt{PySeaWave} package, using the random phase method and a frequency bandwidth of 0-5~rad/s for second-order interactions. The linear wave LF indicator time traces for case 1a were also derived from this package; the additional noise time traces for case 1b were added to this. For case 2, the HF validation dataset of non-linear VBM time traces was generated using the \texttt{PRETTI\_R} program (v19.0.1), which is a non-linear Froude-Krylov solver using linear RAO inputs (in this case from linear frequency-domain potential flow diffraction method \texttt{SEACAL}). The linear VBM LF indicator time traces were generated with \texttt{SEACAL} (v7.2.0) in the zero speed Green's function mode. The ferry hull was described with 3327 panels (a figure of the mesh is included in \cite{VS2025}). For cases 3, 4 and 5, the linear LF vertical bow motions and RWE indicator time traces were generated with \texttt{SEACAL} in the Rankine-source mode. The relative rise velocity per wave event used for case 5 was derived from the RWE time traces with a custom \texttt{Python} script. The Generalised Pareto fit to the ground truth HF validation data for cases 3, 4 and 5 was made using the \texttt{Python} \texttt{scipy.stats.genpareto} package. 

\begin{figure}[t!]
	\centering
        \includegraphics[width=.6\linewidth]{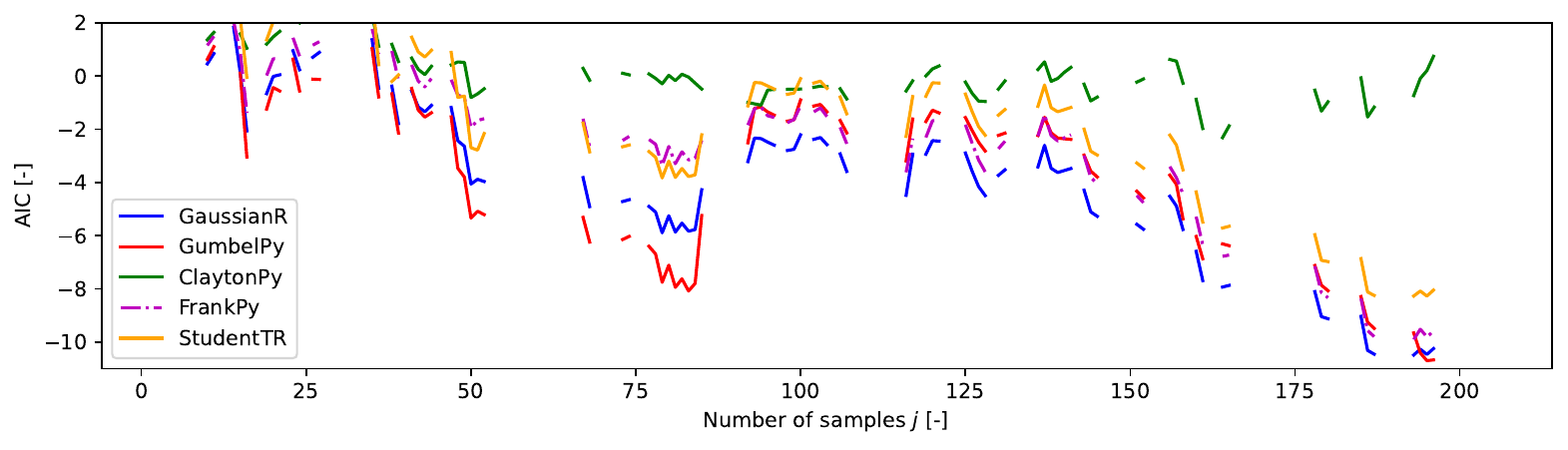}
	\caption{AIC of the five copula candidate models as a function of the number of HF samples, for green water case 3a (which is converged at 47 HF samples).}
	\label{fig:AIC}
\end{figure}

\begin{figure}[t!]
	\centering
        \includegraphics[width=.8\linewidth]{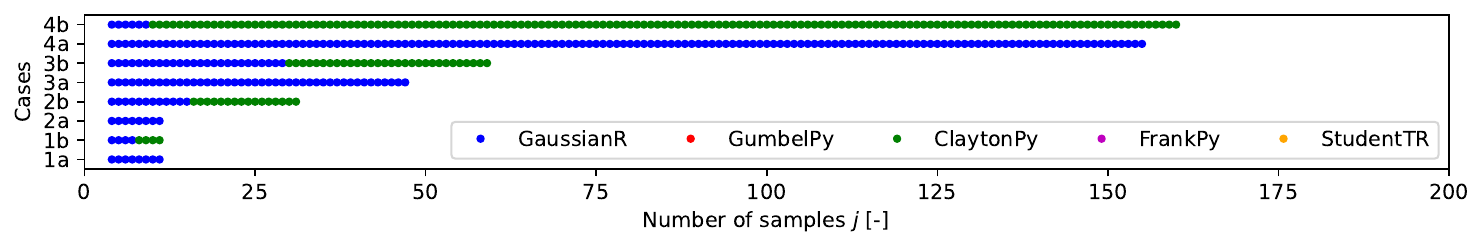}
	\caption{Selected copula model per case and iteration, plotted until convergence in each case.}
	\label{fig:copsel}
\end{figure}

\begin{figure}[t!]
	\centering
    \begin{subfigure}[b]{0.245\textwidth}
        \centering
        \includegraphics[width=\textwidth]{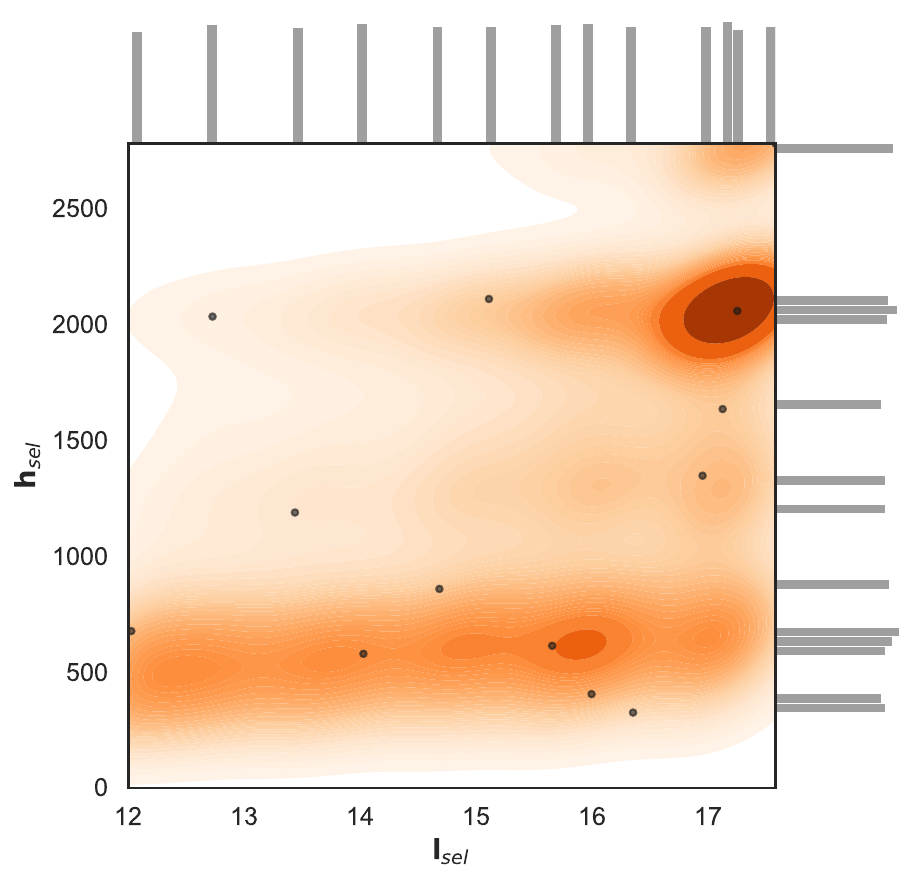}
        \caption{It. 20 (24 HF samples).}
        \label{fig:copselB1}
    \end{subfigure}
    \begin{subfigure}[b]{0.245\textwidth}
        \centering
        \includegraphics[width=\textwidth]{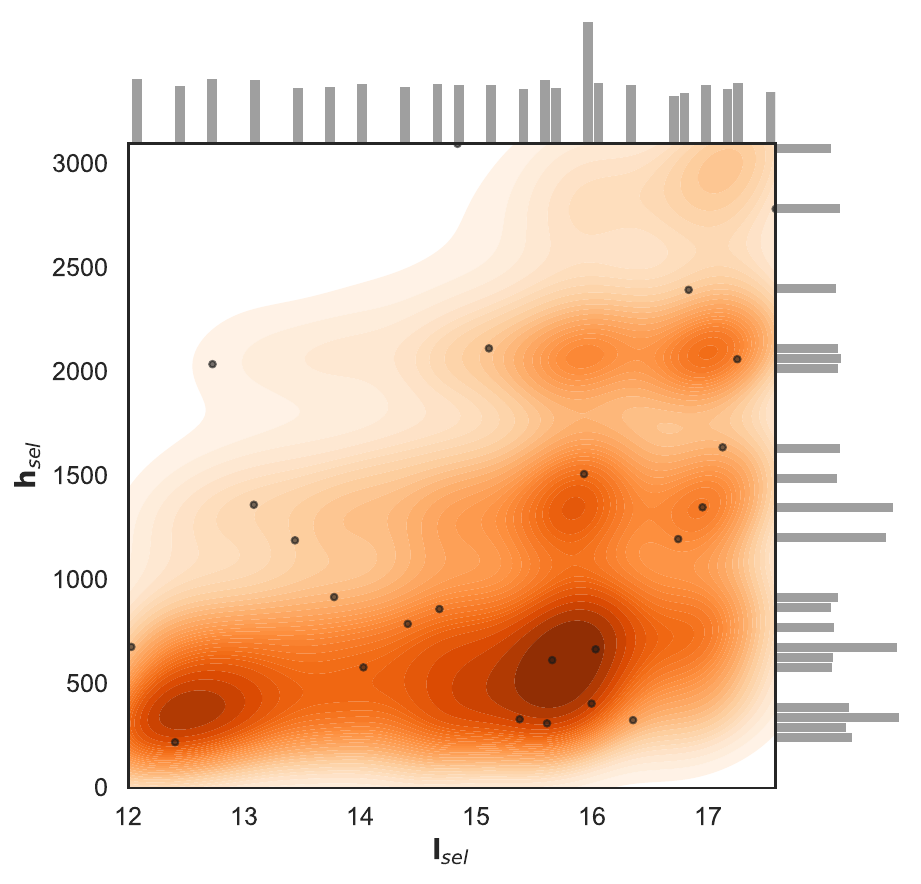}
        \caption{It. 40 (44 HF samples).}
        \label{fig:copselB2}
    \end{subfigure}
    \begin{subfigure}[b]{0.245\textwidth}
        \centering
        \includegraphics[width=\textwidth]{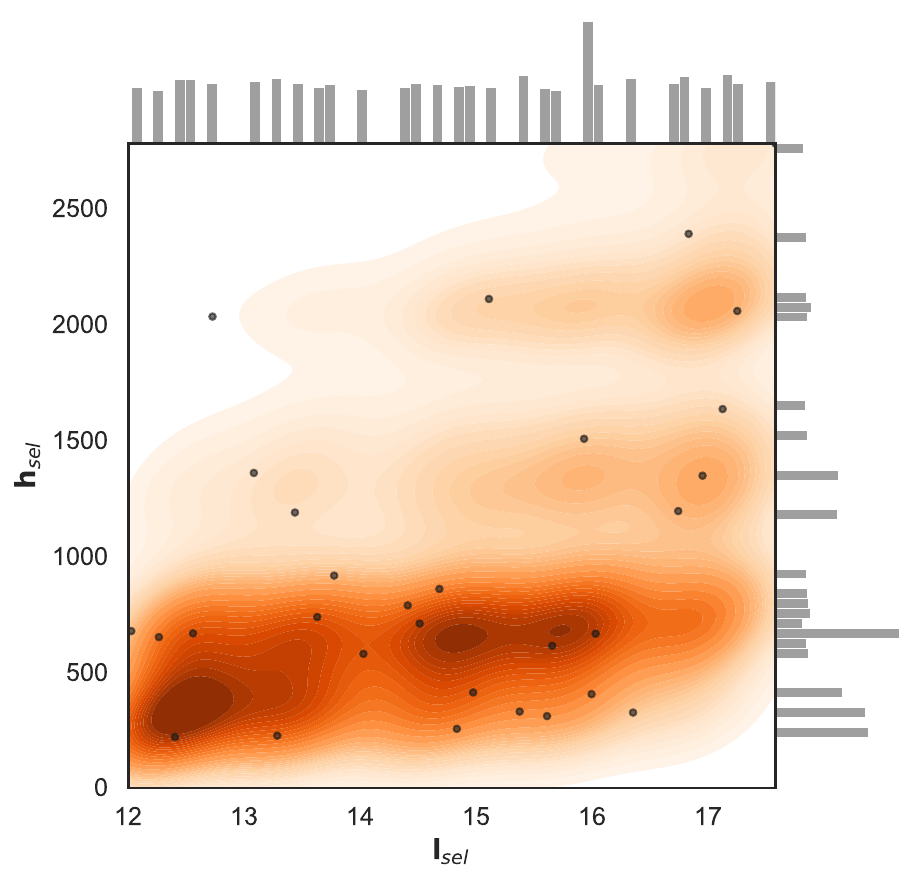}
        \caption{It. 63 (67 HF samples).}
        \label{fig:copselB3}
    \end{subfigure}
    \begin{subfigure}[b]{0.245\textwidth}
        \centering
        \includegraphics[width=\textwidth]{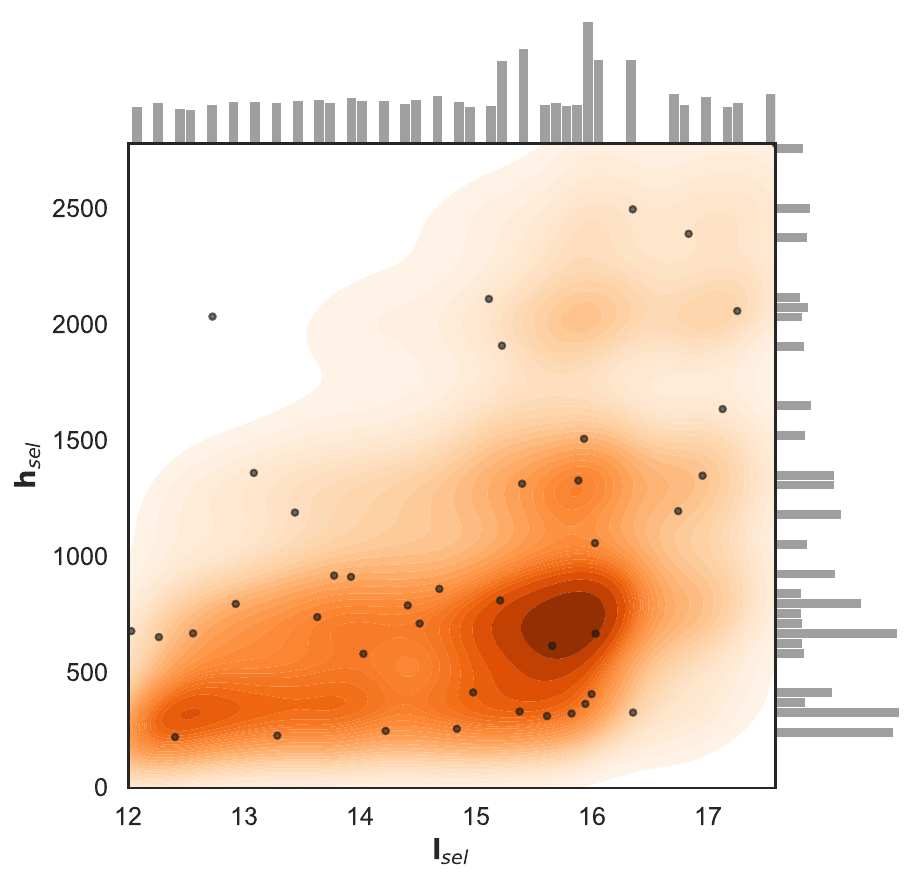}
        \caption{It. 80 (84 HF samples).}
        \label{fig:copselB4}
    \end{subfigure}
    \begin{subfigure}[b]{0.4\textwidth}
        \centering
        \includegraphics[width=\textwidth]{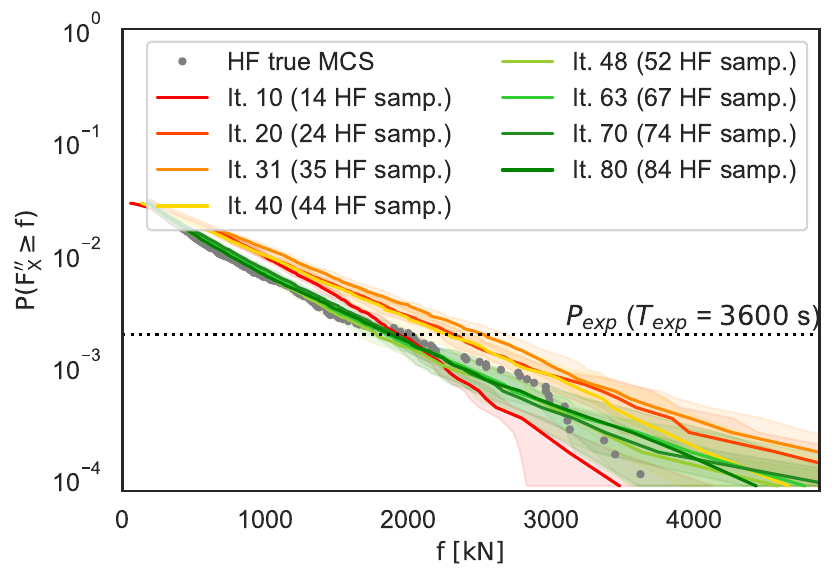}
        \caption{Distribution convergence.}
        \label{fig:distriterations4a}
    \end{subfigure}
    \caption{Convergence of samples and fitted copula models in real space, including histograms of the associated empirical marginals on the sides, at approx. every 20$^{\text{th}}$ iteration (top), and the convergence of the predicted distributions at approx. every 10$^{\text{th}}$ iteration (bottom), both for green water case 3a from PAS. This case was considered converged with 47 HF samples (\Cref{fig:res3a}). The nearest iteration is plotted whenever a false positive sample was found in one of the targeted iterations (it.~=~iteration).}
	\label{fig:copulaconvergence3a}
\end{figure}

%%%%%%%%%%%%%%%%%%%%%%%%%%%%%%%%%%%%%%%%%%%%%%%%%%%%%%%%%%%%%%%%%%%%%%
%%%%%%%%%%%%%%%%%%%%%%%%%%%%%%%%%%%%%%%%%%%%%%%%%%%%%%%%%%%%%%%%%%%%%%
%%%%%%%%%%%%%%%%%%%%%%%%%%%%%%%%%%%%%%%%%%%%%%%%%%%%%%%%%%%%%%%%%%%%%%

\section{Convergence of copula selection process}
\label{app:copulafittingdetails}
Firstly, we evaluate how the iterative copula selection procedure converges for all considered cases. The AIC-based copula selection procedure, described in \Cref{sec:pas_copsel}, aims to identify the candidate copula that best fits the currently available HF samples, rather than to determine which copula performs best on the full validation dataset. Convergence to a single copula model for each case is therefore not required; new insights from the Bayesian adaptive sampling procedure, based on new HF samples, may change which model is most likely. This is visualised in \Cref{fig:AIC}, which shows the AIC for case 3a as it evolves over successive iterations. A lower AIC is better. \Cref{fig:copsel} provides an overview of the selected copula for each case and each iteration (up to a number of iterations past convergence). These figures show that although the Gumbel copula has a slightly lower AIC value when $\sim$5-80 HF samples are available for case 3a, PAS retains the Gaussian copula identified in the first iteration as the most likely model. This is due to the utilised AIC difference threshold of 4.0 (see \Cref{sec:pas_copsel}); only when the AIC difference between models exceeds this threshold is an alternative model considered significantly better. \Cref{fig:AIC} also shows that the Gaussian and Gumbel models are both good models for the full validation dataset. \Cref{fig:res3a} showed that this case was converged and quite accurate at 47 HF samples already. It is therefore not necessary to arrive at a `true copula model' before predicting the distributions; the most likely model with fewer HF samples already produces a sufficient result. \Cref{fig:copsel} shows a consistent preference for the Gaussian copula across cases, likely because it fits very sparse datasets well at low iteration numbers. In later iterations, it is replaced only when a significantly better model is identified ($\Delta$AIC > 4.0). While this may introduce a slight bias in the convergence process (especially as the Gaussian model does not fit all combinations of $\tau$, $\lambda_U$ and $\lambda_L$ in \Cref{fig:scatters} equally well), the results remain acceptable and the procedure is stabilised by the flexibility of the Gaussian copula. When an alternative model is selected, it is most often the Clayton copula, consistent with the generally stronger lower-tail than upper-tail dependence observed, particularly for cases 3 and 4 (see \Cref{fig:scatters}). Finally, \Cref{fig:copulaconvergence3a} shows the evolution of HF samples, fitted copula model and predicted distributions with an increasing number of iterations for case 3a. This shows that the copula model and distributions move somewhat over the first $\sim$40 iterations, but then stabilise to the converged values (as correctly detected by the convergence criterion at 47 samples).

%%%%%%%%%%%%%%%%%%%%%%%%%%%%%%%%%%%%%%%%%%%%%%%%%%%%%%%%%%%%%%%%%%%%%%
%%%%%%%%%%%%%%%%%%%%%%%%%%%%%%%%%%%%%%%%%%%%%%%%%%%%%%%%%%%%%%%%%%%%%%
%%%%%%%%%%%%%%%%%%%%%%%%%%%%%%%%%%%%%%%%%%%%%%%%%%%%%%%%%%%%%%%%%%%%%%

\end{appendices}

\printcredits

%%%%%%%%%%%%%%%%%%%%%%%%%%%%%%%%%%%%%%%%%%%%%%%%%%%%%%%%%%%%%%%%%%%%%%
\section*{Declaration of generative AI / AI-assisted technology in the manuscript preparation process}
During the preparation of this work the authors used ChatGPT in order to rephrase some individual sentences. After using this tool/service, the authors reviewed and edited the content as needed and take full responsibility for the content of the published article.

%%%%%%%%%%%%%%%%%%%%%%%%%%%%%%%%%%%%%%%%%%%%%%%%%%%%%%%%%%%%%%%%%%%%%%
% Load bibliography style file
\bibliographystyle{elsarticle-num.bst}   % Numbers

% Load bibliography database
\bibliography{library_usedinpaperonly}

@article{GS2022,
    author = {Guth, S. and Sapsis, T. P.},
    doi = {10.1016/j.oceaneng.2022.112633},
    journal = {Ocean Eng.},
    pages = {112633},
    title = {{Wave episode based Gaussian process regression for extreme event statistics in ship dynamics: Between the Scylla of Karhunen--Loeve convergence and the Charybdis of transient features}},
    volume = {266},
    year = {2022}}

@article{MS2018,
	Author = {Mohamad, M. A. and Sapsis, T. P.},
	Doi = {10.1073/pnas.1813263115},
	Issn = {0027-8424},
	Journal = {Proc. Natl. Acad. Sci.},
	Title = {{Sequential sampling strategy for extreme event statistics in nonlinear dynamical systems}},
	Volume = {115, 44},
	Year = {2018}}

@inproceedings{BSPFS2018,
	Address = {Madrid, Spain},
	Author = {Bunnik, T. H. J. and Stansberg, C. T. and P{\'{a}}kozdi, C. and Fouques, S. and Somers, L.},
	Booktitle = {37\textsuperscript{th} OMAE Conf.},
	Doi = {10.1115/OMAE2018-78544},
	Publisher = {ASME},
	Title = {{Useful indicators for screening of sea states for wave impacts on fixed and floating platforms}},
	Year = {2018}}

@inproceedings{DSHTKB2020,
	Address = {Virtual, Online},
	Author = {D{\"u}z, B. and Scharnke, J. and Hallmann, R. and Tukker, J. and Khurana, S. and Blanchard, K.},
	Booktitle = {39\textsuperscript{th} OMAE Conf.},
	Doi = {10.1115/OMAE2020-19268},
	Publisher = {ASME},
	Title = {{Comparison of the CFD results to PIV measurements in kinematics of spilling and plunging breakers}},
	Year = {2020}}

@article{VDSJ2021,
    author = {van de Bunt, E. and Dekker, J. and Scharnke, J. and Jaou{\"{e}}n, F.},
    doi = {10.1016/j.oceaneng.2021.108857},
    journal = {Ocean Eng.},
    title = {{Applying force panels for wave impact measurements}},
    volume = {232},
    year = {2021}}

@inproceedings{BSD2019,
	Address = {Glasgow, UK},
	Author = {Bunnik, T. H. J. and Scharnke, J. and de Ridder, E.-J.},
	Booktitle = {38\textsuperscript{th} OMAE Conf.},
	Doi = {10.1115/OMAE2019-95481},
	Publisher = {ASME},
	Title = {{Efficient indicators for screening of random waves for wave impacts on a jacket platform and a fixed offshore wind turbine}},
	Year = {2019}}

@inproceedings{SH2023,
    address = {Melbourne, Australia},
    author = {Scharnke, J. and Helder, J. A.},
    booktitle = {42\textsuperscript{nd} OMAE Conf.},
    doi = {10.1115/OMAE2023-104288},
    publisher = {ASME},
    title = {{Scale effects and variability in wave-in-deck type of impact loading, more insights into the results of the Breakin JIP}},
    year = {2023}}

@article{ED2025,
    author = {Ezeta, R. and D{\"u}z, B.},
    journal = {Ocean Eng.},
    title = {{Predicting the dynamics of a gas pocket during breaking wave impacts using machine learning}},
    year = {2025},
    volume = {321},
    pages = {120321},
    doi = {10.1016/j.oceaneng.2025.120321}}

@article{VS2023,
    author = {van Essen, S. M. and Seyffert, H. C.},
    journal = {J. Offshore Mechanics and Arctic Eng.},
    title = {{Finding dangerous waves -- Review of methods to obtain wave impact design loads for marine structures (OMAE-22-1110)}},
    year = {2023},
    volume = {145},
    number = {6},
    pages = {060801},
    doi = {10.1115/1.4056888}}

@article{VFDKCG2024,
    author = {Erik Vanem and Elias Fekhari and Nikolay Dimitrov and Mark Kelly and Alexis Cousin and Martin Guiton},
    journal = {J. Offshore Mechanics and Arctic Eng.},
    title = {{A joint probability distribution for multivariate wind-wave conditions and discussions on uncertainties (OMAE-23-1131)}},
    year = {2024},
    volume = {146},
    number = {6},
    pages = {061701},
    doi = {10.1115/1.4064498}}

@article{VSS2023,
	Author = {van Essen, S. M. and Scharnke, J. and Seyffert, H. C.},
	Journal = {Marine Struc.},
	Title = {{Required test durations for converged short-term wave and impact extreme value statistics - Part 1: ferry dataset}},
	Year = {2023},
    volume = {90},
    pages = {103410},
    doi={10.1016/j.marstruc.2023.103410}}

@article{SVS2023,
	Author = {Scharnke, J. and van Essen, S. M. and Seyffert, H. C.},
	Journal = {Marine Struc.},
	Title = {{Required test durations for converged short-term wave and impact extreme value statistics - Part 2: deck box dataset}},
	Year = {2023},
    volume = {90},    
    pages = {103411},
    doi={10.1016/j.marstruc.2023.103411}}

@article{GEtAl2020,
	Author = {Gramstad, O. and Agrell, C. and Bitner-Gregersen, E. and Guo, B. and Ruth, E. and Vanem, E.},
	Doi = {10.1016/j.marstruc.2020.102780},
	Issn = {09518339},
	Journal = {Mar. Struct.},
	Title = {{Sequential sampling method using Gaussian process regression for estimating extreme structural response}},
	Volume = {72},
	Year = {2020}}

@article{VMSHKSG2021,
	Author = {van Essen, S. M. and Monroy, C. and Shen, Z. and Helder, J. A. and Kim, D.-H. and Seng, S. and Ge, Z.},
	Doi = {10.1016/j.oceaneng.2021.109218},
	Journal = {Ocean Eng.},
	pages = {109218},
	Title = {{Screening wave conditions for the occurrence of green water events on sailing ships}},
	Volume = {234},
	Year = {2021}}

@article{FSK2007,
	Author = {Forrester, I. J. and Sobester, A. and Keane, A. J.},
	Journal = {Proc. R. Soc. A},
	doi = {10.1098/rspa.2007.1900},
	Pages = {3251-3269},
	Title = {{Multi-fidelity optimization via surrogate modelling}},
	Volume = {463},
	Year = {2007}}

@article{KO2000,
	Author = {Kennedy, M. C. and O'Hagan, A.},
	Journal = {Biometrika},
	Title = {{Predicting the output from a complex computer code when fast approximations are available}},
	Volume = {87, 1},
	Pages = {1-13},
	Year = {2000}}

@article{SD1979,
	Address = {University of Delaware, Newark},
	Author = {Sharma, J. N. and Dean, R. G.},
	Institution = {University of Delaware, Newark},
	Journal = {Ocean Eng. Rep.},
	Title = {{Development and evaluation of a procedure for simulating a random directional second order sea surface and associated wave forces}},
	Volume = {20},
	Year = {1979}}

@book{Ochi1990,
	Author = {Ochi, M. K.},
	Publisher = {John Wiley {\&} Sons},
    address = {Singapore},
	Title = {{Applied Probability and Stochastic Processes in Engineering and Physical Sciences}},
	Year = {1990}}

@inproceedings{JL2018,
    address = {Madrid, Spain},
    author = {Johannessen, T. B. and Lande, {\O}.},
    booktitle = {37\textsuperscript{th} OMAE Conf.},
    doi = {10.1115/OMAE2018-78283},
    publisher = {ASME},
    title = {{Long term analysis of steep and breaking wave properties by event matching}},
    year = {2018}}

@inproceedings{BHV2020,
    address = {Virtual, Online},
    author = {Bandringa, H. and Helder, J. A. and van Essen, S. M.},
    booktitle = {39\textsuperscript{th} OMAE Conf.},
    doi = {10.1115/OMAE2020-18290},
    publisher = {ASME},
    title = {{On the validity of CFD for simulating extreme green water loads on ocean-going vessels}},
    year = {2020}}

@inproceedings{WEtAl1993,
	Address = {Innsbruck, Austria},
	Author = {Winterstein, S. R. and Ude, T. C. and Cornell, C. A. and Bjerager, P. and Haver, S.},
	Booktitle = {Int. Conf. Struct. Saf. Reliab.},
	Month = {aug},
	Title = {{Environmental parameters for extreme response: inverse FORM with omission factors}},
	Year = {1993}}

@article{S2020,
	Author = {Stansberg, C. T.},
	Doi = {10.3390/jmse8050314},
	Journal = {J. Mar. Sci. Eng.},
	Title = {Wave Front Steepness and Influence on Horizontal Deck Impact Loads},
	Volume = {8, 314},
	Year = {2020}}

@techreport{VAOZ2016,
	Author = {Viste-Ollestad, I. and Andersen, T. L. and Oma, N. and Zachariassen, S.},
	Title = {{\href{https://www.ptil.no/contentassets/34ba6b722c0c44a3a137240bae06f623/investigation-report---cosl-drilling---cosl-innovator.pdf}{Investigation report Petroleumstilsynet - Investigation of an accident with fatal consequences on COSLInnovator, 30 December 2015}}},
	Year = {2016}}

@techreport{Havtil2025,
	Author = {{Norwegian Ocean Industry Authority Havtil}},
	Title = {{\href{https://www.havtil.no/contentassets/6e6aeac3d63f410a9845c36bae6da2af/2024_247-rapport-eng-gransking-equinor-asgard-a-innslatt-lugarvindu_oppdatert-mai_skjult-innhold.pdf}{Investigation of incident on 31 January 2024 involving a cabin window on \AA{}sgard A being forced in by green water}}},
	Year = {2025}}

@inproceedings{DG2001,
	Address = {Glasgow, Scotland, UK},
	Author = {Dallinga, R. P. and Gaillarde, G.},
	Booktitle = {Glasgow Marine Fair And Int. Workshop On Safety Of Bulk Carriers},
	Title = {{Hatch cover loads experienced by M.V. Derbyshire during typhoon `Orchid'}},
    city = {Glasgow, Scotland, UK},
	Year = {2001}}

@inproceedings{EK2000,
	Address = {Seattle, USA},
	Author = {Ersdal, G. and Kvitrud, A.},
	Booktitle = {10\textsuperscript{th} ISOPE Conf.},
	Publisher = {Int. Soc. of Offshore and Polar Eng. (ISOPE)},
	Title = {{Green water on Norwegian production ships}},
	Year = {2000}}

@inproceedings{HOELAB2022,
	Address = {Hamburg, Germany},
	Author = {Halsne, M. and Oma N. and Ersdal, G. and Lang{\o}y, M. and Andersen, T. and Bj{\o}rheim, L. G.},
	Booktitle = {41\textsuperscript{th} OMAE Conf.},
	Doi = {10.1115/OMAE2022-81289},
	Publisher = {ASME},
	Title = {{Semisubmersible in service experiences on the Norwegian Continental Shelf}},
	Year = {2022}}

@phdthesis{K2018,
    address = {Delft, The Netherlands},
    author = {Kapsenberg, G. K.},
    school = {Delft University of Technology},
    title = {{On the slamming of ships}},
    type = {{PhD thesis}},
    Doi = {10.4233/uuid:14eac2bb-63ee-47e4-8218-1ba3830a97b4},
    year = {2018}}

@book{OTG14,
    Author = {DNV},
    Title = {\href{https://www.dnv.com/maritime/Offshore/technical-guidance-otg.html}{{Offshore Technical Guidance DNV-OTG-14: Horizontal wave impact loads for column stabilised units}}},
    publisher = {Det Norske Veritas},
    address = {Oslo, Norway},
    Year = {2019}}

@book{DNVGL-CG-0130,
    Author = {DNV},
    Title = {{\href{https://rules.dnv.com/docs/pdf/DNV/CG/2018-01/DNV-CG-0130.pdf}{Class Guideline DNV-CG-0130: Wave loads}}},
    publisher = {Det Norske Veritas},
    address = {Oslo, Norway},
    Year = {2018}}

@book{BV2015,
    address = {Paris, France},
    author = {BV},
    publisher = {Bureau Veritas},
    title = {{\href{https://erules.veristar.com/dy/data/bv/pdf/583-NR_2015-07.pdf}{Rule Note NR583: Whipping and springing assessment}}},
    year = {2015}}

@book{BV2019,
    address = {Paris, France},
    author = {BV},
    publisher = {Bureau Veritas},
    title = {{\href{https://erules.veristar.com/dy/data/bv/pdf/638-NI_2019-02.pdf}{Rule Note NI638: Guidance for long-term hydro-structure calculations}}},
    year = {2019}}

@book{Haver2017,
    address = {Stavanger, Norway},
    author = {Haver, S.},
    number = {May},
    pages = {1-256},
    publisher = {Haver {\&} havet, University in Stavanger, NTNU},
    title = {{Metocean modelling and prediction of extremes}},
    year = {2017}}

@book{ITTC2017Ev,
    author = {ITTC},
    publisher = {International Towing Tank Conf.},
    title = {{\href{https://www.ittc.info/media/8105/75-02-07-023.pdf}{Recommended Practice 7.5-02-07-02.3: Experiments on rarely occurring events}}},
    year = {2017}}

@book{ABS2020,
	Address = {Spring, USA},
	Author = {ABS},
    isbn = {},
	Publisher = {American Bureau of Shipping},
	Title = {{\href{https://pub-rm20.apps.eagle.org/viewer/book-attachment/hnkt3lIKpWEbOWbbm3nqnQ/K8aK9uhB5QaI4RoEjIOnoQ-hnkt3lIKpWEbOWbbm3nqnQ}{Guidance notes on air gap and wave impact analysis for semisubmersibles}}},
	Year = {2020}}

@book{ABS2021,
	Address = {Spring, USA},
	Author = {ABS},
    isbn = {},
	Publisher = {American Bureau of Shipping},
	Title = {{\href{https://pub-rm20.apps.eagle.org/viewer/book-attachment/GwEbw4kR3F46d2hwrKQcLg/fzVGzQ4EAJe2QQ1F~g4QOw-GwEbw4kR3F46d2hwrKQcLg}{Guide for slamming loads and strength assessment for vessels}}},
	Year = {2021}}

@article{KRGV2011,
	Author = {Kingston, G. B. and Rajabali Nejad, M. and Gouldby, B. P. and van Gelder, P. H. A. J. M.},
	doi = {10.1016/j.strusafe.2010.08.002},
	Journal = {Structural Safety},
	Pages = {1},
	Title = {Computational intelligence methods for the efficient reliability analysis of complex flood defence structures},
	Volume = {33},
	Year = {2011}}

@inproceedings{ABL1998,
	Address = {The Hague, The Netherlands},
	Author = {Adegeest, L. and Braathen, A. and L{\o}seth, R.},
	Booktitle = {12\textsuperscript{th} PRADS Conf.},
	Pages = {53--58},
	Title = {{Use of nonlinear sea loads simulations in design of ships}},
	Year = {1998}}

@inproceedings{TFHM2023,
	Address = {Melbourne, Australia},
	Author = {Takami, T. and Fujimoto, W. and Houtani, H. and Matsui, S.},
	Booktitle = {42\textsuperscript{nd} OMAE Conf.},
	doi = {10.1115/OMAE2023-101876},
	Publisher = {ASME},
	Title = {{Extreme wave and vertical bending moment predictions by higher-order spectrum method and FORM}},
	Year = {2023}}

@inproceedings{TDEtAl2023,
	Address = {Melbourne, Australia},
	Author = {Tang, T. and Ding, H. and Dai, S. and Chen, X. and Taylor, P. H. and Zang, J. and Adcock, T. A. A.},
	Booktitle = {42\textsuperscript{nd} OMAE Conf.},
	doi = {10.1115/OMAE2023-102682},
	Publisher = {ASME},
	Title = {{Data informed model test design with machine learning - an example in nonlinear wave load on a vertical cylinder}},
	Year = {2023}}

@article{GJL2023,
	Author = {Gramstad, O. and Johannessen, T. B. and Lian, G.},
	Doi = {10.1016/j.marstruc.2023.103473},
	Journal = {Marine Struc.},
	Title = {{Long-term analysis of wave-induced loads using High Order Spectral Method and direct sampling of extreme wave events}},
	Volume = {103473},
	Year = {2023}}

@article{HANM2006,
	Author = {Huang, D. and Allen, T. T. and Notz, W. I. and Miller, R. A.},
	Doi = {10.1007/s00158-005-0587-0},
	Journal = {Struct. Multidisc. Optim.},
	Title = {{Sequential kriging optimization using multiple-fidelity evaluations.}},
	Volume = {32},
    Pages = {369-382},
	Year = {2006}}

@inproceedings{S2008,
	Address = {Estoril, Portugal},
	Author = {Stansberg, C. T.},
	Booktitle = {27\textsuperscript{th} OMAE Conf.},
	Doi = {10.1115/OMAE2008-57801},
	Publisher = {ASME},
	Title = {{A wave impact parameter}},
	Year = {2008}}

@article{TJH1997,
	Author = {Taylor, P. H. and Jonathan, P. and Harland, L. A.},
	Doi = {10.1115/1.2889772},
	Issn = {1048-9002},
	Journal = {J. Vib. Acoust.},
	Pages = {624--628},
	Title = {{Time domain simulation of jack-up dynamics with the extremes of a Gaussian process}},
	Volume = {119},
	Year = {1997}}

@article{FFN2021,
	Author = {Fuhg, J. N. and Fau, A. and Nackenhorst, U.},
	Journal = {Archives of Computational Methods in Engineering},
	Pages = {2689--2747},
	Title = {State--of--the--Art and Comparative Review of Adaptive Sampling
Methods for Kriging},
	Volume = {28},
    doi = {10.1007/s11831-020-09474-6},
	Year = {2021}}

@inproceedings{PEtAl2022,
	Address = {Hamburg, Germany},
	Author = {P{\'{a}}kozdi, C. and Califano, A. and Akselsen, A. and Croonenborghs, E. and Kim, J. and Peric, M. and Loubeyre, S. and Bouscasse, B. and Ducrozet, G. and Xu-Haihua},
	Booktitle = {41\textsuperscript{st} OMAE Conf.},
	doi = {10.1115/OMAE2022-79152},
	Publisher = {ASME},
	Title = {{Joint-industry effort to develop and verify CFD modeling practice for predicting wave impact}},
	Year = {2022}}

@inproceedings{BHD2015,
	Address = {St. John's, Newfoundland, Canada},
	Author = {Bunnik, T. H. J. and Helder, J. A. and de Ridder, E.-J.},
	Booktitle = {34\textsuperscript{th} OMAE Conf.},
	doi = {10.1115/OMAE2015-41989},
	Publisher = {ASME},
	Title = {{Deterministic simulation of breaking wave impact and flexible response of a fixed offshore wind turbine}},
	Year = {2015}}

@inproceedings{BH2018,
	Address = {Madrid, Spain},
	Author = {Bandringa, H. and Helder, J. A.},
	Booktitle = {37\textsuperscript{th} OMAE Conf.},
	doi = {10.1115/OMAE2018-78089},
	Publisher = {ASME},
	Title = {{On the validity and sensitivity of CFD simulations for a deterministic breaking wave impact on a semi subsmersible}},
	Year = {2018}}

@article{GKS2023,
	Author = {Guth, S. and Katsidoniotaki, E. and Sapsis, T. P.},
	Journal = {Wind Energy},
	Pages = {1--26},
	Title = {{Statistical modeling of fully nonlinear hydrodynamic loads on offshore wind turbine monopile foundations using wave episodes and targeted CFD simulations through active sampling}},
	Volume = {28},
    doi = {10.1002/we.2880},
	Year = {2023}}

@article{GuardianMaud,
	Author = {Reuters},
	Journal = {The Guardian},
	Title = {{\href{https://www.theguardian.com/world/2023/dec/22/norwegian-cruise-ship-ms-maud-loses-power-navigation-system}{Norwegian cruise ship MS Maud loses power in North Sea during storm}}},
	Year = {2023}}

@article{ABCViking,
	Author = {Pereira, I.},
	Journal = {ABC News},
	Title = {{\href{https://abcnews.go.com/International/coast-guard-probing-deaths-injuries-americans-vessels-antarctic/story?id=96864073}{Coast Guard probing deaths, injuries of Americans on vessels in Antarctic waters}}},
	Year = {2023}}

@article{SEXL2022,
	Author = {Sajad {Saraygord Afshari} and Fatemeh Enayatollahi and Xiangyang Xu and Xihui Liang},
	Journal = {{Reliab. Eng. \& System Safety}},
	Title = {{Machine learning-based methods in structural reliability analysis: A review}},
	Year = {2022},
    pages = {108223},
    issn = {0951-8320},
    doi={10.1016/j.ress.2021.108223}}

@article{ZFW2017,
	Author = {Wei Zhao and Feng Fan and Wei Wang},
	Journal = {{Reliab. Eng. \& System Safety}},
	Title = {{Non-linear partial least squares response surface method for structural reliability analysis}},
	Year = {2017},
    volume = {161},
    pages = {69-77},
    issn = {0951-8320},
    doi={10.1016/j.ress.2017.01.004}}

@article{KL2015,
	Author = {Dong Hyawn Kim and Sang Geun Lee},
	Journal = {{Renewable Energy}},
	Title = {{Reliability analysis of offshore wind turbine support structures under extreme ocean environmental loads}},
	Year = {2015},
    volume = {79},
    pages = {161-166},
    doi={0.1016/j.renene.2014.11.052}}

@inproceedings{KJYN2011,
	Author = {Kim, Hyunyul and Jeong, Shinkyu and Yang, Chi and Noblesse, Francis},
	Booktitle = {21\textsuperscript{st} ISOPE Conf.},
	Publisher = {Int. Society of Offshore and Polar Eng.},
	Title = {{Hull form design exploration based on response surface method}},
    pages = {816-825},
	Year = {2011}}

@article{TWB1998,
    author = {Torhaug, R. and Winterstein, S. R. and Braathen, A.},
    doi = {10.5957/jsr.1998.42.1.46},
    journal = {J. Ship Research},
    pages = {46--55},
    title = {{Nonlinear ship loads: stochastic models for extreme response}},
    volume = {42, 1},
    year = {1998}}

@article{J2009,
    author = {Jensen, J. J.},
    doi = {10.1080/17445300802370461},
    journal = {Ships and Offshore Structures},
    pages = {325-333},
    title = {{Extreme value predictions and critical wave episodes for marine structures by FORM}},
    volume = {3, 4},
    year = {2009}}

@article{DLKDBD2025,
    author = {Athanasios Dermatis and Marine Lasbleis and Shinwoong Kim and Guillaume De Hauteclocque and Benjamin Bouscasse and Guillaume Ducrozet},
    doi = {10.1016/j.oceaneng.2024.119919},
    journal = {Ocean Eng.},
    volume = {316},
    pages = {119919},
    title = {{A multi-fidelity approach for the evaluation of extreme wave loads using nonlinear response-conditioned waves}},
    year = {2025}}

@article{DBDBB2025,
  title        = {Prediction of the extreme slow-drift response of moored floating structures using design waves},
  author       = {Athanasios Dermatis and Benjamin Bouscasse and Guillaume Ducrozet and Henrik Bredmose and Harry B. Bingham},
  journal      = {Ocean Eng.},
  volume       = {333},
  pages        = {121456},
  year         = {2025},
  doi          = {10.1016/j.oceaneng.2025.121456}}

@article{AJSB2024,
	Author = {Mohammad Mahdi Abaei and Bernt Johan Leira and Svein Saevik and Ahmad BahooToroody},
	Journal = {{Reliab. Eng. \& System Safety}},
	Title = {{Integrating physics-based simulations with gaussian processes for enhanced safety assessment of offshore installations}},
	Year = {2024},
    volume = {249},
    pages = {110235},
    doi={10.1016/j.ress.2024.110235}}

@article{SRTJ2024,
	Author = {Matthew Speers and David Randell and Jonathan Tawn and Philip Jonathan},
	Journal = {{Ocean Eng.}},
	Title = {{Estimating metocean environments associated with extreme structural response to demonstrate the dangers of environmental contour methods}},
	Year = {2024},
    volume = {311},
    pages = {118754},
    doi={10.1016/j.oceaneng.2024.118754}}

@article{NM2024a,
	Author = {Phong T. T. Nguyen and Lance Manuel},
	Journal = {{Reliab. Eng. \& System Safety}},
	Title = {{Uncertainty quantification in low-probability response estimation using sliced inverse regression and polynomial chaos expansion}},
	Year = {2024},
    volume = {242},
    pages = {109750},
    doi={10.1016/j.ress.2023.109750}}

@article{MI2024a,
	Author = {Amandine Marrel and Bertrand Iooss},
	Journal = {{Reliab. Eng. \& System Safety}},
	Title = {{Probabilistic surrogate modeling by Gaussian process: A review on recent insights in estimation and validation}},
	Year = {2024},
    volume = {247},
    pages = {110094},
    doi={10.1016/j.ress.2024.110094}}

@article{MI2024b,
	Author = {Amandine Marrel and Bertrand Iooss},
	Journal = {{Reliab. Eng. \& System Safety}},
	Title = {{Probabilistic surrogate modeling by Gaussian process: A new estimation algorithm for more robust prediction}},
	Year = {2024},
    volume = {247},
    pages = {110120},
    doi={10.1016/j.ress.2024.110120}}

@article{KOPS2025,
	Author = {Minji Kim and Kevin O'Conner and Vladas Pipiras and Themistoklis Sapsis},
	Journal = {{SIAM/ASA J. on Uncertainty Quantification}},
	Title = {{Sampling low-fidelity outputs for estimation of high-fidelity density and its tails}},
	Year = {2025},
    volume = {13, 1},
    pages = {30-62},
    doi={10.1137/24M1639142}}

@article{LLTL2024,
	Author = {D. Lucio and J.L. Lara and A. Tomas and I.J. Losada},
	Journal = {{Reliab. Eng. \& System Safety}},
	Title = {{Probabilistic assessment of climate-related impacts and risks in ports}},
	Year = {2024},
    volume = {251},
    pages = {110333},
    doi={10.1016/j.ress.2024.110333}}

@article{FXLL2024,
	Author = {Chen Fang and You-Lin Xu and Yongle Li and Jinrong Li},
	Journal = {{Reliab. Eng. \& System Safety}},
	Title = {{Serviceability analysis of sea-crossing bridges under correlated wind and wave loads}},
	Year = {2024},
    volume = {246},
    pages = {110077},
    doi={10.1016/j.ress.2024.110077}}

@book{DNV2019,
	Address = {Oslo, Norway},
	Author = {DNV},
	Publisher = {Det Norske Veritas},
	Title = {{\href{https://www.dnv.com/energy/standards-guidelines/dnv-rp-c205-environmental-conditions-and-environmental-loads/}{Recommended Practice RP-C205: Environmental conditions and environmental loads}}},
	Year = {2019}}

@article{MMJ2025,
   author = {E. B. L. Mackay and Murphy-Barltrop, C.J.R. and Philip Jonathan},
   journal = {J. Offshore Mechanics and Arctic Eng.},
   title = {The {SPAR} model: a new paradigm for multivariate extremes: application to joint distributions of metocean variables (OMAE-24-1018)},
   volume = {147},
   year = {2025},
   doi = {10.1115/1.4065968},
   pages = "011205:1-10"}

@article{HVN2013,
	Author = {Arne Bang Huseby and Erik Vanem and Bent Natvig},
	Doi = {10.1016/j.oceaneng.2012.12.034},
	Journal = {Ocean Eng.},
	pages = {124-135},
	Title = {{A new approach to environmental contours for ocean engineering applications based on direct Monte Carlo simulations}},
	Volume = {60},
	Year = {2013}}

@article{NM2024b,
	Author = {Phong T. T. Nguyen and Lance Manuel},
	Journal = {{Ocean Eng.}},
	Title = {{A bi-fidelity surrogate model for extreme loads on offshore structures}},
	Year = {2024},
    volume = {307},
    pages = {118175},
    doi={10.1016/j.oceaneng.2024.118175}}

@article{CTNCG2015,
	Author = {A.A. Chojaczyk and A.P. Teixeira and L.C. Neves and J.B. Cardoso and C. Guedes Soares},
	Journal = {{Structural Safety}},
	Title = {{Review and application of Artificial Neural Networks models in reliability analysis of steel structures}},
	Year = {2015},
    volume = {52},
    pages = {78-89},
    doi={10.1016/j.strusafe.2014.09.002}}

@article{RMC2018,
	Author = {Atin Roy and Ramkrishna Manna and Subrata Chakraborty},
	Journal = {{Probabilistic Eng. Mechanics}},
	Title = {{Support vector regression based metamodeling for structural reliability analysis}},
	Year = {2018},
    volume = {55},
    pages = {78-89},
    doi={10.1016/j.probengmech.2018.11.001}}

@article{TJG2022,
	Author = {Armin Tabandeh and Gaofeng Jia and Paolo Gardoni},
	Journal = {{Structural Safety}},
	Title = {{A review and assessment of importance sampling methods for reliability analysis}},
	Year = {2022},
    volume = {97},
    pages = {102216},
    doi={10.1016/j.strusafe.2022.102216}}

@article{CPLZW2023,
	Author = {Kai Cheng and Iason Papaioannou and Zhenzhou Lu and Xiaobo Zhang and Yanping Wang},
	Journal = {{Structural Safety}},
	Title = {{Rare event estimation with sequential directional importance sampling}},
	Year = {2023},
    volume = {100},
    pages = {102291},
    doi={10.1016/j.strusafe.2022.102291}}

@article{CGMD2023,
	Author = {Marie Chiron and Christian Genest and Jérôme Morio and Sylvain Dubreuil},
	Journal = {{Reliab. Eng. \& System Safety}},
	Title = {{Failure probability estimation through high-dimensional elliptical distribution modeling with multiple importance sampling}},
	Year = {2023},
    volume = {235},
    pages = {109238},
    doi={10.1016/j.ress.2023.109238}}

@article{BSGT2024,
	Author = {Bao, Y. and Sun, H. and Guan, X. and Tian., Y.},
	Journal = {{Reliab. Eng. \& System Safety}},
	Title = {{An active learning method using deep adversarial autoencoder-based sufficient dimension reduction neural network for high-dimensional reliability analysis}},
	Year = {2024},
    volume = {247},
    pages = {110140},
    doi={10.1016/j.ress.2024.110140}}

@article{ZDF2024,
	Author = {Zhang, Y. and Dong, Y. and Frangopol, D. M.},
	Journal = {{Reliab. Eng. \& System Safety}},
	Title = {{An error-based stopping criterion for spherical decomposition-based adaptive Kriging model and rare event estimation}},
	Year = {2024},
    volume = {241},
    pages = {109610},
    doi={10.1016/j.ress.2023.109610}}

@article{CPS2022,
	Author = {Jianpeng Chan and Iason Papaioannou and Daniel Straub},
	Journal = {{Reliab. Eng. \& System Safety}},
	Title = {{An adaptive subset simulation algorithm for system reliability analysis with discontinuous limit states}},
	Year = {2022},
    volume = {225},
    pages = {108607},
    doi={10.1016/j.ress.2022.108607}}

@article{CFPO2023,
	Author = {Fernandez Castellon, D. and Fenerci, A. and Petersen, {\O}. W. and {\O}iseth, O.},
	Journal = {{Reliab. Eng. \& System Safety}},
	Title = {{Full long-term buffeting analysis of suspension bridges using Gaussian process surrogate modelling and importance sampling Monte Carlo simulations}},
	Year = {2023},
    volume = {235},
    pages = {109211},
    doi={10.1016/j.ress.2023.109211}}

@article{WG2021,
	Author = {David Wilkie and Carmine Galasso},
	Journal = {{Structural Safety}},
	Title = {{Gaussian process regression for fatigue reliability analysis of offshore wind turbines}},
	Year = {2021},
    volume = {88},
    pages = {102020},
    doi={10.1016/j.strusafe.2020.102020}}

@article{HWZHSL2025,
	Author = {Fucheng Han and Wenhua Wang and Xiao-Wei Zheng and Xu Han and Wei Shi and Xin Li},
	Journal = {{Reliab. Eng. \& System Safety}},
	Title = {{Investigation of essential parameters for the design of offshore wind turbine based on structural reliability}},
	Year = {2025},
    volume = {254},
    pages = {110601},
    doi={10.1016/j.ress.2024.110601}}

@article{CLST2011,
	Author = {Giovanni Cuomo and Giorgio Lupoi and {Ken-ihiro} Shimosako and Shigeo Takahashi},
	Journal = {{Coastal Eng.}},
	Title = {{Dynamic response and sliding distance of composite breakwaters under breaking and non-breaking wave attack}},
	Year = {2011},
    volume = {58},
    issue = {10},
    pages = {953-969},
    doi={10.1016/j.coastaleng.2011.03.008}}

@article{ABDRBPD2021,
	Author = {Alessandro Antonini and James Mark William Brownjohn and Darshana Dassanayake and Alison Raby and James Bassit and Athanasios Pappas and Dina D'Ayala},
	Journal = {{Coastal Eng.}},
	Title = {{A Bayesian inverse dynamic approach for impulsive wave loading reconstruction}},
	Year = {2021},
    volume = {168},
    pages = {103920},
    doi={10.1016/j.coastaleng.2021.103920}}

@article{CST2009,
	Author = {Giovanni Cuomo and Ken-ichiro Shimosako and Shigeo Takahashi},
	Journal = {{Coastal Eng.}},
	Title = {{Wave-in-deck loads on coastal bridges and the role of air}},
	Year = {2009},
    volume = {56},
    issue = {8},
    pages = {793-809},
    doi={10.1016/j.coastaleng.2009.01.005}}

@article{ZXZC2025,
	Author = {Nianfan Zhang and Longfei Xiao and Qingping Zou and Cathal Cummins},
	Journal = {{Ocean Eng.}},
	Title = {{Large-scale wave basin experimental study on the spatio-temporal distribution of wave impact loads on a semi-submersible platform}},
	Year = {2025},
    volume = {327},
    pages = {120991},
    doi={10.1016/j.oceaneng.2025.120991}}

@article{ZYMCZ2023,
	Author = {Ting Zhou and Yin Yin and Zhe Ma and Jingjie Chen and Gangjun Zhai},
	Journal = {{Ocean Eng.}},
	Title = {{Numerical investigation of breaking waves impact on vertical breakwater with impermeable and porous foundation}},
	Year = {2023},
    volume = {280},
    pages = {114477},
    doi={10.1016/j.oceaneng.2023.114477}}

@inproceedings{VBS2024,
    address = {Singapore},
    author = {van Essen, S. M. and Bunnik, T. H. J. and Scharnke, J.},
    booktitle = {43\textsuperscript{rd} OMAE Conf.},
    doi = {10.1115/OMAE2024-122486},
    publisher = {ASME},
    title = {{Statistical uncertainty of ship response to waves as a function of test duration}},
    year = {2024}}

@article{VS2025,
	Author = {van Essen, S. M. and Seyffert, H. C.},
	Journal = {{Reliability Eng. \& System Safety}},
	Title = {{Designing for dangerous waves -- a new `Adaptive Screening' method to predict extreme values of non-linear marine and coastal structure responses to waves}},
    pages = {111404},
    volume = {264B},
	Year = {2025}, 
    doi = {10.1016/j.ress.2025.111404}}

@article{VSdata2025,
	Author = {van Essen, S. M. and Seyffert, H. C.},
	Journal = {4TU Repository.},
	Title = {{Scripts and data underlying the publication that defines and applies the new Adaptive Screening method, for extreme value prediction of non-linear wave-induced responses}},
	Year = {2025},
    doi = {10.4121/f1348609-c912-4d06-82b8-197c01f3437b}}

@article{VSdata2025PRADS,
	Author = {van Essen, S. M. and Seyffert, H. C.},
	Journal = {4TU Repository.},
	Title = {{Scripts and data for PRADS publication that varies the acquisition function of the Adaptive Screening method}},
	Year = {2025},
    doi = {10.4121/12777259-c2f6-4b44-b71f-eec5557824d1}}

@inproceedings{VS2025prads,
	Address = {Ann Arbor, Michigan, USA},
	Author = {van Essen, S. M. and Seyffert, H. C.},
	Booktitle = {16\textsuperscript{th} PRADS Conf.},
	Title = {{Designing ships for extreme non-linear responses - the role of the acquisition function in the Adaptive Screening extreme value prediction method}},
	Year = {2025},
    doi = {10.5281/zenodo.17305196}}

@inproceedings{S1959,
  title={Fonctions de r{\'e}partition {\`a} n dimensions et leurs marges},
  author={Sklar, A.},
  booktitle={Publications de l'Institut de statistique de l'Universit{\'e} de Paris},
  volume={8},
  number={3},
  pages={229--231},
  year={1959}}

@article{MECRSLK2024,
	Author = {M. McCann and B. Ebrahimi and G.E. Cinar and W. Renteria and A. Stehno and P. Lynett and J. Kaihatu},
	Journal = {{Coastal Eng.}},
	Title = {{Field observations of Hurricane Ian's wave and surge impact in the areas of Fort Myers Beach and Sanibel Island, USA}},
	Year = {2024},
    volume = {188},
    pages = {104450},
    doi={10.1016/j.coastaleng.2023.104450}}

@incollection{SES2015,
    title = {{Extreme waves: causes, characteristics, and impact on coastal environments and society (Ch.11)}},
    editor = {John F. Shroder and Jean T. Ellis and Douglas J. Sherman},
    booktitle = {Coastal and Marine Hazards, Risks, and Disasters},
    publisher = {Elsevier},
    address = {Boston},
    pages = {307-334},
    year = {2015},
    series = {Hazards and Disasters Series},
    doi = {10.1016/B978-0-12-396483-0.00011-X},
    author = {Jim D. Hansom and Adam D. Switzer and Jeremy Pile}}

@article{XL2026,
    title = {Copula-based conditional reliability analysis of slopes in spatially variable soils},
    Journal = {{Reliability Eng. \& System Safety}},
    volume = {265},
    pages = {111522},
    year = {2026},
    issn = {0951-8320},
    doi = {10.1016/j.ress.2025.111522},
    author = {Yue-Bing Xu and Lei-Lei Liu}}

@article{WWZCJ2025,
    title = {Reliability updating of copula-dependent spatially variable soil slopes based on data-augmented MGPR model along slip surfaces},
    journal = {Engineering Geology},
    volume = {356},
    pages = {108280},
    year = {2025},
    issn = {0013-7952},
    doi = {10.1016/j.enggeo.2025.108280},
    author = {Tao Wang and Changxing Wang and Yifeng Zhou and Hongzhi Cui and Jian Ji}}

@article{ZZYBW2026,
    title = {Reliability-based design optimization method with correlated variables using adaptive conjugate gradient analysis and Copula},
    Journal = {{Reliability Eng. \& System Safety}},
    volume = {265},
    pages = {111469},
    year = {2026},
    issn = {0951-8320},
    doi = {10.1016/j.ress.2025.111469},
    author = {Dequan Zhang and Jingke Zhang and Meide Yang and Shaoqiang Bai and Fang Wang}}

@article{LFFT2025,
    title = {Dimensional synchronous modeling-based enhanced Kriging algorithm and adaptive Copula method for multi-objective synthetical reliability analyses},
    journal = {Chinese Journal of Aeronautics},
    volume = {38},
    number = {9},
    pages = {103652},
    year = {2025},
    issn = {1000-9361},
    doi = {10.1016/j.cja.2025.103652},
    author = {Cheng Lu and Yunwen Feng and Chengwei Fei and Da Teng}}

@article{QXSH2025,
    title = {Seismic reliability assessment for valve hall in converter station based on Copula theory},
    journal = {Structures},
    volume = {77},
    pages = {109002},
    year = {2025},
    issn = {2352-0124},
    doi = {https://doi.org/10.1016/j.istruc.2025.109002},
    author = {Xinzhu Qiao and Qiang Xie and Gaoyang Shi and Jinyu Hu}}

@article{MVM2024,
    title = {A copula-based model to describe the uncertainty of overtopping variables on mound breakwaters},
    journal = {Coastal Eng.},
    volume = {189},
    pages = {104483},
    year = {2024},
    issn = {0378-3839},
    doi = {10.1016/j.coastaleng.2024.104483},
    author = {Patricia Mares-Nasarre and Marcel R.A. {van Gent} and Oswaldo Morales-N{\'a}poles}}

@article{BCLJ2025,
    title = {{Confidence intervals for the reliability of dependent systems: integrating frailty models and copula-based methods}},
    journal = {Computer Modeling in Eng. and Sciences},
    volume = {143},
    number = {2},
    pages = {1401-1431},
    year = {2025},
    issn = {1526-1492},
    doi = {10.32604/cmes.2025.064487},
    author = {Osnamir E. Bru-Cordero and Cecilia Castro and V{\'i}ctor Leiva and Mario C. Jaramillo-Elorza}}

@article{DWLSWZ2024,
    title = {A statistical analysis method for significant wave height and spectral peak frequency considering the random and time-varying effects based on copula function and Bayesian inference},
    journal = {Ocean Modelling},
    volume = {190},
    pages = {102390},
    year = {2024},
    issn = {1463-5003},
    doi = {10.1016/j.ocemod.2024.102390},
    author = {Xiaochuan Duan and Shaoping Wang and Di Liu and Jian Shi and Yinghua Wu and Xiaobao Zhou}}

@article{WZX2022,
    title = {Spatial distribution models of horizontal and vertical wave impact pressure on the elevated box structure},
    journal = {Applied Ocean Research},
    volume = {125},
    pages = {103245},
    year = {2022},
    issn = {0141-1187},
    doi = {10.1016/j.apor.2022.103245},
    author = {Kai Wei and Cong Zhou and Bo Xu}}

@article{SC2011,
    title = {Characterizing impulsive wave-in-deck loads on coastal bridges by probabilistic models of impact maxima and rise times},
    journal = {Coastal Engineering},
    volume = {58},
    number = {9},
    pages = {908-926},
    year = {2011},
    issn = {0378-3839},
    doi = {10.1016/j.coastaleng.2011.05.010},
    author = {Francesco Serinaldi and Giovanni Cuomo}}

@article{GSN2016,
    title = {Extreme large cargo ship panel stresses by bivariate ACER method},
    journal = {Ocean Eng.},
    volume = {123},
    pages = {432-439},
    year = {2016},
    issn = {0029-8018},
    doi = {10.1016/j.oceaneng.2016.06.048},
    author = {Oleg Gaidai and Gaute Storhaug and Arvid Naess}}

@article{BGJM2020,
	Author = {N. Beck and C. Genest and J. Jalbert and M. Mailhot},
	Journal = {{Environmetrics}},
	Title = {{Predicting extreme surges from sparse data using a
copula-based hierarchical Bayesian spatial model}},
	Year = {2020},
    volume = {31},
    pages = {e2616},
    doi={10.1002/env.2616}}

@article{LNDH2025,
    title = {Bayesian model updating with variational inference and Gaussian copula model},
    journal = {Comput. Methods Appl. Mech. Eng.},
    volume = {438},
    pages = {117842},
    year = {2025},
    issn = {0045-7825},
    doi = {10.1016/j.cma.2025.117842},
    author = {Qiang Li and Pinghe Ni and Xiuli Du and Qiang Han}}

@article{PCMW2016,
	Author = {Benjamin Peherstorfer and Tiangang Cui and Youssef Marzouk and Karen Willcox},
	Journal = {{Comput. Methods Appl. Mech. Eng.}},
	Title = {{Multifidelity importance sampling}},
	Year = {2016},
    volume = {300},
    pages = {490--509},
    doi={10.1016/j.cma.2015.12.002}}

@article{KNPVW2019,
	Author = {Boris Kramer and Alexandre Noll Marques and Benjamin Peherstorfer and Umberto Villa and Karen Willcox},
	Journal = {{J. Computational Phys.}},
	Title = {{Multifidelity probability estimation via fusion of estimators}},
	Year = {2019},
    volume = {392},
    pages = {385--402},
    doi={10.1016/j.jcp.2019.04.071}}

@inproceedings{Proppe2019,
    address = {Singapore},
    author = {Carsten Proppe},
    booktitle = {29\textsuperscript{th} ESREL Conf.},
    doi = {10.3850/978-981-11-2724-3 0168-cd},
    publisher = {Research Publishing},
    editor = {Michael Beer and Enrico Zio},
    title = {{Reliability estimation with multi-fidelity simulation methods}},
    year = {2019}}

@article{GM2010,
	Author = {Souparno Ghosh and Bani K. Mallick},
	Journal = {{Environmetrics}},
	Title = {{A hierarchical Bayesian spatio-temporal model for extreme precipitation events}},
	Year = {2010},
    volume = {22},
    pages = {192--204},
    doi={10.1002/env.1043}}

@article{KJ2019,
    title = {Nonparametric estimation of multivariate tail probabilities and tail dependence coefficients},
    journal = {Journal of Multivariate Analysis},
    volume = {172},
    pages = {147-161},
    year = {2019},
    note = {Dependence Models},
    issn = {0047-259X},
    doi = {10.1016/j.jmva.2019.02.013},
    author = {Pavel Krupskii and Harry Joe}}

@article{BHRM2018,
	Author = {Bracken, C. and Holman, K.D. and Rajagopalan, B. and Moradkhani, H.},
	Journal = {{Water Resources Research}},
	Title = {{A Bayesian hierarchical approach to multivariate nonstationary hydrologic frequency analysis}},
	Year = {2018},
    volume = {54},
    pages = {243--255},
    doi={10.1002/2017WR020403}}

@article{DPR2012,
	Author = {A. C. Davison and S. A. Padoan and M. Ribatet},
	Journal = {{Statistical Science}},
	Title = {{Statistical modeling of spatial extremes}},
	Year = {2012},
    volume = {27},
    number = {2},
    pages = {161--186},
    doi={10.1214/11-STS376}}

@article{WGSMV2024,
    title = {Comparison of probabilistic structural reliability methods for ultimate limit state assessment of wind turbines},
    journal = {Structural Safety},
    volume = {111},
    pages = {102502},
    year = {2024},
    issn = {0167-4730},
    doi = {10.1016/j.strusafe.2024.102502},
    author = {Hong Wang and Odin Gramstad and Styfen Sch{\"a}r and Stefano Marelli and Erik Vanem}}

@article{EnvContBenchmark,
    title = {A benchmarking exercise for environmental contours},
    journal = {Ocean Eng.},
    volume = {236},
    pages = {109504},
    year = {2021},
    issn = {0029-8018},
    doi = {10.1016/j.oceaneng.2021.109504},
    author = {Andreas F. Haselsteiner and Ryan G. Coe and Lance Manuel and Wei Chai and Bernt Leira and Guilherme Clarindo and C. {Guedes Soares} and \'{A}sta Hannesd{\'o}ttir and Nikolay Dimitrov and Aljoscha Sander and Jan-Hendrik Ohlendorf and Klaus-Dieter Thoben and Guillaume de Hauteclocque and Ed Mackay and Philip Jonathan and Chi Qiao and Andrew Myers and Anna Rode and Arndt Hildebrandt and Boso Schmidt and Erik Vanem and Arne Bang Huseby}}

@article{DMV2022,
    title = {Quantitative comparison of environmental contour approaches},
    journal = {Ocean Eng.},
    volume = {245},
    pages = {110374},
    year = {2022},
    issn = {0029-8018},
    doi = {10.1016/j.oceaneng.2021.110374},
    author = {Guillaume de Hauteclocque and Ed Mackay and Erik Vanem}}

@article{MZCLGG2026,
    title = {A copula-based transitional markov chain monte carlo method for bayesian model updating},
    Journal = {{Reliability Eng. \& System Safety}},
    volume = {265},
    pages = {111572},
    year = {2026},
    issn = {0951-8320},
    doi = {10.1016/j.ress.2025.111572},
    author = {Pengfei Ma and Yi Zhang and Enjian Cai and Min Luo and Jing Guo and Tong Guo}}

@article{LSHH2020,
    title = {A hybrid Gaussian process model for system reliability analysis},
    Journal = {{Reliability Eng. \& System Safety}},
    volume = {197},
    pages = {106816},
    year = {2020},
    issn = {0951-8320},
    doi = {10.1016/j.ress.2020.106816},
    author = {Meng Li and Mohammadkazem Sadoughi and Zhen Hu and Chao Hu}}

@article{LW2022,
    title = {An active learning reliability analysis method using adaptive Bayesian compressive sensing and Monte Carlo simulation (ABCS-MCS)},
    Journal = {{Reliability Eng. \& System Safety}},
    volume = {221},
    pages = {108377},
    year = {2022},
    issn = {0951-8320},
    doi = {10.1016/j.ress.2022.108377},
    author = {Peiping Li and Yu Wang}}

@article{YSQD2024,
    title = {Adaptive importance sampling approach for structural time-variant reliability analysis},
    journal = {Structural Safety},
    volume = {111},
    pages = {102500},
    year = {2024},
    issn = {0167-4730},
    doi = {10.1016/j.strusafe.2024.102500},
    author = {Xiukai Yuan and Yunfei Shu and Yugeng Qian and Yiwei Dong}}

@article{CPS2025,
    title = {Enhanced sequential directional importance sampling for structural reliability analysis},
    journal = {Structural Safety},
    volume = {114},
    pages = {102574},
    year = {2025},
    issn = {0167-4730},
    doi = {10.1016/j.strusafe.2025.102574},
    author = {Kai Cheng and Iason Papaioannou and Daniel Straub}}

@article{SL2024,
    title = {On the structure dynamic response induced by the dam-break surge impact using multivariate copulas},
    journal = {Ocean Eng.},
    volume = {306},
    pages = {118100},
    year = {2024},
    issn = {0029-8018},
    doi = {10.1016/j.oceaneng.2024.118100},
    author = {Jia Shen and Haijiang Liu}}

@article{YXSZB2025,
    title = {Investigation of methods for the localization and reconstruction of the wave impact on a floating wind turbine pontoon},
    journal = {J. of Ocean Eng. and Science},
    year = {2025},
    issn = {2468-0133},
    doi = {10.1016/j.joes.2025.08.006},
    author = {Zhou Yang and Yuwang Xu and Liying Shi and Chaochao Zhu and Yichen Bao}}

@article{CPA2011,
    title = {Evaluation of wave impact loads on caisson breakwaters based on joint probability of impact maxima and rise times},
    journal = {Coastal Eng.},
    volume = {58},
    number = {1},
    pages = {9-27},
    year = {2011},
    issn = {0378-3839},
    doi = {10.1016/j.coastaleng.2010.08.003},
    author = {Giovanni Cuomo and Rodolfo Piscopia and William Allsop}}

@article{DV2005,
	Author = {D.J. {de Waal} and P.H.A.J.M. {van Gelder}},
	Journal = {{Extremes}},
	Title = {{Modelling of extreme wave heights and periods through copulas}},
	Year = {2005},
    volume = {8},
    pages = {345--356},
    doi={10.1007/s10687-006-0006-y}}

@article{SACM2025,
	Author = {},
	Journal = {{Applied Ocean Research}},
	Title = {{Gaussian copula-based Bayesian Networks for dynamic loads in mooring systems}},
	Year = {2025},
    volume = {165},
    pages = {104809},
    doi={10.1016/j.apor.2025.104809}}

@article{VSdata2025pas,
	Author = {van Essen, S. M. and Seyffert, H. C.},
	Journal = {4TU Repository.},
	Title = {{Scripts and data underlying the paper that introduces and validates the Probabilistic Adaptive Screening method}},
	Year = {2025},
    doi = {10.4121/be3e7819-dabf-4a21-bb29-5b76179ff696}}

@inproceedings{Vprads2019,
	Address = {Yokohama, Japan},
	Author = {van Essen, S. M.},
	Booktitle = {14\textsuperscript{th} PRADS Conf. 2019. Lecture Notes in Civil Engineering, vol 63},
	Doi = {10.1007/978-981-15-4624-2_54},
	Publisher = {Springer, Singapore},
    Editor = {{Okada, T., Suzuki, K., Kawamura, Y.}},
	Title = {{Influence of Wave Variability on Ship Response During Deterministically Repeated Seakeeping Tests at Forward Speed}},
	Year = {2021}}

@phdthesis{B2002,
	Address = {Delft, The Netherlands},
	Author = {Buchner, B.},
	Publisher = {Delft University of Technology},
	School = {Delft University of Technology},
	Title = {{Green water on ship-type offshore structures}},
	Type = {{PhD thesis}},
	Year = {2002}}

@book{DNV-RP-C205,
    Author = {DNV},
    Title = {Det Norske Veritas, Recommended Practice DNV-RP-C205: Environmental conditions and environmental loads},
    publisher = {Det Norske Veritas},
    address = {Oslo, Norway},
    Year = {2019}}

@book{ITTC2017_Glob,
    author = {ITTC},
    publisher = {International Towing Tank Conference},
    title = {{\href{https://www.ittc.info/media/8109/75-02-07-026.pdf}{Recommended Practice 7.5-02-07-02.6: Global loads seakeeping procedure}}},
    year = {2017}}

@article{DS1990,
	Author = {Davidson, A. C. and Smith, R. L.},
	Journal = {{J. Royal Statistical Society, Series B (methodological)}},
	Title = {{\href{https://www.jstor.org/stable/2345667}{Models for exceedances over high thresholds}}},
	Year = {1990},
    volume = {52},
    number = {3},
    pages = {393--442}}

@article{JRWT2020,
	Author = {Jonathan, P. and Randell, D. and Wadsworth, J. and Tawn, J.},
	Journal = {{Ocean Eng.}},
	Title = {{Uncertainties in return values from extreme value analysis of peaks over threshold using the generalised Pareto distribution}},
	Year = {2021},
    volume = {220},
    pages = {107725},
    doi={10.1016/j.oceaneng.2020.107725}}

@incollection{GS2010,
  author    = {G{\"u}dendorf, Gordon and Segers, Johan},
  title     = {Extreme-Value Copulas},
  booktitle = {Copula Theory and Its Applications},
  editor    = {Jaworski, Piotr and Durante, Fabrizio and H{\"a}rdle, Wolfgang Karl and Rychlik, Tomasz},
  series    = {Lecture Notes in Statistics},
  volume    = {198},
  pages     = {127--145},
  publisher = {Springer},
  year      = {2010}}

@article{Lavaud2018,
  author  = {Lavaud, S.},
  title   = {Tail dependence in small samples: from theory to practice},
  journal = {Journal of Operational Risk},
  year    = {2018},
  volume  = {13},
  number  = {1},
  pages   = {15--49},
  doi     = {10.21314/JOP.2017.196}}

@Book{RcopBook,
  title =    {{E}lements of {C}opula {M}odeling with \textsf{R}},
  year =     {2018},
  publisher =    {Springer Use R! Series},
  isbn =         {978-3-319-89635-9},
  url =          {http://www.springer.com/de/book/9783319896342},
  author =   {Marius Hofert and Ivan Kojadinovic and Martin Maechler and Jun Yan}}

@Manual{RcopPackage,
  title = {copula: Multivariate Dependence with Copulas},
  author = {Marius Hofert and Ivan Kojadinovic and Martin Maechler and Jun Yan},
  year = {2016},
  url = {http://cran.r-project.org/package=copula}}

@book{BA2002,
  author    = {Burnham, Kenneth P. and Anderson, David R.},
  title     = {Model selection and multimodel inference: a practical information-theoretic approach},
  edition   = {2},
  year      = {2002},
  publisher = {Springer-Verlag},
  address   = {New York},
  note      = {Chapter~2.7}
}

@article{Akaike1974,
  author  = {Akaike, Hirotugu},
  title   = {A new look at the statistical model identification},
  journal = {IEEE Transactions on Automatic Control},
  year    = {1974},
  volume  = {19},
  number  = {6},
  pages   = {716--723}
}

@inproceedings{HM1970,
	Author = {Hoffman, D. and Maclean, W. M.},
	Booktitle = {Mar. Technol.},
	Title = {{Ship model study of incidence of shipping water forward}},
	Year = {1970}}

@article{HLG1993,
	Author = {Hong, S. Y. and Lee, P. M. and Gong, D. S.},
	Journal = {Sel. Pap. Soc. Nav. Archit. Korea},
	Pages = {37--44},
	Title = {{Experimental study on the deck wetting of a container ship in irregular head waves}},
	Volume = {1},
	Year = {1993}}

@inproceedings{OMTKMH2002,
	Address = {Kitayushu, Japan},
	Author = {Ogawa, Y. and Minumi, M. and Tunizuwu, K. and Kumuno, A. and Mutsunami, R. and Huyushi, T.},
	Booktitle = {12\textsuperscript{th} ISOPE Conf.},
	Publisher = {Int. Soc. of Offshore and Polar Eng. (ISOPE)},
	Title = {{Shipping Water Load due to Deck Wetness}},
	Year = {2002}}

@inproceedings{O1964a,
	Address = {Bergen, Norway},
	Author = {Ochi, M. K.},
	Booktitle = {5\textsuperscript{th} Symp. Nav. Hydrodyn.},
	Publisher = {Office of Naval Research},
	Title = {{Prediction of occurrence and severity of ship slamming at sea ACR112}},
	Year = {1964}}

@inproceedings{VV2014,
    address = {San Francisco, USA},
    author = {{van 't Veer}, A. P. and Vlasveld, E.},
    booktitle = {33\textsuperscript{rd} OMAE Conf.},
    doi = {10.1115/OMAE2014-23915},
    publisher = {ASME},
    title = {{Green water phenomena on a twin-hull FLNG concept}},
    year = {1024}}

@book{N2006,
  author    = {Nelsen, Roger B.},
  title     = {An introduction to copulas},
  series    = {Springer Series in Statistics},
  edition   = {2nd},
  publisher = {Springer-Verlag},
  address   = {New York, NY, USA},
  year      = {2006},
  isbn      = {9780387286594},
  doi       = {10.1007/0-387-28678-0}}

@article{GF2007,
  author  = {Genest, Christian and Favre, Anne-Catherine},
  title   = {Everything you always wanted to know about copula modeling but were afraid to ask},
  journal = {J. Hydrologic Eng.},
  volume  = {12},
  number  = {4},
  pages   = {347--368},
  year    = {2007},
  doi     = {10.1061/(ASCE)1084-0699(2007)12:4(347)}}

@article{HM2007,
	Author = {Hermundstad, O. A. and Moan, T.},
	Doi = {10.1007/s00773-006-0238-1},
	Issn = {0948-4280},
	Journal = {J. Mar. Sci. Technol.},
	Number = {3},
	Pages = {160--182},
	Title = {{Efficient calculation of slamming pressures on ships in irregular seas}},
	Volume = {12},
	Year = {2007}}

\end{document}